\documentclass[12pt,a4paper]{article}
\usepackage{amsmath,amsfonts,epsfig}
\usepackage{amssymb}
\usepackage{mathrsfs}
\usepackage{hyperref}
\newcommand{\beq}{\begin{eqnarray}}
\newcommand{\eeq}{\end{eqnarray}}

\newcommand{\be}{\begin{equation}}
\newcommand{\ee}{\end{equation}}

\newcommand{\bm}{\begin{multline}}
\newcommand{\fm}{\end{multline}}

\begin{document}
\numberwithin{equation}{section}
\setlength{\unitlength}{.8mm}

\begin{titlepage} 
\vspace*{0.5cm}
\begin{center}
{\Large\bf Strong coupling results from the numerical solution of the quantum spectral curve}
\end{center}
\vspace{1.5cm}
\begin{center}
{\large \'Arp\'ad Heged\H us, J\'ozsef Konczer}
\end{center}
\bigskip

\vspace{0.1cm}

\begin{center}
Wigner Research Centre for Physics,\\
H-1525 Budapest 114, P.O.B. 49, Hungary\\ 
\end{center}
\vspace{1.5cm}
\begin{abstract}
In this paper, we solved numerically the Quantum Spectral Curve (QSC) equations 
corresponding to some twist-2 single trace
operators with even spin from the $sl(2)$ sector of $AdS_5/CFT_4$ correspondence.
We describe all technical details of the numerical method which are necessary to 
implement it in C++ language. 

In the $S=2,4,6,8$ cases, our numerical results confirm the analytical 
results, known in the literature for the first 4 coefficients of the strong coupling expansion 
for the anomalous dimensions of twist-2 operators. 
In the case of the Konishi operator, due to the high precision of the numerical data we could
give numerical predictions to the values of two further coefficients, as well.

The  strong coupling behaviour of the coefficients $c_{a,n}$ in the power series representation 
of the ${\bf P}_{\!a}$-functions is also investigated. Based on our numerical data, 
in the regime, where the index of the coefficients is much smaller than $\lambda^{1/4}$,
we conjecture that the coefficients have polynomial index dependence at strong coupling.
This allows one to propose a strong coupling series representation
for the ${\bf P}$-functions being valid far enough from the real short cut.
In the paper 
the qualitative 
strong coupling behaviour of the ${\bf P}$-functions at the branch points is also discussed.
\end{abstract}

\end{titlepage}

\section{Introduction}

 Maldacena's famous AdS/CFT correspondence \cite{adscft1,adscft2,adscft3} is the best elaborated holographic 
 duality conjecture between gauge and string theories. The discovery of integrability on both sides of the
correspondence \cite{Beisert:2010jr}, created a hope to find the exact solution of the theory in the planar limit. 
The mathematical apparatus offered by integrability, proved to be the most efficient in computing
the planar spectrum of anomalous dimensions/string energies. In the large volume limit the spectrum\footnote{In
 this context large volume means: long single trace operators in the super Yang-Mills (SYM) side or equivalently
string states with large $J$-charge in $S^5$.}
was described by the Asymptotic Bethe Ansatz (ABA) equations \cite{Beisert:2005fw} which account for all power-like
corrections in volume, but neglects the exponentially small wrapping corrections. The wrapping corrections \cite{wrapping}
were taken into account by the so-called L\"uscher-formulae \cite{BJ1,BJL,BJHL,BFLZ,BJ3} which are now available
up to the second order in wrapping \cite{defL2,BomL2}. The Thermodynamic Bethe Ansatz (TBA) technique was the first
method which could sum up all wrapping corrections to the ABA in the form of a set of infinite component
nonlinear integral equations \cite{Gromov:2009tv,Bombardelli:2009ns,Arutyunov:2009ur,Gromov:2009bc,AFSe,Cavaglia:2010nm}. Though the TBA equations could provide important results, both in the
weak \cite{AFS,BH10v1,BH10v2} and in the strong \cite{Gromov:2009zb,Frolov:2010wt,Frolov:2012zv,Gromov:2011bz}
 coupling 
regimes\footnote{The strong coupling results came from fitting the
data from the numerical solution of the equations.}, its analytical and numerical treatment proved to be
tedious, due to the cumbersome kernels and the infinite number of unknown functions. Later the FiNLIE method \cite{GKLV10},
which can be considered as an improved finite version of the TBA, allowed one to reach better results in the
perturbative regime \cite{LSV,LV}, but the structure of the equations was still
so complicated that it required reasonable human effort to reach higher and higher orders in the perturbative
regime.

Recently the spectral problem of AdS/CFT (or equivalently the TBA) was reformulated as a nonlinear Riemann-Hilbert
problem for a few unknown functions. The new formulation is called the Quantum Spectral Curve 
(QSC) or ${\bf P}\mu$-system \cite{PmuPRL,PmuLong}.
The efficiency of the QSC method was demonstrated by numerous remarkable analytical and numerical results, the
computation of which seemed to be hopeless in the framework of TBA.

First of all, QSC made it possible to reach in principle arbitrarily high orders in the perturbative regime. In \cite{Marboe:2014gma,Marboe:2014sya}
even 10-loop analytical results were obtained for some operators in the $sl(2)$ sector.
QSC was powerful to get analytical results also in the near-BPS regimes \cite{PmuPRL,Gromov:2014bva}. In \cite{Gromov:2014bva} analytical next-to leading
order results were obtained in the small spin expansion for the anomalous dimensions of twist operators in the $sl(2)$
sector, providing also analytical predictions for the strong coupling expansion coefficients of the anomalous dimensions for some local operators and for the BFKL pomeron intercept. In \cite{Alfimov:2014bwa} leading order BFKL equation was derived by
performing the $S \to -1$ analytical continuation.

Later, in \cite{QSCnum} an efficient  numerical algorithm was proposed for solving the ${\bf P}\mu$-system and it was used
to confirm 2 previously known and to predict several previously unknown coefficients in the weak coupling expansion of the BFKL pomeron intercept.

Recently, analytical expression was obtained for the next-to-next-to leading order of the BFKL pomeron eigenvalue in \cite{QSCnlnlBFKL}, 
and the QSC description of cusped Wilson-lines \cite{QSCcwl} and of the quark-anti-quark potential \cite{QSCqq} were worked out.

In this paper we consider twist-2 operators with even positive integer spin. Using the numerical method of \cite{QSCnum},
we perform the numerical solution of the ${\bf P}\mu$-system for the twist-2 states with $S=2,4,6,8$ in a wide range of the 
t'Hooft coupling. 

Though analytical strong coupling results are available in the literature for the anomalous dimensions of the states
under consideration, they come from small spin results matched with classical and quasi classical string-theory
 results \cite{Gromov:2014bva} and not directly from the strong coupling solution of the
${\bf P}\mu$-system. This is why the aim of the paper is to gain a deeper insight into the strong coupling behaviour of the
solutions of the ${\bf P}\mu$-system.

In the $S=2,4,6,8$ cases, our accurate numerical results confirmed the analytical predictions of 
\cite{Gromov:2014bva} for the first 4 coefficients of the strong coupling expansion for $\Delta$.
In the case of the Konishi operator, due to the high precision of the numerical data, we could
give numerical predictions to the values of two further coefficients.

Beyond the numerical investigation of the anomalous dimensions, we investigated numerically the 
strong coupling behaviour of the coefficients $c_{a,n}$ in the power series representation 
of the ${\bf P}_a$-functions.
Based on our high precision numerical data, in the regime, where the index of the coefficients is
much smaller than $\lambda^{1/4}$,
we conjectured that the coefficients have polynomial index dependence at strong coupling. 
This allowed us to propose a strong coupling series representation
for the ${\bf P}_a$-functions being valid far enough from the real short cut.
To get some insight into the behaviour of ${\bf P}_a$ close to the real branch cut,
we also investigated 
the qualitative strong coupling behaviour of the ${\bf P}$-functions at the branch points.

The paper is organized as follows:
In sections 2. and 3. we recall the ${\bf P}\mu$-  and ${\bf Q}\omega$-descriptions 
of the states under consideration and explain, how the
free parameters coming from the symmetries of the QSC are fixed. The next section contains the detailed
description of the numerical method together with all necessary technical subtleties which make it possible
to implement the numerical code in C++ programming language. The analysis of the numerical data is presented
in sections 5. and 6. The paper is closed by the summary of our results. Some technical details of the numerical
 method and some tables of numerical data are placed into the appendices of the paper.

\section{Preliminaries}

In this paper adapting the method of \cite{QSCnum}, we solve numerically the QSC equations for some twist-2
operators in the $sl(2)$-sector of the theory. The corresponding operators can be schematically represented as:
\begin{equation}
{\mathcal O}=\mbox{Tr}(D^S \, Z^L)+\dots,
\end{equation}
where $Z$ is a complex scalar field of the theory, $D$ denotes the light-cone covariant derivative, $L$ is the twist,
and $S$ is the spin of the state.
Here we investigate the case when $L=2$ and $S$, the spin of the state, is even. The reason for this choice 
is to avoid 
treating null vectors in the internal linear problems of the numerical method (See remark at the end of subsection \ref{4pont2}).

 So that we could use the high order perturbative results of
 \cite{Marboe:2014gma} as initial values for the numerical iterative algorithm, 
we parametrized the ${\bf P}$-functions
and fixed the symmetries of the ${\bf P}\mu$-system in the same way as it was done in \cite{Marboe:2014gma}. 

Now, we recall the most necessary equations and relations of the QSC framework. 
The QSC method \cite{PmuPRL,PmuLong} describes the full planar spectrum of $AdS_5/CFT_4$ by the solutions of a set of nonlinear Riemann-Hilbert
equations. The fundamental objects of QSC are the eight $\bf P$- and ${\bf Q}$-functions which separately form a basis on the $2^8$ element
of the Q-system of $AdS_5/CFT_4$. 
In the $sl(2)$ sector, due to the left-right symmetry of the T-hook, 
one can describe the whole $Q$-system by only four ${\bf P}_a, a=1,..,4$ 
or four ${\bf Q}_i, i=1,..4$-functions, such that
the other four (upper indexed) components are simple linear combinations of them:
\begin{eqnarray} \label{chiP}
{\bf P}^a=\chi^{a b} \,{\bf P}_b, \qquad {\bf P}^a \,{\bf P}_a=0, \qquad a=1,...,4  \label{chiP}\\
{\bf Q}^i=-\chi^{i j} \,{\bf Q}_j, \qquad {\bf Q}^i \,{\bf Q}_i=0, \qquad i=1,...,4,  \label{chiQ}
\end{eqnarray}
 where $\chi$ is a constant matrix:
 
 {\begin{equation} \label{chi}
\chi=\left(
\begin{array}{cccc}
0 & 0 & 0 & -1 \\
0 & 0 & 1 & 0  \\
0 & -1 & 0 & 0  \\
1 & 0 & 0 & 0 
\end{array}
\right).
\end{equation} }
 The ${\bf P}_a$ and ${\bf Q}_i$ functions are analytic in the spectral parameter $u$ with branch cuts.
 The positions of the branch points depend on the 't Hooft coupling: $\lambda$ and they may be located at $u=\pm 2g+i {\mathbb Z}$, 
 where $g=\frac{\sqrt{\lambda}}{4 \pi}$. All branch points are assumed to be of square root type. This means that,
 the result of two subsequent analytical continuations around a branch point is an identity transformation.
 The advantage of the choice of ${\bf P}_a$s or ${\bf Q}_i$s as basis is their very simple discontinuity structure.
 On the complex $u$-plane, ${\bf P}_a$  has a single short cut, 
 while ${\bf Q}_i$ has only a single long cut, such that the discontinuities lie on the real axis.

\subsection{The ${\bf P}\mu$-system and the H-symmetry fixing}

Since the states we study lie in the left-right symmetric $sl(2)$ sector of the theory, we specify the presentation of the Riemann-Hilbert equations of the QSC for this sector.
For any function $f(u)$, denote $\tilde{f}(u)$ the analytical continuation around the branch point $\pm 2g$ and for 
short $f^{[\pm n]}(u)$ stands for $f(u\pm i \frac{n}{2})$. 
Then the ${\bf P}\mu$-equations take the form \cite{PmuPRL}: 
\begin{eqnarray}
\mu_{ab}-\tilde{\mu}_{ab}=\tilde{\bf P}_a \, {\bf P}_b-\tilde{\bf P}_b \, {\bf P}_a, \label{qsc1}\\
\tilde{\bf P}_a=(\mu \chi)_a^{\, \, \,b} \, {\bf P}_b, \label{qsc2} \\
\tilde{\mu}_{ab}=\mu^{[2]}_{ab}, \label{qsc3}
\end{eqnarray}
where $\mu_{ab}=-\mu_{ba}$ and $(\mu \chi)_a^{\, \, \,b}=\mu_{ac} \chi^{cb}$.
The equations are valid in the strip $0<\mbox{Im} u<1$, and elsewhere by their analytical continuations.
In this representation $\mu_{ab}$ has infinitely many short cuts and as a consequence of (\ref{qsc1}-\ref{qsc3}),
it satisfies the Pfaffian-relation:
\begin{equation}
\mbox{Pf}(\mu)\equiv \mu_{12} \mu_{34}-\mu_{13} \mu_{24}+\mu_{14} \mu_{23}=1.
\end{equation}
In the $sl(2)$ sector $\mu_{14}=\mu_{23}$. 
For twist-$L$ states, the large $u$ behaviour of ${\bf P}_a$ and $\mu_{ab}$ is fixed to  \cite{PmuPRL}:
\begin{eqnarray}\label{asympt1}
{\bf P}_1 \simeq A_1 \, u^{-\frac{L+2}{2}}, \,\, \ \ {\bf P}_2 \simeq A_2 \, u^{-\frac{L}{2}}, \,\, \ \ {\bf P}_3 \simeq A_3 \, u^{\frac{L-2}{2}}, \,\, \ \
{\bf P}_4 \simeq A_4 \, u^{\frac{L}{2}}, \; \nonumber \\
\mu_{12} \sim u^{\Delta-L}, \,\, \mu_{13} \sim u^{\Delta-1}, \,\, \mu_{14}=\mu_{23} \sim u^{\Delta}, \,\, \mu_{24} \sim u^{\Delta+1},  \,\, \mu_{34} \sim u^{\Delta+L},
\end{eqnarray}
where $S$ is the spin of the state and $\Delta$ is its conformal dimension. In addition the  prefactors are constrained by the relations:
\begin{eqnarray}\label{AA}
A_1A_4&=& \frac{[(L-S+2)^2-\Delta^2] [(L+S)^2-\Delta^2]}{16i L (L+1)}\,,  \nonumber  \\
A_2A_3&=& \frac{[(L+S-2)^2-\Delta^2] [(L-S)^2-\Delta^2]}{16iL (L-1) }\,.
\end{eqnarray}
Following the lines of \cite{Marboe:2014gma} we also introduce the ${\bf p}_a$ functions by a rescaling of the original ${\bf P}_a$s;
\begin{equation}\label{smallP}
{\bf p}_a \equiv (g \, x)^{\frac{L}{2}} \, {\bf P}_a. 
\end{equation}
Here $x\equiv x_s(u/g)$, where
\begin{equation} \label{xs}
x_s(u)=\frac{u}{2}\left( 1+\sqrt{1-\frac{4}{u^2}}\right), \qquad |x_s(u)|>1,
\end{equation}
is the short cut solution of the equation $x+\frac{1}{x}=u$.
By the introduction of ${\bf p}_a$, the sign ambiguity arising in the cases of odd $L$ can be eliminated.
In addition to the previously listed equations and properties, analyticity constraints are also imposed on the
possible solutions of (\ref{qsc1}-\ref{qsc3}). Namely, in the QSC formulation of the spectral problem of $AdS_4/CFT_5$
correspondence, it is postulated \cite{PmuPRL} that  ${\bf P}_a$ and $\mu_{ab}$ have no poles on the first sheet and their
absolute value is bounded at the branch points.

The ${\bf P}\mu$-system (\ref{qsc1}-\ref{qsc3}) is invariant under the linear redefinitions (H-symmetry \cite{Marboe:2014gma}):
\begin{equation}
{\bf P}_a\to H_{a}{}^{b}\,{\bf P}_b\,,\ \ \mu_{ab}\to H_{a}{}^{c}H_{b}{}^{d}\mu_{cd}\,,\ \ \chi^{ab}\to \chi^{cd}(H^{-1})_{c}{}^{a}(H^{-1})_{d}{}^{b}\,,
\end{equation}
where $H$ is a constant matrix with $\mbox{det}H=1$. In principle $H$ might have 15 components, but if one
would like to preserve the prescriptions (\ref{asympt1}) for the large $u$ asymptotics, then only 6 non-zero
elements remain to be fixed. These elements can be fixed by fixing the values of $A_1$ and $A_2$ and by imposing
the value of 4 other coefficients in the large $u$ expansion of ${\bf p}_a$. 
In our numerical framework, we used the  H-symmetry fixing conditions of \cite{Marboe:2014gma}. 
The requirements are as follows:
\begin{itemize}
\item $A_1\equiv g^2$ and $A_2 \equiv 1$,
\item ${\bf p}_2$ has no term proportional to $u^{-1}$ in its large $u$ expansion,
\item ${\bf p}_3$ has no term proportional to $u^{0}$ in its large $u$ expansion,
\item ${\bf p}_4$ has no terms proportional to $u^{0}$ and $u^{-1}$ in its large $u$ expansion.
\end{itemize}
We used this H-symmetry fixing scheme, so that we could use the high order perturbative results of 
\cite{Marboe:2014gma} as initial values for our numerical iterative algorithm.
Nevertheless, since we study left-right symmetric states, also parity symmetries can be imposed on the
first sheet.
For the twist-2 case, we required that on the first sheet:
\begin{itemize}
\item ${\bf P}_1$ is even and real{\footnote {Here we call $f$ real, if $f(u)^*=f(u^*).$}} function of $u$.
\item ${\bf P}_2$ is odd and real function of $u$.
\item ${\bf P}_3$ is even and imaginary{\footnote {Here we call $f$ imaginary, if $f(u)^*=-f(u^*).$}} function of $u$.
\item ${\bf P}_4$ is odd and imaginary function of $u$.
\end{itemize}
These conditions allow us to use the following series representations for the 
${\bf p}_a$-functions at $L=2$:
\begin{equation}\label{p12}
{\bf p}_1=\frac{g}{x}+\sum\limits_{n=1}^{\infty} \, \frac{c_{1,n}}{x^{2n+1}}, \qquad
{\bf p}_2=1+\sum\limits_{n=1}^{\infty} \, \frac{c_{2,n}}{x^{2n}},
\end{equation}
\begin{equation}\label{p34}
{\bf p}_3=A_3 \, u+\sum\limits_{n=0}^{\infty} \, \frac{c_{3,n}}{x^{2n+1}}, \qquad
{\bf p}_4=A_4 \, u^2 +\sum\limits_{n=1}^{\infty} \, \frac{c_{4,n}}{x^{2n}}.
\end{equation}
The coefficients $c_{a,n}$ are functions of the coupling constant $g$. In our case $c_{1,n}$ and $c_{2,n}$ are real,
while $c_{3,n}$ and $c_{3,n}$ are pure imaginary{\footnote{We note that in accordance with the H-symmetry fixing conditions
 and (\ref{smallP}), (\ref{p12}), (\ref{p34}), by definition $c_{1,0}\equiv g, \quad c_{2,0}\equiv 1, \quad c_{4,0}\equiv 0$. }}.
In (\ref{p12}) the leading terms of the $1/x$ expansion are fixed by the H-symmetry fixing conditions $A_1=g^2$ and
$A_2=1$. In (\ref{p34}) $A_3$ and $A_4$ are considered as functions of $\Delta$ and $g$, if we express them
by the fixed $A_1=g^2$ and $A_2=1$ coefficients through (\ref{AA}). 
These series representations automatically satisfy all the symmetry requirements discussed above and converge
on the entire $u$ plane \cite{QSCnum}. The radius of convergence in $1/x$  is $R=|x_s(2+\frac{i}{g})|$.
As a consequence $\tilde{\bf P}_a$ can also be represented by the analytical continuation ($x \to 1/x$) of the
series (\ref{p12}) and (\ref{p34}), but it is not convergent on the entire $u$ plane. Its convergence is
restricted to a oval domain lying around the real short cut of ${\bf p}_a$ \cite{QSCnum}. 

Thus, the parameters to be determined by the numerical solution of the ${\bf P}\mu$-system are as follows:
\begin{itemize}
\item The coefficients: $c_{1,n}, \qquad n=1,..., \qquad c_{1,n}\in {\mathbb R} $,
\item The coefficients: $c_{2,n}, \qquad n=1,...,  \qquad c_{2,n}\in {\mathbb R}$,
\item The coefficients: $c_{3,n}, \qquad n=0,...,  \qquad c_{3,n}\in i\,{\mathbb R}$,
\item The coefficients: $c_{4,n}, \qquad n=1,...,  \qquad c_{4,n}\in i\,{\mathbb R}$,
\item The anomalous dimension: $\Delta \in {\mathbb R}$.
\end{itemize}
In the numerical solution, ${\bf p}_a$s are represented as truncated versions of (\ref{p12}) and
(\ref{p34}), thus only a finite number of coefficients are to be determined.

The concrete numerical solution of QSC \cite{QSCnum} is implemented through the ${\bf P}\omega$-system. 
This means that starting from the ${\bf P}_a$ functions, one should determine the ${\bf Q}_i$
functions of the ${\bf Q}\omega$-system and the coefficients are determined from the
discontinuity equations of the ${\bf Q}\omega$-system. To do so, we have to recall the
${\bf Q}\omega$-system and its relation to the ${\bf P}\mu$-system.

\section{The ${\bf Q}\omega$-system and its relation to the ${\bf P}\mu$-system}

The nonlinear Riemann-Hilbert equations for the ${\bf Q}\omega$-system are very similar to
those of the ${\bf P}\mu$-system \cite{PmuLong}:

\begin{eqnarray}
\omega_{ij}-\tilde{\omega}_{ij}=\tilde{\bf Q}_i \, {\bf Q}_j-\tilde{\bf Q}_j \, {\bf Q}_i, \label{Qsc1}\\
\tilde{\bf Q}_i=-(\omega \chi)_i^{\, \, \,j} \, {\bf Q}_j, \label{Qsc2} \\
{\omega}_{ij}=\omega^{[2]}_{ij}, \label{Qsc3}
\end{eqnarray}
where $\omega_{ij}=-\omega_{ji}$ and $(\omega \chi)_i^{\, \, \,j}=\omega_{ik} \chi^{kj}$.
The equations are valid in the strip $0<\mbox{Im} u<1$, and elsewhere by their analytical continuations.
In this representation $\omega_{ij}$ has infinitely many short cuts and as a consequence of (\ref{Qsc1}-\ref{Qsc3}),
it satisfies the Pfaffian-relation:
\begin{equation}
\mbox{Pf}(\omega)\equiv \omega_{12} \omega_{34}-\omega_{13} \omega_{24}+\omega_{14} \omega_{23}=1.
\end{equation}
In the $sl(2)$ sector $\omega_{14}=\omega_{23}$. For large $u$, $\omega_{ij}$ tends to a constant and the
large $u$ asymptotics of ${\bf Q}_i$ is governed by the global charges of $AdS_5$ \cite{PmuLong}:
\begin{equation}\label{asympt2}
{\bf Q}_i\sim(B_1 \,u^{\frac{\Delta-S}{2}},B_2 \,u^{\frac{\Delta+S-2}{2}}, B_3 \,u^{-\frac{\Delta+S}{2}}, B_4\, u^{\frac{-\Delta+S-2}{2}}).
\end{equation}
In the $sl(2)$-sector, the prefactors $B_i$ satisfy an equation similar to (\ref{AA}):
\begin{equation}\label{BB}
\begin{split}
B_1 B_4\! &= \! \frac{i \, (-2+L+S-\Delta)(L+S-\Delta)(L-S+\Delta)(2+L-S+\Delta)}
{16 (-1+S) \Delta (1-S+\Delta)} \,,   \\
B_2 B_3 \!&= \! \frac{i \, (-2 - L + S + \Delta) (-L + S + \Delta) (-2 + L +
      S +\Delta) (L + S + \Delta)}{16 (-1 + 
     S) \Delta (-1 + S + \Delta)} \,.  
\end{split}
\end{equation}
This means that fixing two of the coefficients $B_i$ is in our hand.
For the sake of brevity, we introduce the vectors \cite{QSCnum}:
\begin{equation}\label{Mt}
\tilde{M}_a=\left\{ \frac{L}{2}+1, \frac{L}{2}, -\frac{L}{2}+1, -\frac{L}{2}  \right\},
\end{equation}
\begin{equation}\label{Mh}
\hat{M}_i=\left\{ \frac{\Delta-S}{2}+1, \frac{\Delta+S}{2}, -\frac{\Delta+S}{2}+1, \frac{-\Delta+S}{2}  \right\}.
\end{equation}
Then the large $u$ asymptotics can be given by the short formulae:
\begin{equation}\label{PQasympt}
{\bf P}_a\sim A_a u^{-\tilde M_a}, \ \ {\bf Q}_i\sim B_iu^{\hat M_i-1}, \ {\bf P}^a\sim A^au^{\tilde M_a-1},\
	{\bf Q}^i\sim B^iu^{-\hat M_i}.
\end{equation}
The ${\bf Q}$-functions can be constructed from the ${\bf P}$-functions in the following way.
First, one should find 16 upper half plane analytic functions ${\cal Q}_{a|i}$ as 
 solutions of a set of homogeneous linear difference equations:
\begin{equation}\label{Qai}
{\cal Q}_{a|i}(u+\tfrac i2)-{\cal Q}_{a|i}(u-\tfrac i2)=-
{\bf P}_a(u)
{\bf P}^b(u) \,
{\cal Q}_{b|i}(u+\tfrac i2)\ \qquad a,i\in\{1,2,3,4 \}.
\end{equation}
The index $i$ of ${\cal Q}_{a|i}$ labels the 4 linearly independent solutions of (\ref{Qai}).
Then the ${\bf Q}$-functions are defined by the formula: 
\begin{equation}\label{Qidef}
{\bf Q}_i(u)=-{\bf P}^a(u)\;{\cal Q}_{a|i}(u+i/2)\ \qquad \mbox{Im} u>0 .
\end{equation}
Since ${\cal Q}_{a|i}$ is upper half plane analytic, the determination of $\tilde{\bf Q}_i$ 
is simple:
\begin{equation}\label{Qitilde}
\tilde{\bf Q}_i(u)=-\tilde{\bf P}^a(u)\,{\cal Q}_{a|i}(u+i/2) .
\end{equation}
As a consequence, (\ref{Qai}) can be rephrased as follows:
\begin{equation}\label{QaiPQ}
{\cal Q}_{a|i}(u+\tfrac i2)-{\cal Q}_{a|i}(u-\tfrac i2)=
{\bf P}_a(u) \, {\bf Q}_i(u).
\end{equation} 
From this equation the leading order large $u$ behaviour of ${\cal Q}_{a|i}$ can be
determined \cite{PmuLong}:
\begin{equation}
{\cal Q}_{a|i} \simeq B_{a|i} \, u^{-\tilde{M}_a+\hat{M}_i} ,  
\qquad B_{a|i}= \frac{-i \, A_a \, B_i \,}{-\tilde{M}_a+\hat{M}_i}.
\end{equation}

\subsection{The brief description of the numerical method}

The strategy of the numerical method is as follows \cite{QSCnum}.
One starts from the series representations (\ref{p12},\ref{p34}) of ${\bf P}_a$ and the goal is
to compute numerically $\Delta$ and those coefficients of the series, which are left undetermined
after fixing the symmetries of QSC. 

Then from the representations (\ref{p12},\ref{p34}), $\tilde{\bf P}_a$ can be determined by an $x \to 1/x$
transformation. This representation of  $\tilde{\bf P}_a$ is convergent in an oval shaped region containing
entirely the branch cut on the real axis.

The next step is to solve the recursion for ${\cal Q}_{a|i}$. This is done in two steps:
first it is solved in the large $u$ limit, and then the recurrence relations (\ref{Qai})
are used to pull back the solution to the real axis.
Then ${\bf Q}_i$ and $\tilde{\bf Q}_i$ are constructed from (\ref{Qidef},\ref{Qitilde}). 

In order to exploit the ${\bf Q}\omega$-equations, one has to determine $\omega_{ij}$, as well.
 It is computed from ${\bf Q}_i$ and $\tilde{\bf Q}_i$ by an integral expression derived
from $(\ref{Qsc1})$ and $(\ref{Qsc3})$ (See (\ref{omega}) later).

All the quantities computed so far, are considered as functions of $\Delta$ and the unknown coefficients of the series (\ref{p12},\ref{p34}). This discrete set of variables is determined by imposing the equations (\ref{Qsc2}).

In practice the whole process goes iteratively. One starts from a "good" approximation for the unknown 
coefficients and $\Delta$, and goes through the steps discussed above.
 By the solution of (\ref{Qsc2}), one gets the new initial values for the unknowns and the procedure is repeated
until convergent result is obtained.

In the next section we describe the numerical method in detail, this is why the reader, who is interested in only
the numerical results, might skip the next section.

\section{The numerical method}

In this section we describe our implementation of the numerical solution of QSC equations.
We try to write down all important details and subtleties, in order to give help to
those, who would like to solve numerically QSC equations in a fundamental programming language
 like C++ or Fortran. The technical details, 
we are going to write down, help to reduce each step of the numerical method to 
solving linear equations and to summations.
The numerical implementation of these two simple mathematical problems is quite 
straightforward in any fundamental programming language.

\subsection{Initial values and the discretization}

In the previous section we described the set of unknown coefficients to be determined
by the numerical method. The H-symmetry of the ${\bf P}\mu$-system was partly fixed by fixing the values
 of $A_1=g^2$ and $A_2=1$. Then $A_3$ and $A_4$ are given by (\ref{AA}) and they depend on $A_1,A_2$ and
$\Delta$, provided $L$ and $S$  are fixed previously. As we mentioned, this choice of
H-symmetry fixing was made to be able to use the perturbative results of \cite{Marboe:2014gma} as initial values. 
Thus, for the twist-2 states with even $S$, in the weak coupling regime, where $g\lesssim \frac14$, 
we used the six-loop perturbative results of \cite{Marboe:2014gma} for the unknowns as initial values for the iterations.
According to our experience beyond the radius of convergence of the perturbative series (i.e. $g=1/4$), the  
perturbative results were not good initial values for the iterations anymore. For $\tfrac14 \lesssim g $,
 the numerical method failed to converge if we used the high loop perturbative results of \cite{Marboe:2014gma}
as initial values. 
For higher values of the coupling constant $g$, the initial values of the unknowns should be made out of
the numerical data belonging to smaller values of $g$. This means that beyond $g \simeq 1/4$, one should increase $g$
in small steps, and the initial values should be determined as appropriate compositions of the previously computed data.
In our concrete numerical studies, we increased $g$ with $\Delta g=0.1,0.05,0.02,0.01$ and the initial values
were given by a 4, 5, or 6 order Taylor-series composed of the previously computed numerical data. This construction
of initial values is given in appendix \ref{appA}.

Since the numerical method uses also the ${\bf Q}\omega$-system, we have further freedom to fix 2 of the coefficients $B_i$.
We fixed the values of $B_1$ and $B_2$, then $B_3$ and $B_4$ are completely determined by (\ref{BB}).
For the sake of simplicity, for small $g$ we used the choice:
\begin{equation} \label{B12ch}
B_1=1, \qquad B_2=1.
\end{equation}
For higher values of $g$, the choice of these coefficients play important role in the convergence of the numerical 
algorithm. Our experience suggests decreasing their values as $g$ is increased. For example, in case of the
Konishi operator ($S=2$) the $B_1=B_2=1/g^2$ choice was necessary{\footnote{In case one insisted on not decreasing
$\Delta g$ below $0.05$.}} to reach satisfying convergence in the regime $g>2$.

So far we explained, how to fix the "free" coefficients and how to construct good initial values for the iterative
numerical algorithm. The next step is to choose the discretization points for our functions. The final equation
(\ref{Qsc2}) is imposed on the short cut of the real axis, this is why we need to give an appropriate discretization
of the interval $[-2g,2g]$. The discretization should be dense enough to be able to compute the integral expressions
for $\omega_{ij}$ with high enough numerical precision. Since all functions in the QSC framework have square root-type
behaviour at the branch points, it is plausible to choose the discretization points as zeros of the
 Chebyshev-polynomials. The reason is that on the interval $[-1,1]$ the Chebyshev-polynomials of the second kind form
an orthonormal basis with respect to the square-root type weight function $\sqrt{1-u^2}$. 
A summary on the necessary properties and identities of the 
Chebyshev-polynomials is given in appendix \ref{appB}.  

In order to be able to use the advantages of formulae (\ref{bn}) and (\ref{us}),  
the discretization points are chosen to be the zeros of the
appropriately scaled{\footnote{Scaling means only a $u \to \frac{u}{2g}$ scaling of the argument, such that
the polynomial to be defined on $[-2g,2g]$ instead of the usual interval of definition $[-1,1]$.}} $l_c$th 
Chebyshev-polynomial of the first kind ($T_{l_c}(\frac{u}{2g})$). The integer number $l_c$ measures, how dense 
the discretization is.
Then the formula for our discretization points reads as{\footnote{The same set of discretization points were chosen in \cite{QSCcwl}.}}:
\begin{equation}
u_A=-2 \, g\, \cos\left( \frac{\pi \, (A-\frac12)}{l_c}\right), \, \qquad  T_{l_c}(\frac{u_A}{2g})=0, \qquad A=1,..,l_c. \end{equation}

\subsection{The determination of ${\cal Q}_{a|i}$}\label{4pont2}

The necessary values: ${\cal Q}_{a|i}(u_A+\tfrac i2), \quad A=,...,l_c$ are determined by (\ref{Qai}) in two steps.
In the first step, (\ref{Qai}) is solved analytically for large $u$ in the context of a $1/u$ expansion.  
One introduces an integer index cutoff $N_I$, such that the first $N_I$ terms of the $1/u$ series 
are computed. 
Then another integer truncation index $N_u$ is introduced, 
such that at the points $u'_A=u_A+i \,(N_u+\frac12)$, the series representation of ${\cal Q}_{a|i}$
truncated at $N_I$, should approximate ${\cal Q}_{a|i}(u'_A)$ within the required numerical accuracy.
Then, in the second step, the desired discrete values ${\cal Q}_{a|i}(u_A+\tfrac i2)$, are 
computed from ${\cal Q}_{a|i}(u'_A)$ by the successive application of
the recurrence relation (\ref{Qai}).

In the large $u$ regime the following series representations are used:
\begin{equation}\label{Qaisor}
{\cal Q}_{a|i}(u) \simeq B_{a|i} \, u^{-\hat{M}_i+\tilde{M}_a} \, \sum\limits_{n=0}^{\infty} \, \frac{b_{a|i,n}}{u^{2n}}, \qquad b_{a|i,0}\equiv 1,
\end{equation} 
\begin{equation}\label{Pasor}
{\bf P}_{a}(u) \simeq A_{a} \, u^{-\tilde{M}_a} \, \sum\limits_{n=0}^{\infty} \, \frac{k_{a,n}}{u^{2n}}, \qquad k_{a,0}\equiv 1,
\end{equation}
\begin{equation}\label{Pfasor}
{\bf P}^{a}(u) \simeq A^{a} \, u^{\tilde{M}_a-1} \, \sum\limits_{n=0}^{\infty} \, \frac{k^{\, \, \, a}_{n}}{u^{2n}}, \qquad k^{\, \, \, a}_{0}\equiv 1.
\end{equation}
As a consequence of the parity symmetries of ${\bf P}_a$, only even powers of $u$ appear in the sums. From (\ref{chiP})
it follows that: $A^a=\chi^{ab} \, A_b$ and $k^{\, \, \, a}_{n}=|\chi^{ab}| \, k_{b,n}$.
The relation among the coefficients of the $1/u$ (\ref{Pasor},\ref{Pfasor}) and the $1/x$ 
(\ref{p12},\ref{p34}) expansions can be computed by the $x \leftrightarrow u$ relation:
\begin{equation}\label{xun}
x^{-n}=\left(\frac{g}{u}\right)^{n} \, \sum\limits_{s=0}^{\infty} \, \kappa_{s}^{(n)} \,
\left(\frac{g}{u}\right)^{2 \, s},
\end{equation}
where
\begin{equation}\label{kappans}
\kappa^{(n)}_s=\left\{ \begin{array}{lll}
(-1)^{s+1} \,\frac{n}{s} \, \binom{n+2 \, s-1}{s-1} & n+2 \, s \leq 0  \quad 
\mbox{and} \quad n \neq 0, \\
\frac{n}{s} \, \binom{n+2 \, s-1}{s-1} &   n+2 \, s > 0, \\
\delta_{s,0} & n = 0.
\end{array} \right.
\end{equation}

Formulae (\ref{xun}) and (\ref{kappans}) are valid for non-integer values of $n$, as well. 
In the twist-2 case the concrete forms of the $k_{a,n} \leftrightarrow c_{a,n}$ relations read as follows:
 \begin{equation}\label{k1m}
k_{1,m}=\frac{g}{A_1} \, \sum\limits_{n=0}^m \, c_{1,n} \, (\sigma_1)_{n,m}, \qquad c_{1,0}\equiv g=\frac{A_1}{g},
\end{equation}
\begin{equation}\label{k2m}
k_{2,m}=\frac{1}{A_2} \, \sum\limits_{n=0}^m \, c_{2,n} \, (\sigma_2)_{n,m}, \qquad c_{2,0}\equiv 1=A_2,
\end{equation}
\begin{equation}\label{k3m}
k_{3,m}= \kappa^{(1)}_m \, g^{2m} \,+\,\Theta(m-1) \frac{g}{A_3} \, \sum\limits_{n=0}^{m-1} \, c_{3,n} \, (\sigma_1)_{n,m-1},
\end{equation}
\begin{equation}\label{k4m}
k_{4,m}= \kappa^{(1)}_m \, g^{2m} \,+\,\Theta(m-1) \frac{1}{A_4} \, \sum\limits_{n=0}^{m-1} \, c_{4,n} \, (\sigma_2)_{n,m-1}, \qquad c_{4,0}\equiv 0,
\end{equation}
where $\Theta$ is the unit-step function and
\begin{equation}
(\sigma_1)_{n,m}=g^{2 \, m}\, \sum\limits_{s=0}^{m-n} \, \kappa_s^{(1)} \, \kappa_{m-n-s}^{(2 \, n+1)},
\qquad 
(\sigma_2)_{n,m}=g^{2 \, m}\, \sum\limits_{s=0}^{m-n} \, \kappa_s^{(1)} \, \kappa_{m-n-s}^{(2 \, n)}.
\end{equation}
Substituting the series representations (\ref{Qaisor},\ref{Pasor},\ref{Pfasor}) into (\ref{Qai}), a coefficient
$b_{a|i,m}$ is determined by such a $4 \times 4$ linear problem, whose matrix ${\cal T}^{a\, b|i}_m$ depend only on 
$L,S,\Delta$, while its source vector ${\cal F}^{a|i}_m$ depends on $b_{a|i,m'}$ with $m'<m$. Starting with $m=1$, 
this fact allows the successive determination of $b_{a|i,m}$. The linear problem determining $b_{a|i,m}$ takes the form:
\begin{equation}\label{4x4lin}
\sum\limits_{b=1}^4 \,{\cal T}^{a\, b|i}_m \, b_{a|i,m}={\cal F}^{a|i}_m, \qquad m=1,2,...,N_I
\end{equation}
where 
\begin{equation}
{\cal T}^{a\, b|i}_m=A_a \, A^b \, B_{b|i}- i \, \delta_{ab} \, B_{a|i} \, (-\alpha_{a|i}+2 \, m), 
\end{equation}
with $\alpha_{a|i}=-\tilde{M}_a+\hat{M}_i$. The source term is the difference of two terms:
\begin{equation}
{\cal F}^{a|i}_m={\cal F}^{a|i}_{1,m}-{\cal F}^{a|i}_{2,m},
\end{equation}
with
\begin{equation}
\begin{split}
{\cal F}^{a|i}_{1,m}&=- i \,B_{a|i} \, \left\{ \binom{\alpha_{a|i}}{2 \, m+1} \, \left(\frac {-1}{4} \right)^{m}
+\sum\limits_{n=1}^{m-1} \, b_{a|i,n} \, \binom{-2\,n }{2 \, m-2 \, n +1} \, \left(\frac {-1}{4} \right)^{m-n} 
\right. \\
&+\sum\limits_{n=1}^{2 \, m-1} \, \binom{\alpha_{a|i}}{2 \, m+1-n} \, \sum\limits_{k=1}^{[n/2]} \,
b_{a|i,k} \, \binom{-2 \,k}{n-2 \, k} \, \left(\frac {-1}{4} \right)^{m-k} \\ 
&+\alpha_{a|i} \left. \, \sum\limits_{k=1}^{m-1} \, b_{a|i,k} \, \binom{-2 \, k}{2 \, m-2 \, k} \, 
\, \left(\frac {-1}{4} \right)^{m-k} \right\},
\end{split}
\end{equation}
\begin{equation}
\begin{split}
{\cal F}^{a|i}_{2,m}&=A_a\,A^b \, B_{b|i} \, 
\left\{ \sum\limits_{n=0}^{m} \, \, q_{m-n}^{ab}\left[  \binom{\alpha_{b|i}}{ 2\, n} \left(\frac{-1}{4}\right)^n 
+\sum\limits_{j=1}^{n-1} \, b_{b|i,j} \, \binom{-2 \,j}{2 \, n-2 \, j} \, \left(\frac{-1}{4}\right)^{n-j}
\right.   \right. \\
&+ 
\left. \left. \sum\limits_{k=1}^{2 \, n -1} \, \sum\limits_{j=1}^{[k/2]} \, b_{b|i,j} \, \binom{-2 \, j}{k-2 \, j} \,
\left(\frac{-1}{4}\right)^{n-j} \! \! \binom{\alpha_{b|i}}{2\, n-k} \right] + \sum\limits_{n=1}^{ m-1} \, b_{b|i,n} \, q_{m-n}^{ab} \right\},
\end{split}
\end{equation}
where 
\begin{equation}
q^{ab}_{n}=\sum\limits_{l=0}^{n} \, k_{a,n-l} \, k_{l}^{\, \, \, b} \qquad q^{ab}_{0}\equiv 1,
\end{equation}
and in the summation limits $[...]$ stands for integer part.
To avoid any confusion, we note that throughout the paper, in case the letter $i$ stands for an index, than 
it denotes a positive integer number running from 1 to 4. In any other cases it denotes the imaginary unit
i.e. $i^2=-1$.
The solution of (\ref{4x4lin}) for $m=1,..,N_I$ ,through (\ref{Qaisor}), gives a numerically accurate approximation of 
${\cal Q}_{a|i}(u_A+ i (N_u+ \tfrac 12))$. Then ${\cal Q}_{a|i}(u_A+\tfrac i2)$ is computed by the
successive application of the recurrence relation (\ref{Qai}):
\begin{equation}\label{QaiA}
{\cal Q}_{a|i}(u_A+\tfrac i2)=\left[ U(u_A+i) \, U(u_A+2 \, i)...U(u_A+i \, N_u)\right]_a{}^b \, {\cal Q}_{b|i}(u_A+ i (N_u+\tfrac 12)),
\end{equation}  
where the $4 \times 4$ matrix $U(u)$ is given by \cite{QSCnum}:
\begin{equation} \label{U}
U(u)_a{}^b=\delta_{a}{}^b+{\bf P}_a(u) \, {\bf P}^b(u).
\end{equation}
With the help of (\ref{Qidef}) and (\ref{Qitilde}) it is easy to determine ${\bf Q}_i$ and $\tilde{\bf Q}_i$
at the discretization points:
 \begin{equation}\label{QiA}
{\bf Q}_i(u_A)=-{\bf P}^a(u_A+i\, 0)\;{\cal Q}_{a|i}(u_A+i/2),
\end{equation}
\begin{equation}\label{QiAtilde}
\tilde{\bf Q}_i(u_A)=-\tilde{\bf P}^a(u_A+i \, 0)\,{\cal Q}_{a|i}(u_A+i/2) .
\end{equation}
The $+i \, 0$ prescription is to avoid the evaluation of functions on their branch cuts.
When one takes the series representations (\ref{p12},\ref{p34})
at $u_A+i \, 0$, it is better to use the mirror $x$, the long cut version of $x$,
 since it is regular in $[-2 g, \, 2 g]$:
\begin{equation} \label{xm}
x\to x_s((u_A+i \, 0)/g)=1/x_m(u_A/g), \quad \mbox{with} \quad x_m(u)=\tfrac u2 -\tfrac i2 \, \sqrt{4-u^2}.
\end{equation}

We close this subsection with a remark, which explains why we choose even integer values for $S$ in the numerical 
studies. The reason is that in case of left-right symmetric states: $\mbox{det} \, {\cal T}^{a\, b|i}_m \sim S\pm 2 \, m- 1$, which{\footnote{Here the sign $\pm$ means that for $i=3,4$ the $+$, and for $i=1,2$ the $-$ sign
should be meant.}} means that for odd values of $S$, one should take care of the zero modes of ${\cal T}^{a\, b|i}_m$.
This problem is absent in the even $S$ case.

\subsection{The computation of $\omega_{ij}$}

For the numerical algorithm we need to determine $\omega_{ij}$ at the positions $u_A+i \, 0$. From (\ref{Qsc1}) and (\ref{Qsc3})
the following integral representation can be derived \cite{QSCnum}:
\begin{equation}\label{omega}
\omega_{ij}(u)=\omega_{ij}^{(0)}(u)+\omega_{ij}^{c},
\end{equation}
where $\omega_{ij}^{(0)}$ accounts for the discontinuity relations and periodicity,
\begin{equation}\label{omega0}
\omega_{ij}^{(0)}(u)=\frac{i}{2} \, \int\limits_{-2 \,g}^{2 \, g} dv \, \coth\left[\pi \,(u-v) \right] \,
\left[ \tilde{{\bf Q}}_i(v) \, {\bf Q}_j(v) -{\bf Q}_i(v) \, \tilde{{\bf Q}}_j(v) \right],
\end{equation}
and $\omega_{ij}^{c}$ is a constant  matrix to fulfill (\ref{Qsc2}) close to infinity \cite{QSCnum}:
\begin{equation}\label{omegac}
\omega_{ij}^c=i \, I_{ij} \, \cot(\pi \, \hat{M}_j), \qquad I_{ij}=\frac{i}{2} \, \int\limits_{-2 \,g}^{2 \, g} dv  \,
\left[ \tilde{{\bf Q}}_i(v) \, {\bf Q}_j(v) -{\bf Q}_i(v) \, \tilde{{\bf Q}}_j(v) \right].
\end{equation}
In the $sl(2)$-sector, the antisymmetry of $\omega_{ij}^c$ is ensured by $I_{12}=I_{21}=I_{14}=I_{41}=I_{23}=I_{32}=I_{24}=I_{42}\equiv 0$.
In \cite{QSCnum}, it was explained that for numerical purposes, instead of using (\ref{Qsc2}) as a final equation to fix the unknown coefficients,
it is better to use a more regular version:
\begin{equation}\label{Qreg}
\tilde{\bf Q}_i(u)=\omega_{ij}^{reg}(u) \, {\bf Q}^j(u),
\end{equation}
where $\omega_{ij}^{reg}(u)=\frac{1}{2}(\omega_{ij}(u)+\tilde{\omega}_{ij}(u))$ has no branch cut along the real axis.
Our task is to compute $\omega_{ij}^{reg}(u_A), \quad A=1,...,l_c$ from the, so far computed, discrete set of ${\bf Q}_i(u_A)$ and 
$\tilde{\bf Q}_i(u_A)$.

The strategy goes as follows.
Since ${\bf Q}_i$ and $\tilde{\bf Q}_i$ are bounded at the branch points $\pm 2 g$, their antisymmetric combination
can be represented as:
\begin{equation}\label{QQrho}
\tilde{\bf Q}_i(u) \, {\bf Q}_j(u)-{\bf Q}_i(u) \, \tilde{\bf Q}_j(u)=\sqrt{4 \, g^2 - u^2} \, \rho_{ij}(u),
\qquad u \in [-2 g,2 g],
\end{equation}
where $\rho_{ij}(u)$ is a smooth bounded function on the real short cut. This allows one to represent $\rho_{ij}(u)$ as 
a convergent series with respect to some sequence of orthogonal polynomials. 

For practical purposes explained in appendices \ref{appB} and \ref{appC}, we choose 
the Chebyshev-polynomials of the second kind $U_n(\frac{u}{2g})$ as basis for this expansion:
\begin{equation}\label{rho-U}
\rho_{ij}(u)=\sum\limits_{n=0}^{\infty} \, a_{ij}^{(n)} \, U_n(\tfrac{u}{2g}).
\end{equation}
As a consequence of the convergence of this series, the coefficients quite fast tend to zero. Thus,
$\rho_{ij}$ can be computed very accurately from the appropriately truncated version of (\ref{rho-U}).
If the first $l_c$ terms are left from (\ref{rho-U}) after truncation,
than the coefficients can be computed from well known formulae for the Chebyshev-polynomials.
First, we introduce the matrix:
\begin{equation}\label{Cf}
{\cal C}_{k,i}=\cos\left( \frac{\pi \, (k-\tfrac 12) (i-1)}{l_c} \right), \qquad k,i=1,...,l_c.
\end{equation}
Then we compute the expansion coefficients with respect to the Chebyshev-polynomials of the first kind:
\begin{equation}\label{bijn}
b_{ij}^{(n)}=\frac{2}{l_c} \, \sum\limits_{A=1}^{l_c} \, \frac{\tilde{\bf Q}_i(u_A) \, {\bf Q}_j(u_A)-{\bf Q}_i(u_A) \, \tilde{\bf Q}_j(u_A)}
{\sqrt{4 \, g^2-u^2_A}} \, {\cal C}_{l_c-A+1,n+1}, \qquad n=0,1,...,l_c-1,
\end{equation}
and finally using the identity (\ref{TU}), the coefficients of (\ref{rho-U}) are given by:
\begin{equation}\label{aijdet}
\begin{split}
&a_{ij}^{(n)}=\frac{b_{ij}^{(n)}-b_{ij}^{(n+2)}}{2}, \qquad \qquad 0\leq n \leq l_c-3, \\
&a_{ij}^{(n)}=\frac{b_{ij}^{(n)}}{2}, \qquad \qquad \qquad n=l_c-2,l_c-1.
\end{split}
\end{equation}
 Using the results of appendix \ref{appC}, $\omega_{ij}$ and $\omega_{ij}^{reg}$ can be expressed in 
terms of the coefficients $a_{ij}^{(n)}$ by the formulae:
\begin{equation}\label{omegaija}
\begin{split}
\omega_{ij}(u_A)\simeq i \, g\, \sum\limits_{n=0}^{l_c-1} \, a_{ij}^{(n)} \, \left\{
x_m(\tfrac{u_A}{g})^{n+1}+\sum\limits_{k=1}^{\infty} \, \left(
\frac{1}{x_s(\tfrac{u_A-i \, k}{g})^{n+1}}+\frac{1}{x_s(\tfrac{u_A+i \, k}{g})^{n+1}}
\right) \right\}+\omega_{ij}^c,
\end{split}
\end{equation}
\begin{equation}\label{omegaregija}
\begin{split}
\omega_{ij}^{reg}(u_A)\simeq i \, g\, \sum\limits_{n=0}^{l_c-1} \, a_{ij}^{(n)} \, \left\{T_{n+1}(\tfrac{u_A}{2g})
+\sum\limits_{k=1}^{\infty} \, \left(
\frac{1}{x_s(\tfrac{u_A-i \, k}{g})^{n+1}}+\frac{1}{x_s(\tfrac{u_A+i \, k}{g})^{n+1}}
\right) \right\}+\omega_{ij}^c,
\end{split}
\end{equation}
where $T_n$ denotes $n$th Chebyshev-polynomial of the first kind, and the expression of $I_{ij}$ entering $\omega_{ij}^c$
is also simple in terms of $a_{ij}^{(n)}$:
\begin{equation}\label{Iij}
I_{ij}=i \, g^2 \, \pi \, a_{ij}^{(0)}.
\end{equation}
One can recognize that in (\ref{omegaregija}) the multiplier of $a_{ij}^{(n)}$ depend on only $g$ and the discretization points $u_A$.
This is why it is useful to compute it at the beginning of the numerical method. 
The computation of the quantity:
\begin{equation}\label{OMEGA}
\Omega_{A,n}(g)=\sum\limits_{k=1}^{\infty} \, \left(
\frac{1}{x_s(\tfrac{u_A-i \, k}{g})^{n}}+\frac{1}{x_s(\tfrac{u_A+i \, k}{g})^{n}}
\right), \qquad A,n=1,.. l_c
\end{equation}
involves an infinite sum. The numerical method for computing it within a given
numerical accuracy, is described in appendix \ref{appD}.

The coefficients of (\ref{p12},\ref{p34}) and $\Delta$ are determined by imposing the equations:
\begin{equation}\label{final}
F_i(u_A)\equiv \tilde{\bf Q}_i(u_A)-\omega_{ij}^{reg}(u_A)\, {\bf Q}^j(u_A)=0, \qquad i,j=1,...4,\quad A=1,..,l_c.
\end{equation}
Instead of solving numerically (\ref{final}) as an equation, \cite{QSCnum} proposed to solve it as an
optimization problem. This means that one tries to find the numerical solution of (\ref{final}) by
minimizing the quantity:
\begin{equation} \label{SS0}
{\cal S}=\sum\limits_{i=1}^4 \,\sum\limits_{A=1}^{l_c} \, |F_i(u_A)|^2.
\end{equation}
This is performed by the Levenberg-Marquardt algorithm described in detail in the next subsection.

\subsection{The Levenberg-Marquardt algorithm}

The minimization of ${\cal S}$ is achieved via the Levenberg-Marquardt algorithm.
To describe it, we put all unknowns into a single vector ${\bf c}$. In our case certain unknowns
are real{\footnote{The coefficients of ${\bf p}_1$ and ${\bf p}_2$ and $\Delta$ are the real ones.},
 while others are pure imaginary{\footnote{The coefficients of ${\bf p}_3$ and ${\bf p}_4$ are the imaginary ones.}.
The real unknowns are put into the first $\Lambda_1$ components of ${\bf c}$, while the other components
are the imaginary ones:
$$ {\bf c}_k \in {\mathbb R}, \qquad k=1,...,\Lambda_1,$$
$$ {\bf c}_k \in i \,{\mathbb R}, \qquad k=\Lambda_1+1,...,\Lambda.$$
If we truncate the sums in (\ref{p12}) at $N_0$th term, then the number of real unknowns is $\Lambda_1=2\,N_0+1$. 
The reason is  that the number of coefficients in the truncated versions of (\ref{p12}) is $2 \, N_0$, plus $1$, 
because $\Delta$ is also a real unknown. 
If the sums in (\ref{p34}) are also truncated at the $N_0$th term, 
then the number of imaginary components is $\Lambda-\Lambda_1=2 \, N_0$. 
Thus, if all infinite sums are truncated at the $N_0$th term, 
then ${\bf c}$ is a $\Lambda=4 \, N_0+1$ component vector.

For short, we introduce the multi-index $I=(i,A), \quad i=1,..,4, \quad A=1,...,l_c$ and  denote 
${\cal F}_I=F_i(u_A)$. In this notation (\ref{SS0}) takes the form:
\begin{equation} \label{SS1}
{\cal S}({\bf c})=\sum\limits_{I=1}^{4 \, l_c} \, {\cal F}_I({\bf c}) \,  {\cal F}^*_I({\bf c}),
\end{equation}
and our task is to find the vector $\tilde{\bf c}$, which minimizes ${\cal S}({\bf c})$.
Assuming that ${\bf c}$ is close to $\tilde{\bf c}$, ${\cal S}({\bf c})$ can be linearized around
the minimum and the minimization process consists of subsequent iterative minimizations of the
linearized approximations of ${\cal S}({\bf c})$.

To expand (\ref{SS1}) around the minimum one needs to compute the derivative matrix:
\begin{equation}
{\cal J}_{Ik}({\bf c})=\frac{\partial {\cal F}_I({\bf c})}{\partial {\bf c}_k}, \qquad I=1,..,4 \, l_c, \quad
k=1,...,\Lambda.
\end{equation} 
In practice it is done with the help of a second order formula for the first derivative:
$f'(u)= \frac{f(u+h)-f(u-h)}{2 \,h}+O(h^2)$ with $h$ being a small number. 
Thus ${\cal J}_{Ik}({\bf c})$ is numerically approximated by the formula:
\begin{equation}
{\cal J}_{Ik}({\bf c})\approx\left\{ \begin{array}{ll}
\frac{{\cal F}_I(\{ {\bf c}_j+h \, \delta_{jk} \})-{\cal F}_I(\{ {\bf c}_j-h \, \delta_{jk} \})}{2 \, h},
\qquad \quad k=1,...,\Lambda_1 \\
\frac{{\cal F}_I(\{ {\bf c}_j+i\, h \, \delta_{jk} \})-{\cal F}_I(\{ {\bf c}_j- i\, h \, \delta_{jk} \})}{2 \, h \, i},
\qquad k=\Lambda_1+1,...,\Lambda.
\end{array} \right.
\end{equation}
It is worth to introduce its sign modified conjugate:
\begin{equation}
\tilde{\cal J}^*_{Ik}({\bf c})=\left\{ \begin{array}{ll}
{\cal J}^*_{Ik}({\bf c}), \qquad \quad 1\leq k \leq \Lambda_1, \\
-{\cal J}^*_{Ik}({\bf c}), \qquad \Lambda_1< k \leq \Lambda.
\end{array} \right.
\end{equation}
If ${\bf c}$ is close to the minimum $\tilde{\bf c}$ of ${\cal S}({\bf c})$, then
using a linear approximation:  
\begin{equation}
{\cal S}(\tilde{\bf c})\approx \sum\limits_{I=1}^{4 \, l_c} \, 
\left[ {\cal F}_I({\bf c})-\sum\limits_{k=1}^\Lambda \, {\cal J}_{Ik}({\bf c}) \, ({\bf c}_k-\tilde{\bf c}_k)  \right]\cdot
\left[ {\cal F}^*_I({\bf c})-\sum\limits_{k=1}^\Lambda \, \tilde{\cal J}^*_{Ik}({\bf c}) \, ({\bf c}_k-\tilde{\bf c}_k)  \right]
\end{equation}
and imposing the minimum condition $\frac{{\cal S}(\tilde{\bf c})}{\partial \tilde{\bf c}_k}=0$,  one gets a
set of linear equations for the components of the minimum vector:
\begin{equation}\label{cck}
\tilde{\bf c}_k={\bf c}_k-\sum\limits_{j=1}^{\Lambda} \,{\cal M}^{-1}_{kj}({\bf c}) \, v_j({\bf c}), \qquad k,j=1,...,\Lambda,
\end{equation}
where
\begin{equation} \label{vj}
v_j({\bf c})=\sum\limits_{I=1}^{4 \, l_c} \, \left\{{\cal J}_{Ij}({\bf c}) \, {\cal F}^*_I({\bf c})+\, \tilde{\cal J}^*_{Ij}({\bf c})
\, {\cal F}_I({\bf c}) \right\}, \qquad j=1,..,\Lambda,
\end{equation}
\begin{equation} \label{Mjk}
{\cal M}_{jk}({\bf c})=\sum\limits_{I=1}^{4 \, l_c} \,
 \left\{ {\cal J}_{Ij}({\bf c}) \, \tilde{\cal J}^*_{Ik}({\bf c}) + 
 \tilde{\cal J}^*_{Ij}({\bf c}) \, {\cal J}_{Ik}({\bf c})
 \right\}, \qquad j,k=1,...,\Lambda.
\end{equation}
In practice, during the iteration, equation (\ref{cck}) determines the new values of the unknowns from the
old ones. Namely, if ${\bf c}^{(n)}$ denotes the value of ${\bf c}$ after the $n$th iteration, then
 its value after the $n+1$st iteration is given by:
\begin{equation}\label{ccit}
{\bf c}^{(n+1)}_k={\bf c}^{(n)}_k-\sum\limits_{j=1}^{\Lambda} \,{\cal M}^{-1}_{kj}({\bf c}^{(n)}) \, v_j({\bf c}^{(n)}), \qquad k=1,...,\Lambda.
\end{equation}
The iterational prescription (\ref{ccit}) works very well, if the initial value of ${\bf c}$ is very close
to the exact solution. Otherwise, it does not define a convergent iteration. In such cases the
Levenberg-Marquardt (LM) modification of (\ref{ccit}) is needed to decrease the difference
 $|{\bf c}^{(n+1)}-{\bf c}^{(n)}|$
 at each step of the iteration \cite{QSCnum}, and so to slow down and stabilize the iteration process. 
In the Levenberg-method, equation (\ref{ccit}) is modified by adding a unit-matrix multiplied with an iteration 
number dependent number to $\cal M$. In case of the Marquardt-method the unit-matrix is changed to the diagonal
part of ${\cal M}$:
\begin{equation}
\begin{split}
&{\cal M}_{kj}({\bf c}^{(n)}) \to {\cal M}_{kj}({\bf c}^{(n)})+\lambda^{(n)} \, \delta_{kj}, \qquad \qquad
\quad \qquad \mbox{Levenberg-method}, \label{Lm} \\
&{\cal M}_{kj}({\bf c}^{(n)}) \to {\cal M}_{kj}({\bf c}^{(n)})+\lambda^{(n)} \, {\cal M}_{kk}({\bf c}^{(n)}) \, \delta_{kj}, \qquad
\mbox{Marquardt-method}, 
\end{split}
\end{equation}
where $\lambda^{(n)}$ is an iteration number dependent number. The main drawback of the Levenberg-Marquardt
modification is that, it defines a quite stable, but very slowly converging algorithm. To find the minimum of
${\cal S}({\bf c})$ within practically acceptable amount of time, the term proportional to $\lambda$ should be
switched off after a few number of iterations. Here, we have to mention, another important property of the LM-algorithm,
namely the larger the value of $\lambda$, the slower the convergence is. This is why, it is also desirable to
decrease the value of $\lambda$ at each step of the iteration.

Taking into account the facts and experiences above, we used the LM-algorithm in the following way:

First, we choose a not too large initial value for $\lambda^{(0)}$ and a divisor $\nu>1$.
For the states under consideration we took $\lambda^{(0)}=2.1$ and $\nu=2.0$. 
At the $n$th step starting from ${\bf c}^{(n)}$, we go through the whole iteration process
 with $\lambda^{(n)}$ and get the new vector ${\bf c}^{(n+1)}$. If
${\cal S}({\bf c}^{(n+1)})<{\cal S}({\bf c}^{(n)})$, then we decrease the value of $\lambda$
by dividing it by $\nu$, i.e. $\lambda^{(n+1)}=\frac{\lambda^{(n)}}{\nu}$. Otherwise we increase
the value of $\lambda$ by multiplying it by $\nu$: $\lambda^{(n+1)}=\lambda^{(n)}\, \nu$ and the new iteration
starts from the old initial values i.e. ${\bf c}^{(n+1)}={\bf c}^{(n)}$.
After a certain number of such iterations, when ${\cal S}({\bf c}^{(n)})$ becomes small enough ($\sim 1$), 
the action of $\lambda$ is switched off and the further iterations are done with the $\lambda^{(n)}\equiv 0$
formula (\ref{ccit}). We note that in our concrete numerical computations we used the 
Marquardt-type (\ref{Lm}) modification of (\ref{ccit}) and
in practice we do not compute the inverse of ${\cal M}$, but solve the following 
set of linear equations for ${\bf c}^{(n+1)}$:
\begin{equation}\label{ccitlin}
\sum\limits_{j=1}^{\Lambda} \,{\cal M}_{kj}({\bf c}^{(n)}) \, ({\bf c}^{(n)}_j-{\bf c}^{(n+1)}_j)=
 v_k({\bf c}^{(n)}), \qquad k=1,...,\Lambda.
\end{equation}

\subsection{The complete algorithm}

In this subsection we write down the process of the numerical algorithm.

\begin{itemize}
\item First, initial values are chosen for ${\bf c}^{(0)}, \, \lambda^{(0)},$ and $\nu$.
\item Going through the process described in the previous subsections, we compute ${\cal F}_I({\bf c}^{(0)})$.
\item To compute the derivative ${\cal J}_{jk}$, one does the same computation another $2 \, \Lambda$ times,
 but starting from the 1-component shifted initial value vectors:\newline
 ${\bf c^{ (0)}_{k\pm}}=\{{\bf c}^{(0)}_1,...,{\bf c}^{(0)}_{k-1},
{\bf c}^{(0)}_{k}\pm H,{\bf c}^{(0)}_{k+1},..,{\bf c}^{(0)}_{\Lambda}\}$, where $H=\pm h$ or $H=\pm \, i\, h$
depending on the properties of ${\bf c}_k$ under complex conjugation. 
\item Then the quantities ${\cal J}_{jk},\, \tilde{\cal J}_{jk}, {\cal M}_{jk}, \,v_j$ and ${\cal S}({\bf c}^{(0)})$
are computed.
\item The corrected values of the unknowns (i.e. ${\bf c}^{(1)}$) are computed by the Marquardt-version of (\ref{ccit}).
\item ${\cal S}({\bf c}^{(1)})$ is computed from ${\bf c}^{(1)}$. 
\item The initial values of the next iteration are chosen by the rule: \newline
If ${\cal S}({\bf c}^{(1)})<{\cal S}({\bf c}^{(0)})$, then $\lambda^{(1)}=\frac{\lambda^{(0)}}{\nu}$ and the
next iteration starts from ${\bf c}^{(1)}$. Otherwise $\lambda^{(1)}=\lambda^{(0)}\, \nu$ and the new iteration
starts from the old initial values i.e. ${\bf c}^{(1)}={\bf c}^{(0)}$.
\item The whole process starts from the beginning...
\item After several such iterations $\lambda$ is set to be zero, and (\ref{ccit}) determines the
new approximations for the unknowns. 
\end{itemize}

\section{Numerical results for the Konishi operator}

The Konishi operator is the most studied element of the set of single trace operators in the ${\cal N}=4$
super Yang-Mills (SYM) theory. The set of twist-2 operators with even spin also includes it as the $L=S=2$ special case.
In this section we summarize our numerical results obtained for the Konishi operator. 

We solved the
QSC equations in the range $g\in [0.1 , 7.0]$ and by fitting the numerical data, we determined numerically
the first few coefficients of the large $g$ expansion of some important quantities.
Previous numerical investigations \cite{Gromov:2009zb,Frolov:2010wt,Frolov:2012zv,Gromov:2011bz} could determine the first few coefficients of the large $g$ series
of the anomalous dimension $\Delta$. Now, beyond the numerical determination of the coefficients of the strong coupling
series of $\Delta$, we also determine the large $g$ behaviour of the coefficients of the $1/x$ series in 
(\ref{p12},\ref{p34}).
We also study the strong coupling behaviour of the ${\bf p}_a$ functions around the branch points $u=\pm 2 g$.

We note that the numerical data for $\Delta(g)$ and $c_{a,n}(g)$ are available in the corresponding text 
file{\footnote{The name of the corresponding text file is: {\textit L2S2data.txt}.}}
uploaded together with the paper. The pure numerical data can be read in a Mathematica notebook with the
help of the {\textit DATAIN.nb} notebook file{\footnote{It is also uploaded with this paper.}}, where it is also explained, how to get a required quantity
out of the huge array of numerical data.

\subsection{Numerical results for $\Delta$}

We are interested in the coefficients of the strong coupling expansion of $\Delta$: 
\begin{equation}\label{deltaki}
\Delta = \Delta^{(0)} \lambda^\frac{1}{4} + \Delta^{(1)} \lambda^{-\frac{1}{4}}  + \Delta^{(2)} \lambda^{-\frac{3}{4}} + \Delta^{(3)} \lambda^{-\frac{5}{4}} + 
+ \Delta^{(4)} \lambda^{-\frac{7}{4}}+ \Delta^{(5)} \lambda^{-\frac{9}{4}}+\dots
\end{equation}
For the twist-L operators in the $sl(2)$ sector, there are analytical predictions for the first four coefficients of (\ref{deltaki}).
The coefficients depend on $L$ and $S$ and take the form \cite{Gromov:2014bva}:
\begin{equation}\label{D01}
\Delta^{(0)}=\sqrt{2 \, S}, \qquad \Delta^{(1)}=\frac{2 \, L^2+S(3 \, S-2)}{4 \, \sqrt{2 S}},  
\end{equation}
\begin{equation}\label{D2}
\Delta^{(2)} = \frac{-21\,S^4 +(24-96\,\zeta_3) S^3+4 \left(5 L^2-3\right) S^2+8 L^2 S -4 L^4}{64 \sqrt{2}\,S^{3/2}},
\end{equation}
\begin{eqnarray}\label{D3}
\Delta^{(3)} &=& \frac{187\,S^6 + 6\,(208\,\zeta_3 + 160\,\zeta_5-43)\,S^5 +\left(-146\,L^2 - 4\,(336\,\zeta_3-41)\right)S^4 }{512 \sqrt{2}\,S^{5/2}} + \nonumber \\
	&+& \frac{\left(32\,(6\,\zeta_3+7)\,L^2-88\right)S^3 + \left(-28\,L^4 + 40\,L^2\right) S^2 - 24\,L^4 S + 8\,L^6}{512 \sqrt{2}\,S^{5/2}}.
\end{eqnarray}
The first two coefficients in (\ref{D01}) can be determined either from Basso's slope function \cite{Basso:2011rs} or from
semi-classical computations in string theory \cite{Gromov:2011de,Roiban:2011fe,Vallilo:2011fj}. The next two coefficients were determined by matching
the $O(S^2)$ term of the small spin expansion with classical and semi-classical results \cite{Gromov:2014bva}.

To determine numerically the coefficients in (\ref{deltaki}), we computed $\Delta$ numerically in the 
range $g\in[0.1,7.0]$ range with approximately
20 digits of accuracy and in the range $g\in[4.6,7]$ we fitted the numerical data with a power series of
 the form of (\ref{deltaki}).

The fitting method went as follows. We fitted a power series of type (\ref{deltaki}) to the numerical data. We 
increased the order of the truncation of the series until the numerical values of the coefficients stabilized. 
First, we concentrated on the first coefficient $\Delta^{(0)}$. We experienced that it is very close to the exact 
value (\ref{D01}). This is why we assumed that its value is
equal to the analytical prediction. Then we subtracted $\Delta^{(0)} \lambda^\frac{1}{4}$ from the numerical data and 
fitted the new set of data with a truncated power series of type 
$ \Delta^{(1)} \lambda^{-\frac{1}{4}}  + \Delta^{(2)} \lambda^{-\frac{3}{4}} +... $. 
Again, we increased the order of the truncation of the series until the numerical values of the coefficients 
stabilized. Then we concentrated on the coefficient $\Delta^{(1)}$. 
We experienced that, the fitted value of the coefficient $\Delta^{(1)}$ is very close to the analytical prediction
 given by (\ref{D01}). Again, we assumed that the exact value of $\Delta^{(1)}$ is given by (\ref{D01}), 
and we subtracted also the second term of (\ref{deltaki}) from the numerical data. Then to get $\Delta^{(2)}$,
 we fitted the new set of data with a series starting at of order $\lambda^{-\frac{3}{4}}$ etc.

Our results for the fitted values of the coefficients of (\ref{deltaki}) are shown in table \ref{t1}.
The numerical data confirms with high precision the analytical predictions for the
$n=0,1,2,3$ cases. Table \ref{t1} contains fitted values for the $n=4,5$ cases as well.
Since so far there are no available analytical predictions for these coefficients, we gave
numerical estimations for further two previously unknown coefficients of the strong coupling
expansion of the anomalous dimension for the Konishi state .

In table \ref{t1}. $\delta_{rel} \Delta^{(n)}$ denotes the relative error defined by $\big| 
\frac{\Delta^{(n)}_{exact}-\Delta^{(n)}_{fitted}}{\Delta^{(n)}_{exact}} \big|$ }. 
For $n=4,5$ in the
lack of analytical results, $\delta_{rel} \Delta^{(n)}$ was computed as
the ratio of the estimated error for $\Delta^{(n)}_{fitted}$ and $\Delta^{(n)}_{fitted}$.

\begin{table}
\begin{center}
\begin{tabular}{|c||c|c|c|}
\hline
$n$ & $\Delta^{(n)}_{exact}$ & $\Delta^{(n)}_{fitted}$ & $\delta_{rel} \Delta^{(n)}$ \\
\hline
0 & 2.0 & 1.999999999999898 & $5.0 \cdot 10^{-14}$  \\
\hline
1 & 2.0 & 1.999999999995831 & $2.8 \cdot 10^{-12}$  \\
\hline
2 & -3.106170709478783 & -3.106170709557684 & $2.5 \cdot 10^{-11}$  \\
\hline
3 & 15.48929958253284 & 15.48929957822780 & $2.8 \cdot 10^{-10}$  \\
\hline
4 & - & -91.97602372540774 & $8.2 \cdot 10^{-9}$  \\
\hline
5 & - & 758.5146133674111 & $1.1 \cdot 10^{-6}$  \\
\hline
\end{tabular}
\bigskip
\caption{Comparison of the analytical predictions and the fitted values for $\Delta^{(n)}$. 
$\delta_{rel} \Delta^{(n)}$ denotes the relative error.
\label{t1}}
\end{center}
\end{table}
\normalsize

Apart from fitting the coefficients of the strong coupling expansion of $\Delta$, we also constructed
a Pade-approximation like formula for $\Delta$. According to our estimation, our 
approximation formula gives the values of $\Delta$ with 14-digits of accuracy in the range of available numerical 
data i.e $g\in[0.1,7.0]$ and 
with at least 9-digits of accuracy for $g>7.0$. The actual form of the Pade-approximation like formula for 
the anomalous dimension of the Konishi state can be found in appendix \ref{appE}.

\subsection{The strong coupling behaviour of ${\bf p}_a$}

In this subsection the strong coupling behaviour of the ${\bf p}_a$-functions is studied
through the investigation of the strong coupling behaviour of the coefficients of 
the series (\ref{p12}) and (\ref{p34}). First, let us see, how the coefficients $c_{a,n}(g)$, 
look as functions of $n$ at fixed $g.$ Since the coefficients decay exponentially 
fast with a rate determined by the radius of convergence $R(g)=|x_s(2+\tfrac{i}{g})|$ of the
problem, for demonstrational purposes it is worth to introduce $\hat{c}_{a,n}(g)$
by the definition:
\begin{equation}\label{cahat}
\hat{c}_{a,n}(g)=c_{a,n}(g) \, R(g)^{2 \, n+d_{a}}, \qquad d_a=\delta_{a,1}+\delta_{a,3}.
\end{equation}
In order for the readers to get a taste about the $n$-dependence of $\hat{c}_{a,n}(g)$, we show
$\hat{c}_{1,n}(g)$ at $g=4.4$ in figure \ref{c1nabra}. 
 \begin{figure}[htb]
\begin{flushleft}
\hskip 15mm
\leavevmode
\epsfxsize=120mm
\epsfbox{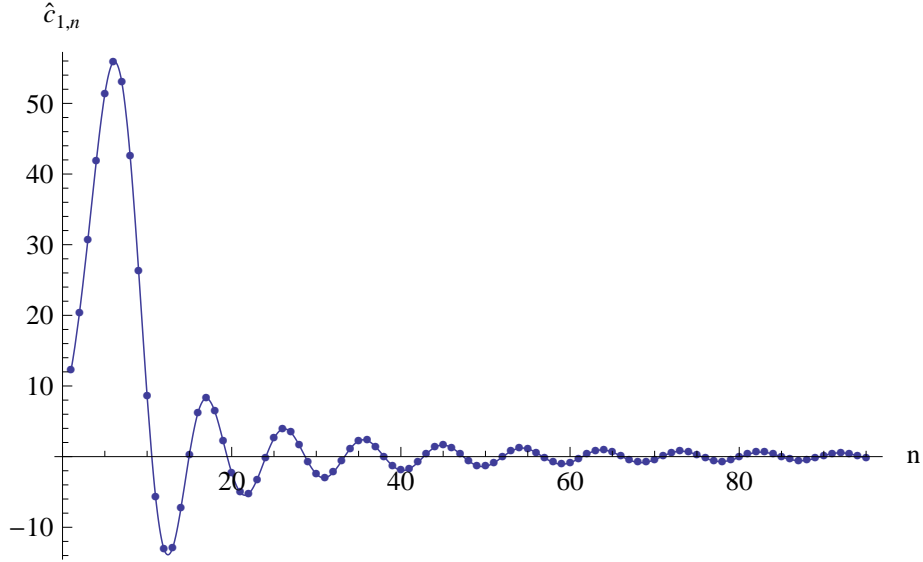}
\end{flushleft}
\caption{{\footnotesize
The plot of $\hat{c}_{1,n}$ at $g=4.4$. The data points are connected by an interpolating function only for
 demonstrational purposes. 
}}
\label{c1nabra}
\end{figure} 
In the other $a=2,3,4$ cases,  
the picture is structurally very similar. The most important properties of $\hat{c}_{a,n}(g)$ at fixed $g$, 
can be summarized as follows:
\begin{itemize}
\item The enveloping curve of $\hat{c}_{a,n}(g)$ has a power like decay with an exponent being
close to 1.5. I.e. $\hat{c}_{a,n}(g)\sim n^{-\epsilon_a(g)}$, where $\epsilon_a(g)\sim 1.5 \pm 0.2.$
\item If $\hat{c}_{a,n}(g)$ is considered as a continuous function of $n$, then it has infinitely
many zeros.
\item In the large $n$ regime the zeros are located periodically, such that the characteristic
wavelength of this periodicity $\Lambda_a(g)\sim {\mathtt a}_0 \sqrt{g}$ at strong coupling, with
${\mathtt a}_0 \sim 4.4.$
\end{itemize}

One can recognize another interesting property of the coefficients, if one 
plots $\hat{c}_{a,n}(g)$ at all available values of $g$ on the same plot.
They all have very similar shape, which suggests that in the strong coupling limit 
they can be transformed into a universal $g$-independent function with some scale transformation.
Indeed, figures \ref{Skaled12} and \ref{Skaled34} show that the transformed coefficients $g^{-\hat{n}_a}\,\hat{c}_{a,\sqrt{g} \nu}$ with
$(\hat{n}_1,\hat{n}_2,\hat{n}_3,\hat{n}_4)=(1,0,3,2)$ tend to universal $g$-independent functions 
${\cal K}_a(\nu)$ at 
strong coupling. For later purposes, we write it down in a formula as well:
\begin{equation}\label{chatK}
g^{-\hat{n}_a} \, \hat{c}_{a,\sqrt{g} \nu}={\cal K}_a(\nu)+\dots,
\end{equation}
where the dots stand for negligible terms for $g \to \infty.$

This fact shows that the in the strong coupling limit the relevant scale of the problem is given by 
$\sqrt{g}$ or equivalently $\lambda^{\tfrac 14}$ as it is expected from the strong coupling
behaviour of the anomalous dimension.

\begin{figure}[htbp]
\begin{center}
\begin{picture}(260,70)  \epsfxsize=70mm
\put(0,15) {\epsfbox{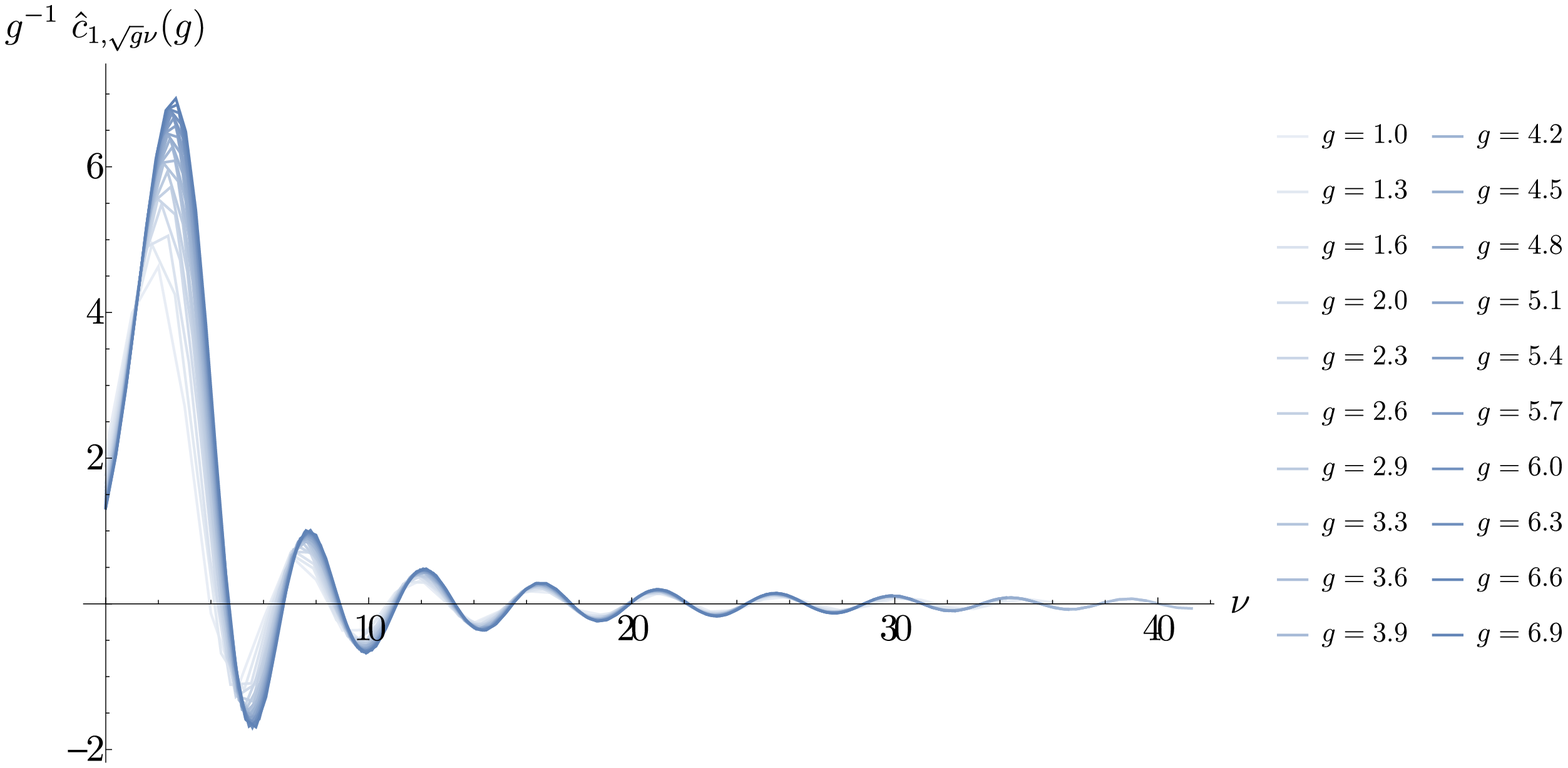}}
\put(95,15) {\epsfbox{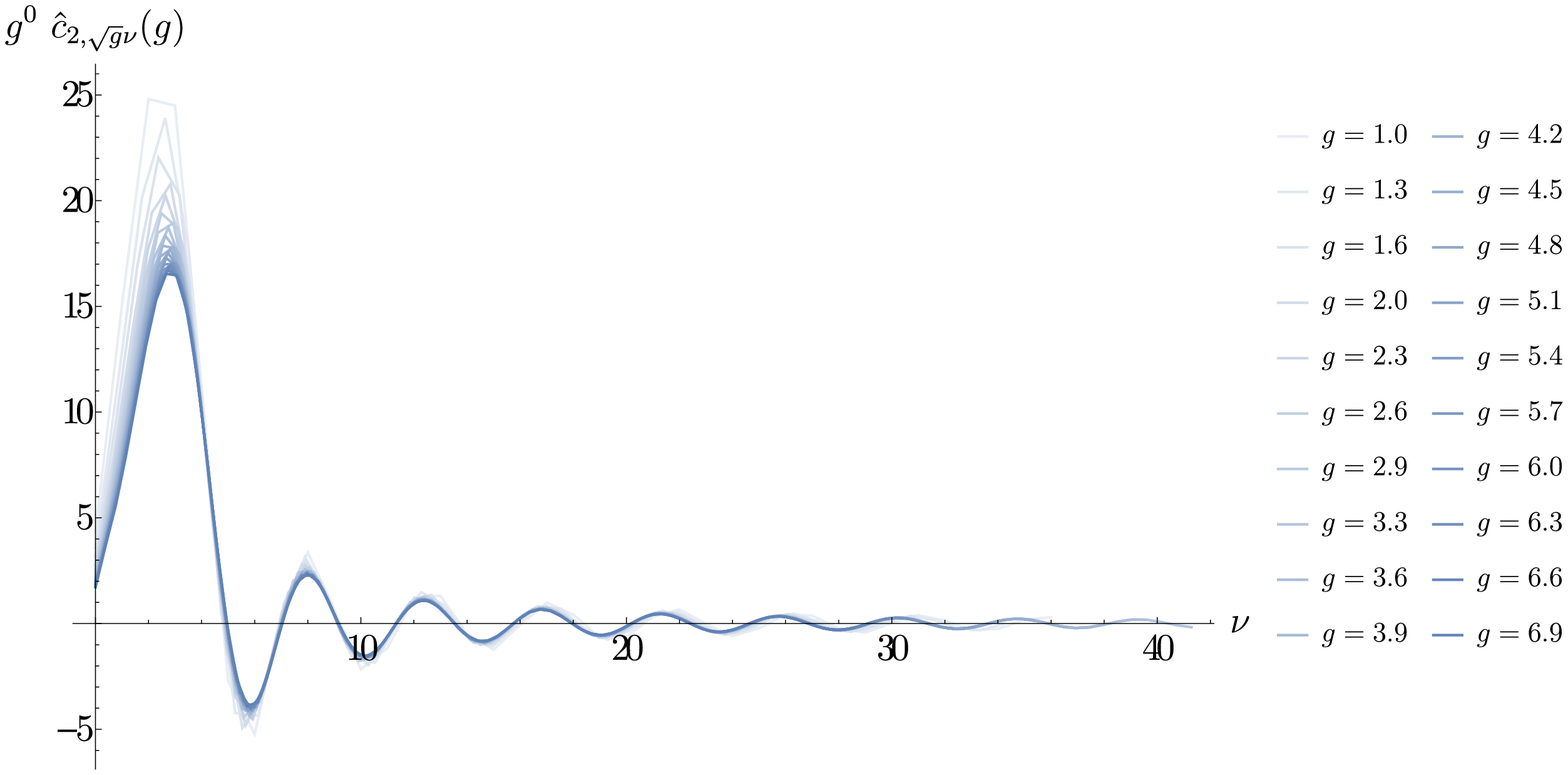}}
\put(0,0) {\parbox{150mm}
{\caption{ \label{Skaled12}\protect {\small
 The demonstration of the strong coupling scaling property of $\hat{c}_{a,n}(g)$ for $a=1$ (left) and
 $a=2$ (right) cases.}}}}
\end{picture}
\end{center}
\end{figure}
\begin{figure}[htbp]
\begin{center}
\begin{picture}(260,70)  \epsfxsize=70mm
\put(0,15) {\epsfbox{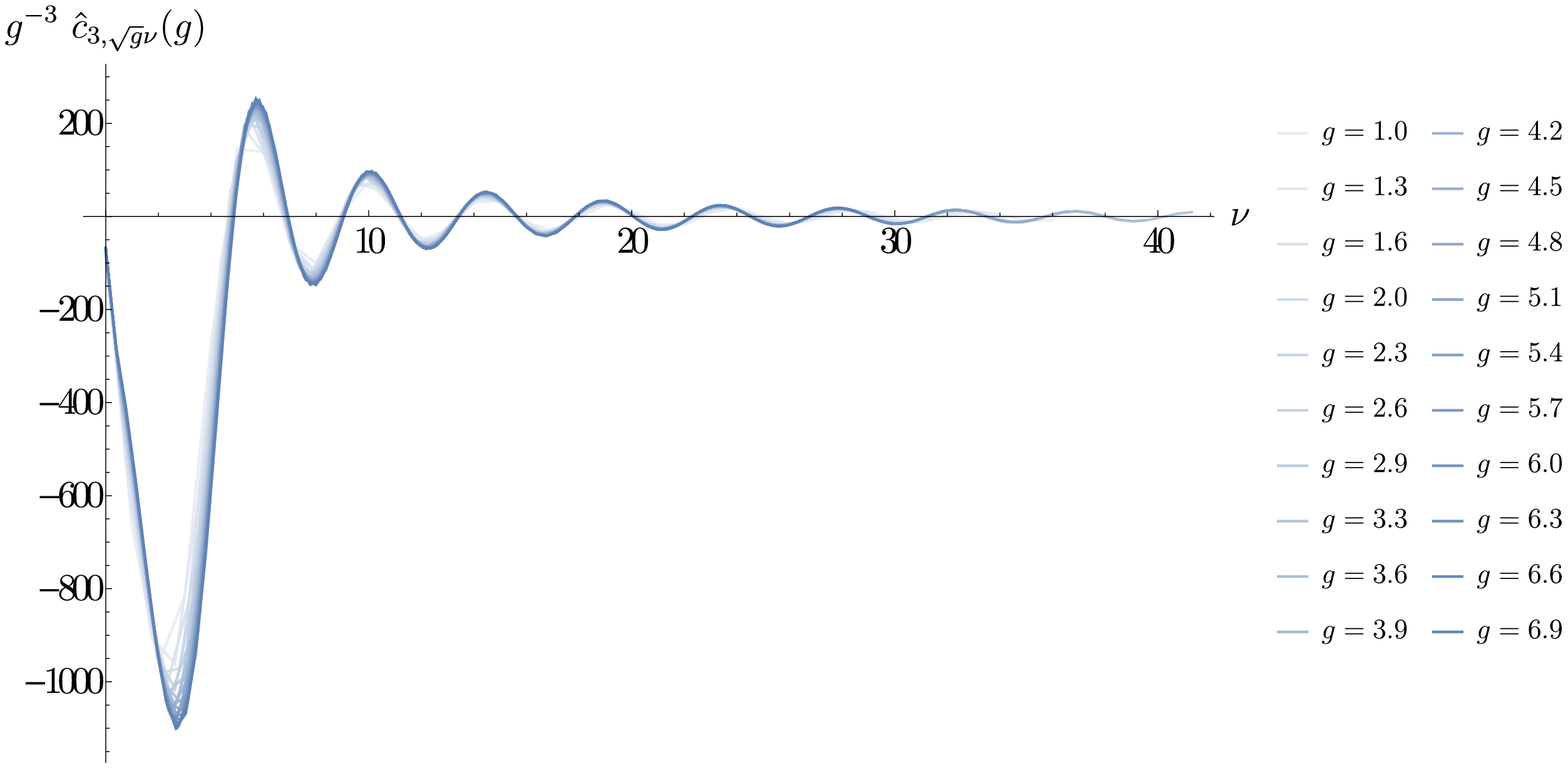}}
\put(95,15) {\epsfbox{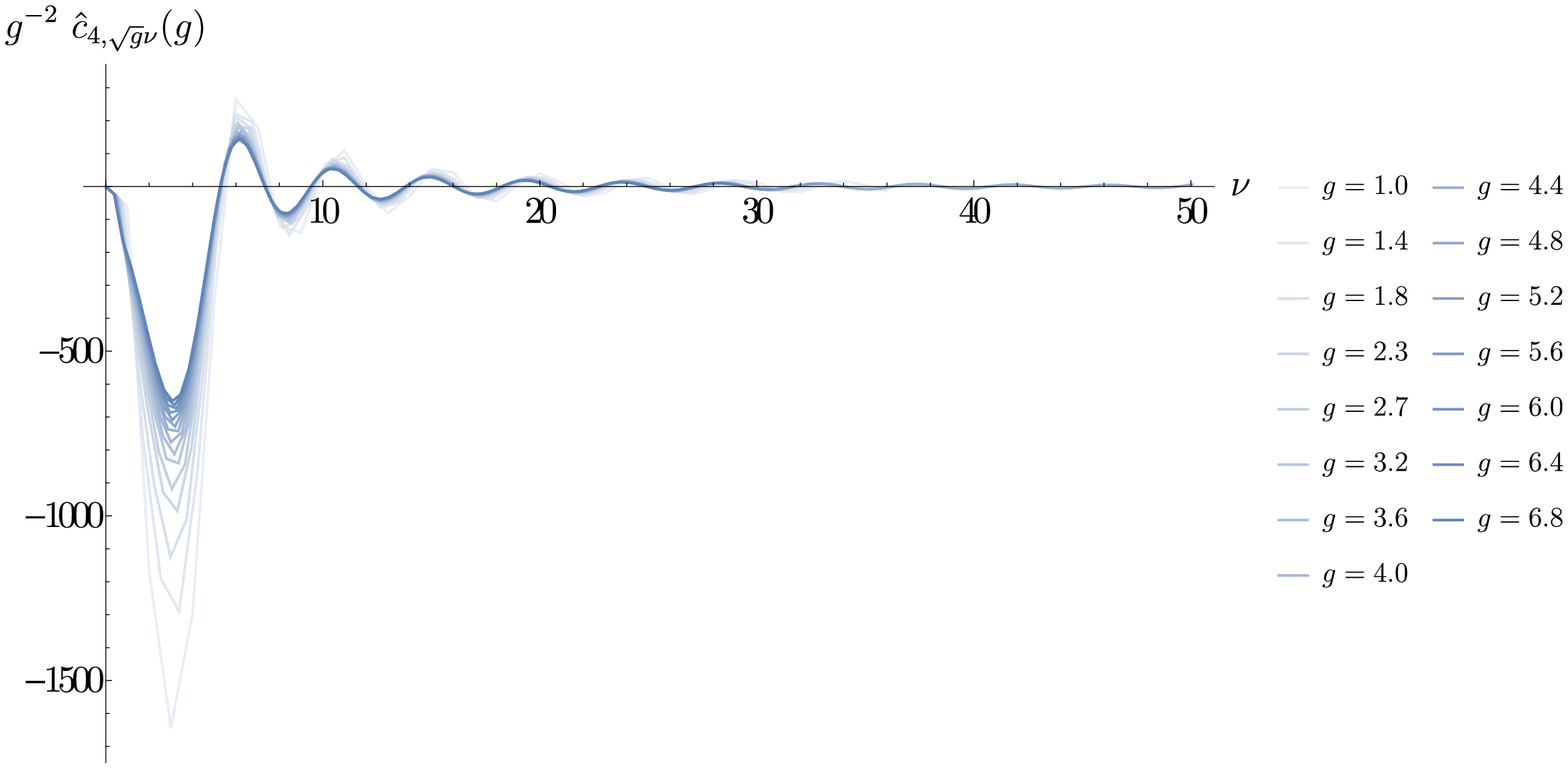}}
\put(0,0) {\parbox{150mm}
{\caption{ \label{Skaled34}\protect {\small
 The demonstration of the strong coupling scaling property of $\hat{c}_{a,n}(g)$ for $a=3$ (left) and
 $a=4$ (right) cases.}}}}
\end{picture}
\end{center}
\end{figure}

\subsubsection{Strong coupling behaviour of $c_{a,n}$ for fixed $n$}

In this subsection we investigate, how the coefficients of the series (\ref{p12}) and (\ref{p34})
behave at strong coupling, if we fix the value of the index $n$. 
We considered the first 12 or 14 coefficients of the series (\ref{p12}) and (\ref{p34}). I.e.
$c_{a,n}$ with $a=1,..,4$ and $n=0,...,14$. Then in the range $g\in[4.6,7.0]$ we fitted the
numerical data with a series{\footnote{We tried to fit other types of series in $g$, like series in $1/\sqrt{g}$ etc., 
but only the $1/g$ case gave numerically stable coefficients. }} in $1/g$.
Our numerical data was consistent with the series expansions as follows:
\begin{equation}\label{Cserpg}
c_{a,n}(g)=g^{n_a} \, \sum\limits_{k=0}^{\infty} \, \frac{{\mathfrak c}_{a,n}^{(k)}}{g^k},
\end{equation}
where the integer leading power $n_a$ and the numerical values of ${\mathfrak c}_{a,n}^{(k)}$ were determined 
from the fitting process. The best fits yield the following values for the leading
powers{\footnote{We note that $n_a=\hat{n}_a$ of (\ref{chatK}) for $a=1,2,3,4$.}}:
 \begin{equation}\label{na}
(n_1,n_2,n_3,n_4)=(1,0,3,2).
\end{equation}
For $a=1$ and $a=2$ we know from our H-symmetry fixing conditions that $c_{1,0}\equiv g$ and
$c_{2,0}\equiv 1$ exactly. For $a=1,2$, (\ref{na}) shows that at large $g$ in leading order all coefficients 
behave in the same way, and this leading order power behaviour is determined by the H-symmetry fixing
condition. The situation is very similar in the $a=3,4$ cases. 
There the leading powers are the same
as those of $A_3 \, u=A_3 \, g \, (x+\tfrac1x)$ and $A_4 \, u^2=A_3 \, g^2 \, (x+\tfrac1x)^2$ with $x$ being fixed.
 From (\ref{AA}) and (\ref{deltaki}) it follows that, 
at large $g$: $\, A_3\sim g^2/A_2=g^2$, i.e. $g \, A_3 \sim g^3 \Rightarrow n_3=3$. Similarly: $A_4\sim g^2/A_1=1$,
 i.e. $g^2 \, A_4 \sim g^2 \Rightarrow n_4=2$.

Next, we can concentrate on the first, leading order coefficients{\footnote{We just recall
that ${\mathfrak c}_{a,n}^{(k)}$ are real for $a=1,2$ and are pure imaginary for $a=3,4$.}}
 ${\mathfrak c}_{a,n}^{(0)}$ in (\ref{Cserpg}).
Table \ref{t2}. shows their fitted values. 
\begin{table}
\begin{center}
\begin{tabular}{|c||c|c|c|c|}
\hline
$n$ & ${\mathfrak c}_{1,n}^{(0)}$ & ${\mathfrak c}_{2,n}^{(0)}$ & $\mbox{Im}{\mathfrak c}_{3,n}^{(0)}$ & $\mbox{Im}{\mathfrak c}_{4,n}^{(0)}$ \\
\hline
0 & 1 & 1 & -52.637890142265 & 0  \\
\hline
1 & 0.999999999978 & 1.33333333332 & -131.594725354130 & -8.77298169101892  \\
\hline
2 & 0.999999999972 & 1.33333333330 & -131.594725352303 & -35.091926761981  \\
\hline
3 & 0.999999999975 & 1.33333333330 & -131.594725351127 & -35.091926761524  \\
\hline
4 & 0.999999999981 & 1.33333333331 & -131.594725350389 & -35.091926761099  \\
\hline
5 & 0.999999999989 & 1.33333333331 & -131.594725349952 & -35.091926760721  \\
\hline
6 & 0.999999999997 & 1.33333333332 & -131.594725349753 & -35.091926760392  \\
\hline
7 & 0.999999999997 & 1.33333333333 & -131.594725349575 & -35.091926760235  \\
\hline
8 & 0.999999999923 & 1.33333333332 & -131.594725353134 & -35.091926759769  \\
\hline
9 & 0.999999991211 & 1.3333333316 & -131.59472498873 & -35.09192675312  \\
\hline
10 & 0.999999696177 & 1.3333332528 & -131.59470707428 & -35.09192702373  \\
\hline
11 & 0.999994934595 & 1.3333316031 & -131.59437309066 & -35.09193282608  \\
\hline
12 & 0.999948649172 & 1.3333119412 & -131.59080568467 & -35.09196891042  \\
\hline
13 & 0.999643526630 & 1.3331583442 & -131.56546821836 & -35.09189673341  \\
\hline
14 & 0.998177159531 & 1.3323036598 & -131.43562861485 & -35.08931535172  \\
\hline
\end{tabular}
\bigskip
\caption{The numerical values of ${\mathfrak c}_{a,n}^{(0)}$.
\label{t2}}
\end{center}
\end{table}
\normalsize
Looking at the data, one can recognize the remarkable fact that for fixed values of
the index $a$, and for $n \geq 1+\delta_{a,4}$ the coefficients ${\mathfrak c}_{a,n}^{(0)}$ seem 
to be $n$-independent. The difference between the numerical values of the columns are supposed to 
be the consequence of numerical  errors. Then, it is tempting to guess the exact values of 
${\mathfrak c}_{a,n}^{(0)}$ from the available numerical data of table \ref{t2}.

It is not hard to make good proposals for the cases $a=1,2$:
\begin{equation}\label{guess12}
{\mathfrak c}_{1,n}^{(0)}=1, \qquad {\mathfrak c}_{2,n}^{(0)}=\frac 43, \qquad n=1,2,...
\end{equation}
To guess the exact values of ${\mathfrak c}_{a,n}^{(0)}$ for $a=3,4$ seem to be more difficult, but the
following train of thoughts leads to reasonable proposals. One can recognize that based on (\ref{guess12}), 
in the case of $a=1,2$, in (\ref{p12}) all $1/x$ powers has the same coefficient{\footnote{In leading order for 
large $g.$}}. Then one can suspect that the same thing might happen for the cases $a=3,4$. 
Such an assumption gives analytical predictions for the differences ${\mathfrak c}_{3,1}^{(0)}-{\mathfrak c}_{3,0}^{(0)}$
and ${\mathfrak c}_{4,2}^{(0)}-{\mathfrak c}_{4,1}^{(0)}$. The leading order expressions for $A_3$ and $A_4$
can be computed from (\ref{AA}) and the H-symmetry fixing conditions by exploiting (\ref{deltaki},\ref{D01}):
\begin{equation}\label{A34largeg}
A_3=-8\, \pi^2 \, g^2 \, i+... \, \qquad A_4=-\tfrac{8}{3} \, \pi^2 \, g \, i+.... 
\end{equation}
Then substituting $u \to g(x+\tfrac 1 x)$ into (\ref{p34}) and imposing that the coefficients of each $1/x$ power are equal,
one gets the analytical predictions: 
\begin{equation}\label{c31-c30}
{\mathfrak c}_{3,1}^{(0)}-{\mathfrak c}_{3,0}^{(0)}=-8\, \pi^2 \, i,
\end{equation}
\begin{equation}\label{c42-c41}
{\mathfrak c}_{4,2}^{(0)}-{\mathfrak c}_{4,1}^{(0)}=-\tfrac{8}{3}\, \pi^2 \, i.
\end{equation}
Using the data of table \ref{t2}, one can check that (\ref{c31-c30}) and (\ref{c42-c41}) are satisfied 
with high precision. 
Now, (\ref{c31-c30}) and (\ref{c42-c41}) suggests that $\frac{{\mathfrak c}_{3,n}^{(0)}}{\pi^2}$ and 
$\frac{{\mathfrak c}_{4,n}^{(0)}}{\pi^2}$ 
are simple fractions. This assumption and further analysis of the numerical data of table \ref{t2}., led us to the 
following proposals for the exact values of the coefficients:
\begin{eqnarray}\label{guess34}
{\mathfrak c}_{3,0}^{(0)}=-\tfrac{16}{3} \pi^2 \, i, \qquad {\mathfrak c}_{3,n}^{(0)}=-\tfrac{40}{3} \pi^2 \, i,
 \qquad n=1,2,... \nonumber \\
{\mathfrak c}_{4,1}^{(0)}=-\tfrac{8}{9} \pi^2 \, i, \qquad {\mathfrak c}_{4,n}^{(0)}=-\tfrac{32}{9} \pi^2 \, i,
 \qquad n=2,3,...
\end{eqnarray}
At the points $n=1,2,3$ (\ref{guess34}) agrees with the numerical values of table \ref{t2} with at about 9-digits
of precision. As $n$ increases the deviation from (\ref{guess34}) also increases. The increasing deviation from 
(\ref{guess34}) is due to the fact that the numerical errors increase as $n$-increases. 
Nevertheless, for larger values of $n$, there are still so many digits of agreement between (\ref{guess34}) and 
the numerical values of table \ref{t2}. that we have very little doubt about that (\ref{guess12}) and (\ref{guess34}) 
give the analytical values for ${\mathfrak c}_{a,n}^{(0)}$.
If we accept (\ref{guess12}) and (\ref{guess34}) as the exact analytical values for
${\mathfrak c}_{a,n}^{(0)}$, we can sum up the emerging geometrical series and give
analytical formulae for the leading order large $g$ behaviour of the functions ${\bf p}_a$. 
The results of the summations take the forms:
\begin{equation}\label{p12largeg}
{\bf p}_1=g \, \frac{x}{x^2-1}\,\left(1+O(\tfrac1g) \right), \qquad {\bf p}_2=1+ \frac43 \, \frac{1}{x^2-1}+O(\tfrac1g),
\end{equation}
\begin{equation}\label{p3largeg}
{\bf p}_3=-i \,g^3 \, \left\{ 8\, \pi^2 \,x+\frac{40 \pi^2}{3} \, \frac{x}{x^2-1} \right\}
\,\left(1+O(\tfrac1g) \right), 
\end{equation}
\begin{equation}\label{p4largeg}
{\bf p}_4= -i \, g^2 \, \left\{   \frac{8 \pi^2}{3}  \, x^2+\frac{16\pi^2}{3}  +\frac{32 \pi^2}{9} 
  \,\frac{1}{x^2-1}  \right\} \,\left(1+O(\tfrac1g)\right).
\end{equation}
The above formulae has the common property that they have poles at $x=\pm1$. The positions of these poles 
are in accordance with the $g \to \infty$ limit of the radius of convergence $R$. Nevertheless,
there are two facts, which indicate that (\ref{p12largeg},\ref{p3largeg},\ref{p4largeg}) cannot 
be good approximations of the functions ${\bf p}_a$ on the entire $u$-plane at strong coupling.

First, in (\ref{p12largeg},\ref{p3largeg},\ref{p4largeg}) the neglected terms are $O(1/g)$ with respect to the
leading ones, in case the multipliers of $1/g$ in the correction terms are bounded functions of $u$
with $g$ independent upper and lower bounds. We will see in the next subsection that this is not the case.

Another problem, which indicates the restricted validity of (\ref{p12largeg},\ref{p3largeg},\ref{p4largeg}),
 emerges when one would like to compute $\tilde{\bf p}_a$ at strong coupling.
Naively, it can be done by a simple $x \to 1/x$ transformation in (\ref{p12largeg},\ref{p3largeg},\ref{p4largeg}). 
But the result does not account for the
the $\tilde{\bf p}_a(u) \sim u^{4 \, \sqrt{\pi \, g}+...}$ large $u$ asymptotics expected from (\ref{asympt1})
 and (\ref{deltaki},\ref{D01}).

The main reason for these discrepancies is that the coefficients $c_{a,n}(g)$ depend on $n$ and $g$. This is why
the result of the $g \to \infty$ limit depends on the relative magnitude of these two variables.

In the expansion (\ref{Cserpg}) we considered the limit, when $n\sim1$ and $g \to \infty$. 
To be more precise, we will see later that, the $n\ll \sqrt{g}$ limit is the one, 
which corresponds to the expansion (\ref{Cserpg}).

\subsubsection{Terms beyond the leading order}

From the available numerical data, one can fit further coefficients in (\ref{Cserpg}), as well.
We determined numerically the coefficients ${\mathfrak c}_{a,n}^{(k)}$ for $n\in\{1,...,12 \}$ and
 $k\in\{1,...,8\}$.
In this range of $k$ the fitted coefficients are $n$-dependent. 
The scaling property (\ref{chatK}) implies that ${\mathfrak c}_{a,n}^{(k)} \sim n^{2k}$ at large $n$.
The simplest function, which accounts for this behaviour is a polynomial of order $2k.$ 
 Indeed, table \ref{tcanka1k1}. and the tables of appendix \ref{appF}. show that the numerical values of ${\mathfrak c}_{a,n}^{(k)}$
 can be perfectly described by polynomials of order $2k.$
\begin{table}
\begin{center}
\begin{tabular}{|c||c|c|c|}
\hline
$n$ & ${\mathfrak c}_{1,n}^{(1)}$ & $\alpha_{1,1}^{(n)}$ &  $\Delta P_{rel}$ \\
\hline
1 & -0.7288876650125799 & 0 & 0   \\
\hline
2 & -1.868353108854596 & -0.7288876650125799 & 0   \\
\hline
3 & -3.418396331525119 & -0.4105777788294359 & $2.5 \cdot 10^{-13}$   \\
\hline
4 & -5.379017333025561 & - & $2.7 \cdot 10^{-13}$    \\
\hline
5 & -7.750216113353829 & - & $4.2 \cdot 10^{-13}$   \\
\hline
6 & -10.53199267250888 & - & $7.7 \cdot 10^{-13}$   \\
\hline
7 & -13.72434701051003 & - & $2.7 \cdot 10^{-13}$   \\
\hline
8 & -17.32727912734954 & - & $1.4 \cdot 10^{-12}$   \\
\hline
9 & -21.34078902283901 & - & $6.2 \cdot 10^{-12}$   \\
\hline
10 & -25.76487669904681 & - & $6.2 \cdot 10^{-11}$   \\
\hline
11 & -30.59954215432269 & - & $1.1 \cdot 10^{-10}$   \\
\hline
12 & -35.84478537778597 & - & $1.4 \cdot 10^{-10}$   \\
\hline
13 & -41.50060635466954 & - & $9.4 \cdot 10^{-10}$   \\
\hline
14 & -47.56700499419227 & - & $4.0 \cdot 10^{-9}$   \\
\hline
\end{tabular}
\bigskip
\caption{Numerical values of ${\mathfrak c}_{1,n}^{(1)}$ and the estimated values of the
 coefficients $\alpha_{1,1}^{(n)}$ of the
 polynomial Ansatz (\ref{PAR}). $\Delta P_{rel}$ is the relative error measuring, how
 precise the polynomial description of the various coefficients.
\label{tcanka1k1}}
\end{center}
\end{table}
\normalsize
This is why, we make the following {\bf conjecture:}
\begin{itemize}
\item {\it The coefficients ${\mathfrak c}_{a,n}^{(k)}$ are polynomials of order $2k$ in $n$.}
\end{itemize}
As a consequence, the polynomials can be given by $2 k \!+\!1$ $n$-independent parameters, 
which, for practical purposes, we parametrized as follows:
\begin{equation}\label{PAR}
{\mathfrak c}_{a,n}^{(k)}=\sum\limits_{m=1}^{2k+1} \, \alpha_{a,k}^{(m)} \, c_{n}^{(m,a)}, \qquad 
n\geq 1+\delta_{a,4}, \qquad k=0,1,2,...,
\end{equation}
where:
\begin{equation}\label{cmnalt}
c_{n}^{(1,a)}\equiv 1,\quad \mbox{and} \quad
c_{n}^{(m,a)}=\tfrac{\prod\limits_{j=1}^{m-1}\, (n-j+d_a) }{(m-1)!}, \qquad d_a=\delta_{a,1}+\delta_{a,3}, \quad
m=2,3,...
\end{equation}
The symbols $c_{n}^{(m,a)}$ are chosen to account for the pure pole terms at $x=\pm 1$ 
arising in ${\bf p}_a$ of $x$:
\begin{equation}\label{cnmasum}
\frac{x^{d_a}}{(x^2-1)^m}=\sum\limits_{n=1-d_a}^{\infty} \, \frac{c_{n}^{(m,a)}}{x^{2n+d_a}}. 
\end{equation}
We note that in the $k=0$ special case, by definition 
$\alpha_{a,0}^{(m)}={\mathfrak c}_{a,a}^{(0)}\, \delta_{m,1}$ 
 and that (\ref{PAR}) can be used only when $n\geq 1+\delta_{a,4}$. 

The conjectured (\ref{PAR}) representation of ${\mathfrak c}_{a,n}^{(k)}$ implies the following series representation
for ${\bf p}_a(x)$ at strong coupling:
\begin{equation}\label{p.a.}
\begin{split}
{\bf p}_a(x)=\!\delta_{a,2}+\! \delta_{a,3} \left(\,g  A_3(g) \, x+ \frac{{\cal A}_3(g)}{x}\right)+
\! \delta_{a,4} \left(\,g^2 A_4(g) \, (x^2+2) + \frac{{\cal A}_4(g)}{x^2} \right)+\! \\
+g^{n_a} \! \left( \sum\limits_{k=1}^{\infty} \frac{1}{g^k}\! \sum\limits_{m=1}^{2k+1} 
\alpha_{a,k}^{(m)} \frac{x^{\delta_{a,1}+\delta_{a,3}}}{(x^2-1)^m}
   \right),
\end{split}
\end{equation}
where ${\cal A}_3(g)$ and ${\cal A}_4(g)$ admit the strong coupling series representations:
\begin{equation}\label{calA34}
{\cal A}_3(g)=g^2 \sum\limits_{k=0}^{\infty} \, \frac{{\cal A}_3^{(k)}}{g^k}, \qquad {\cal A}_4(g)=g \sum\limits_{k=0}^{\infty} \, \frac{{\cal A}_4^{(k)}}{g^k}.
\end{equation}
The first few values of ${\cal A}_3(g)$ and ${\cal A}_4(g)$ are given in the table \ref{tcalA34}.
All elements of table \ref{tcalA34} are small numbers, lying in the range of numerical errors.
This fact suggests us to make the following {\bf conjecture:}
\begin{itemize}
\item {\it ${\cal A}_3^{(k)}$ and ${\cal A}_4^{(k)}$ of (\ref{calA34}) are zero for all $k\geq 0.$ }
\end{itemize}
As a consequence ${\cal A}_3(g)= {\cal A}_4(g) \equiv 0$, which implies that besides of the 
$\tfrac{1}{(x^2-1)^m}$ type of terms, there are no $\tfrac{1}{x}$ or $\tfrac{1}{x^2}$ terms present
 in the strong coupling series (\ref{p.a.}).
\begin{table}
\begin{center}
\begin{tabular}{|c||c|c|}
\hline
$k$ & $\mbox{Im}{\cal A}_{3}^{(k)}$ & $\mbox{Im}{\cal A}_{4}^{(k)}$  \\
\hline
0 & $3.7 \cdot 10^{-11}$ & $6.4 \cdot 10^{-11}$    \\
\hline
1 & $-4.6 \cdot 10^{-8}$ & $-1.28 \cdot 10^{-8}$    \\
\hline
2 & $2.0 \cdot 10^{-5}$ & $1.45 \cdot 10^{-6}$    \\
\hline
\end{tabular}
\bigskip
\caption{The first three numerical values of ${\cal A}_{3}^{(k)}$ and ${\cal A}_{4}^{(k)}.$
All values are in the magnitude of the numerical errors.
\label{tcalA34}}
\end{center}
\end{table}
\normalsize

The formula (\ref{p.a.}) indicates that in the $a=1$ case there is some simplification due to the H-symmetry fixing
condition $c_{1,0}\equiv g$. This implies that in the large $x$ expansion of (\ref{p.a.}) the coefficient
of $\tfrac 1 x$ does not get $\tfrac 1 g$ corrections. As a consequence: $\alpha_{1,k}^{(1)}\equiv 0$ for
$k \geq 1$. This means that in the $a=1$ case only $2k$ parameters describe the conjectured polynomials 
of order $2k$. This fact was built in the polynomial fits as it is demonstrated by table \ref{tcanka1k1}.

Reshuffling the series part of (\ref{p.a.}), it can be written as a series in $\tfrac{1}{g(x^2-1)^2}$: 
\begin{equation}\label{fuzet}
\begin{split}
{\bf p}_a(x)=\!\delta_{a,2}\!+\! \delta_{a,3} \left(g  A_3(g)  x+\! \frac{{\cal A}_3(g)}{x}\right)+
\! \delta_{a,4} \! \left(g^2 A_4(g) (x^2+2) \!+ \! \frac{{\cal A}_4(g)}{x^2} \right)\!+\!
{\bf p}_{a}^{series}(x), \\
{\bf p}_{a}^{series}(x)=g^{n_a} \, x^{\delta_{a,1}+\delta_{a,3}}\, \left\{
\frac{1}{x^2-1} \, \sum\limits_{n=0}^{\infty} \, \frac{\alpha_{a,n}^{(2n+1)}}{[g(x^2-1)^2]^n}
\left(1+\sum\limits_{k=1}^{\infty} \frac{1}{g^k} \, \frac{\alpha_{a,n+k}^{(2n+1)}}{\alpha_{a,n}^{(2n+1)}} \right)\right.+\\
\left.
\sum\limits_{n=0}^{\infty} \, \frac{\alpha_{a,n}^{(2n)}}{[g(x^2-1)^2]^n}
\left(1+\sum\limits_{k=1}^{\infty} \frac{1}{g^k} \, \frac{\alpha_{a,n+k}^{(2n)}}{\alpha_{a,n}^{(2n)}} \right)
\right\}.
\end{split}
\end{equation}
Now, we are in the position to discuss the regime of validity of (\ref{fuzet}) in the rapidity plane.  
Formula (\ref{fuzet}) implies that at strong coupling the variable $z=\tfrac{1}{g(x^2-1)^2}$ becomes relevant 
and within the range of convergence, apart from sum trivial factors, 
${\bf p}_{a}^{series}(x)$ can be represented as a sum of functions of $z$,
such that each function is suppressed with an inverse power of $g$:
\begin{equation}\label{fuzetformal}
\begin{split}
{\bf p}_{a}^{series}(x)=g^{n_a} \, x^{\delta_{a,1}+\delta_{a,3}}\, \left\{
\frac{1}{x^2-1} \,\sum\limits_{k=0}^{\infty} \,\frac{1}{g^k} \, {\mathfrak f}_{a,k}^{odd}(z)+
\sum\limits_{k=0}^{\infty} \,\frac{1}{g^k} \,{\mathfrak f}_{a,k}^{even}(z)
\right\},
\end{split}
\end{equation}
To study the range of validity of (\ref{fuzet}), one has to determine the radius of convergence of the series
representations of ${\mathfrak f}_{a,0}^{odd}(z)$ and ${\mathfrak f}_{a,0}^{even}(z).$ We just recall:
\begin{equation}\label{f_0}
\begin{split}
{\mathfrak f}_{a,0}^{odd}(z)=\sum\limits_{n=0}^{\infty} \,\alpha_{a,n}^{(2n+1)} \,z^n,
\qquad
{\mathfrak f}_{a,0}^{even}(z)=\sum\limits_{n=0}^{\infty} \,\alpha_{a,n}^{(2n)} \,z^n.
\end{split}
\end{equation}
The radius of convergence of these series is determined by the large $n$ behaviour of the coefficients.
Our numerical data suggests that:
$$\alpha_{a,n}^{(2n+1)}\sim 4^{2n}, \qquad \alpha_{a,n}^{(2n)} \sim 4^{2n} $$ for large $n$. 
This implies that the radius of convergence of ${\mathfrak f}_{a,0}^{odd/even}(z)$ is $\tfrac 14.$
Thus one can conclude that the validity of the series representation (\ref{fuzet}) is 
restricted by the inequality:
\begin{equation}\label{ineqfuzet}
\begin{split}
\frac{4}{g(x^2-1)^2}<1.
\end{split}
\end{equation}
In the strong coupling limit, (\ref{ineqfuzet}) may fail, if $x$ is close to $\pm 1.$ 
In the language of the rapidity{\footnote{Throughout this section, we use the convention, 
when the branch points are scaled to be located at $\pm 2.$}} $u$, 
this means that $u$ is close to the branch points $\pm 2.$
Using the series representation:
\begin{equation}\label{x2v}
x_s(2+v)=1+\sqrt{v}+\tfrac{v}{2}+O(v^{3/2}),
\end{equation}
one obtains that (\ref{fuzet}) is convergent if:
\begin{equation}\label{fuzetvalid}
\begin{split}
\frac{4}{g|v|}<1 \quad \Rightarrow  \quad \frac{4}{g}<|v|, \qquad u\!=\!\pm 2 \!+\!v.
\end{split}
\end{equation}
Thus, naively one might conclude that the series representation (\ref{fuzet}) gives the correct strong coupling
approximation of ${\bf p}_a$ in the domain where, the distance of the rapidity $u$ from the branch 
points is larger than $\tfrac 4 g.$ Unfortunately the situation is a bit worse. The series (\ref{fuzet})
will be an appropriate strong coupling approximation for ${\bf p}_a(u)$ only outside of an oval region
containing the real short cut $[-2,2]$, such that the horizontal dimension of the oval region is 4 plus a
number of order $\tfrac 1 g$, and its vertical dimension is of order $\tfrac{1}{\sqrt{g}}.$ See figure \ref{Oval}. 
 \begin{figure}[htb]
\begin{flushleft}
\hskip 15mm
\leavevmode
\epsfxsize=120mm
\epsfbox{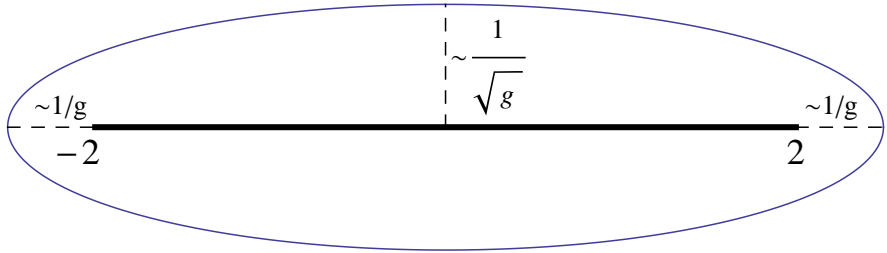}
\end{flushleft}
\caption{{\footnotesize
The oval region outside of which the strong coupling series representation (\ref{fuzet}) accounts for all power like contributions in $\tfrac{1}{g}$. 
}}
\label{Oval}
\end{figure} 

The reason is as follows. Rephrasing (\ref{Cserpg}) one obtains that:
\begin{equation}\label{canscalingK}
c_{a,n}(g)=g^{n_a} \, {\cal K}_a\left(\tfrac{n}{\sqrt{g}}\right)\cdot(1+O(\tfrac{1}{\sqrt{g}})).
\end{equation}
The $O(\tfrac{1}{\sqrt{g}})$ magnitude of the corrections is a consequence of (\ref{PAR}).
From (\ref{canscalingK}) it follows that the $n=\mbox{fixed}, \, g \to \infty$ limit
corresponds to the $\tfrac{n}{\sqrt{g}}\to 0$ limit. This implies that the strong coupling
series representation (\ref{Cserpg}) of the coefficients is a good approximation until $n \ll \sqrt{g}.$
 (\ref{canscalingK}) also implies that, at strong coupling a typical sum appearing in ${\bf p}_a$ can be roughly 
estimated by an integral:
\begin{equation}
\begin{split}
\sum\limits_{n} c_{a,n} \, x^{-2 n} \sim g^{n_a} \, \sum\limits_{n} {\cal K}_a (\tfrac{n}{\sqrt{g}}) \,
x^{-2 \,n} \sim g^{n_a+\tfrac12} \, \sum\limits_{n} \tfrac{1}{\sqrt{g}} {\cal K}_a (\tfrac{n}{\sqrt{g}}) \,
x^{-2 \sqrt{g} \,\tfrac{n}{\sqrt{g}}} \\
\sim g^{n_a+\tfrac12} \,\int dz' \, {\cal K}_a(z') \, e^{-2 \,z'\, \sqrt{g} \,\ln x}. 
\end{split}\label{nonumber}
\end{equation}
The strong coupling series (\ref{fuzet}) was obtained by inserting the series (\ref{Cserpg}) into 
(\ref{p12}) and (\ref{p34}) and evaluating the sums from 1 to infinity. In this representation
the strong coupling corrections go as inverse powers of $g.$ 
Since the validity of (\ref{Cserpg}) is restricted to $n \ll \sqrt{g}$,
(\ref{fuzet}) can be appropriate representation
of ${\bf p}_a$, if the neglected contributions coming from the $\sqrt{g} \lesssim n$ region are exponentially small
in $g.$ As (\ref{nonumber}) shows, the exponentially small corrections grow up to power like in the regime, where
 $\sqrt{g} \, \ln x$ or equivalently $|x|^{-\sqrt{g}}$ becomes of order $1.$
Now we will show that this can happen in an appropriate neighborhood of the real short cut of the $u$-plane. 

At the branch points, $x$ is given by (\ref{x2v}), therefore $\sqrt{g}\, \ln x \sim 1$, when 
 $u$ lies within a circle of radius $\sim\tfrac 1g$, whose center is located at the branch points $\pm 2.$

On the other hand $x$ is a pure phase on the real cut, i.e. $|x|=1.$ If $u_0 \in [-2,2]$, then $\ln x(u)$ 
can be expanded in a regular Taylor-series around $u_0$. This yields that 
$|x(u_0+\delta u)|^{-\sqrt{g}} \sim 1$ if $\delta u \sim \tfrac{1}{\sqrt{g}}$.

To summarize, the contributions of the $\sqrt{g} \lesssim n$ terms are not negligible in (\ref{p12}) and (\ref{p34})
if $u$ lies in an oval domain containing the real short cut $[-2,2]$, such that the horizontal 
dimension of the oval region is 4 plus a
number of order $\tfrac 1 g$, and its vertical dimension is of order $\tfrac{1}{\sqrt{g}}.$ (See figure \ref{Oval}.)
This is the region, where the strong coupling formula (\ref{fuzet}) becomes invalid.
To be more precise, the neglected contributions of the $\sqrt{g} \lesssim n$ terms are exponentially small outside
of this oval domain, and become  
power-like inside the domain.

Now, we have shown that conjecture (\ref{p.a.}}) cannot be an appropriate approximation
of ${\bf p}_a$ close to the real short cut, this is why we also studied the behaviour of ${\bf p}_a$ close to the
branch points in the context of a series expansion in the deviation from the branch points.

\subsubsection{Series expansion around the branch points}

Now, we study the behaviour of ${\bf p}_a$ at the branch points. Inserting the power series{\footnote{Its infinite order version.}} 
(\ref{x2v}) into the series representations (\ref{p12}) and (\ref{p34}), one ends
up with the expansions:
\begin{equation}\label{pabranch}
{\bf p}_a(2+v)=\sum\limits_{k=0}^{\infty} \, \beta_{a,k}(g) \, v^{k/2},
\end{equation}
where we use the convention, when the rapidity is scaled, such that the branch points are at $\pm 2$ and
$v$ denotes the deviation from them. The coefficients $\beta_{a,k}(g)$ are certain linear combinations
of the momenta{\footnote{Here, by momentum we mean sums like: $\sum\limits_{n=0}^{\infty} \,n^k \, c_{a,n}(g), 
\mbox{with} \, k\in {\mathbb N}$.}} 
of the coefficients $c_{a,n}(g)$. For example the first coefficient is just the sum of the
coefficients $c_{a,n}(g)$, i.e. $\beta_{a,0}(g)=\sum\limits_{n=0}^{\infty} \, c_{a,n}(g)$. 

We fitted the coefficients $\beta_{a,k}(g)$ by a power series in $\sqrt{g}$. The coefficients of the
numerical fits proved to be stable with respect to increasing the truncation index of the series,
in case the following $g$ dependence was assumed:
\begin{equation}\label{betafit0}
\beta_{a,k}(g)=g^{n_a+1/2+k/2} \, \sum\limits_{n=0}^{\infty} \, \frac{\gamma_{a,k}^{(n)}}{g^n}.
\end{equation}
The numerical values of the first few coefficients $\gamma_{a,k}^{(n)}$ can be found in tables \ref{tgamma0h}.
and \ref{tgamma1h}.
\begin{table}
\begin{center}
\begin{tabular}{|c||c|c|c|c|}
\hline
$k$ & $\gamma^{(0)}_{1,k}$ & $\gamma^{(0)}_{2,k}$ & $\mbox{Im}\gamma^{(0)}_{3,k}$ & $\mbox{Im}\gamma^{(0)}_{4,k}$ \\
\hline
0 & 1.9168(4) & 2.5549(6) & -252.1(1) & -67.34(2)  \\
\hline
1 & -4.603(1) & -6.133(3) & 605.9(1) & 160.9(4)  \\
\hline
2 & 8.517(3) & 11.34(1) & -1120.5(5) & -297(1)  \\
\hline
3 & -13.079(6) & -17.44(1) & 1720(1) & 474(6)  \\
\hline
4 & 17.27(1) & 23.04(2) & -2270(3) & -633(15)  \\
\hline
5 & -20.00(1) & -26.66(2) & 2628(3) & 726(16)  \\
\hline
\end{tabular}
\bigskip
\caption{Numerical values of the first few $\gamma^{(0)}_{a,k}$. 
\label{tgamma0h}}
\end{center}
\end{table}
\normalsize
\begin{table}
\begin{center}
\begin{tabular}{|c||c|c|c|c|}
\hline
$k$ & $\gamma^{(1)}_{1,k}$ & $\gamma^{(1)}_{2,k}$ & $\mbox{Im}\gamma^{(1)}_{3,k}$ & $\mbox{Im}\gamma^{(1)}_{4,k}$ \\
\hline
0 & -0.774(4) & -0.507(3) & 135.8(8) & 37.49(6)  \\
\hline
1 & 2.98(1) & -1.88(2) & -223.8(8) & 167(2)  \\
\hline
2 & -7.18(2) & 4.53(5) & 527(4) & -401(7)  \\
\hline
3 & 13.32(5) & -8.11(5) & -968(8) & 596(42)  \\
\hline
4 & -20.6(1) & 12.4(1) & 1473(16) & -910(104)  \\
\hline
5 & 27.3(2) & -16.4(2) & -1941(21) & 1323(115)  \\
\hline
\end{tabular}
\bigskip
\caption{Numerical values of the first few $\gamma^{(1)}_{a,k}$. 
\label{tgamma1h}}
\end{center}
\end{table}
\normalsize
Concentrating on only the leading order behaviour of (\ref{pabranch}), the following pattern arises:
 \begin{equation}\label{pabranchlead}
{\bf p}_a(2+v)=g^{n_a+1/2} \, \sum\limits_{k=0}^{\infty} \, \gamma_{a,k}^{(0)} \, (g \,v)^{k/2}
+g^{n_a-1/2} \, \sum\limits_{k=0}^{\infty} \, \gamma_{a,k}^{(1)} \, (g \,v)^{k/2}
+...,
\end{equation}
where dots mean terms negligible for large $g$. 

As a consequence we can conclude that for large $g$, close to the branch points
${\bf p}_a$ behaves like a function of $g v$ and the sub-leading corrections are suppressed
by positive integer powers of  $\tfrac{1}{g}$: 
\begin{equation}\label{branchbehaviour}
{\bf p}_a(2+v)=g^{n_a+1/2} \, \left( f_a^{(0)}(g \,v)+\tfrac{1}{g} f_a^{(1)}(g \,v)+....\right).
\end{equation}

\section{Higher spin results}

In this section we publish the numerical results obtained in the $S=4,6,8$ cases.
For these higher spin values, we could not reach as large values of the coupling
constant $g$ as it was done in the case of the Konishi operator. The reason for this, is
that increasing the spin, the numerical algorithm becomes more and more sensible to
the choice of initial values. This fact forced us to increase $g$ in very small $\Delta g\sim 0.02$
steps. As a consequence, we needed to run 50 jobs subsequently in order to increase $g$ with one single unit.
Unfortunately, this process proved to be very time consuming. By increasing $S$, also the internal precision
of the computations must have been increased, in order to get convergence and reach the required 
precision for $\Delta$ and $c_{a,n}$.
For example at strong coupling $g \gtrsim 2.7$, the $S=4,6,8$ cases required 60-, 80- and 100-digits of precision
respectively. The necessity of the application of such high precisions  made also the runtime of the jobs very long.

Because of these difficulties, in the $S=4,6,8$ cases, the numerical results we obtained were less accurate than
those of the Konishi state. This is why, in the higher spin cases, we restricted our numerical work to 3 types of
investigations. Namely,
\begin{itemize}
\item Numerical determination of the first 4 coefficients $\Delta^{(n)}$ of the strong coupling series of $\Delta$.

\item Numerical determination of the coefficients ${\mathfrak c}_{a,n}^{(0)}$ of (\ref{Cserpg}).

\item Investigation of the qualitative strong coupling behaviour of the ${\bf p}_a$-functions at the branch points. 
\end{itemize}

The fitted values of the coefficients in (\ref{deltaki}) at different values of $S$ can be found in tables \ref{t1/4}.,
\ref{t1/6}.,and \ref{t1/8}. The  numerical estimations of the first coefficients beyond the analytical prediction
 (i.e. $\Delta^{(4)}$) are also presented, but only 
to "give a taste" about their magnitude. 
\begin{table}
\begin{center}
\begin{tabular}{|c||c|c|c|}
\hline
$n$ & $\Delta^{(n)}_{exact}$ & $\Delta^{(n)}_{fitted}$ & $\delta_{rel} \Delta^{(n)}$ \\
\hline
0 & 2.828427125 & 2.828428230 & $3.9 \cdot 10^{-7}$  \\
\hline
1 & 4.242640687 & 4.242592283 & $1.1 \cdot 10^{-5}$  \\
\hline
2 & -13.91210165 & -13.91277126 & $4.8 \cdot 10^{-5}$  \\
\hline
3 & 113.9955688 & 113.9696603 & $2.3 \cdot 10^{-4}$  \\
\hline
4 & - & -1279.745751 & $1.8 \cdot 10^{-3}$  \\
\hline
\end{tabular}
\bigskip
\caption{Comparison of the analytical predictions and the fitted values for $\Delta^{(n)}$ at $S=4$. 
\label{t1/4}}
\end{center}
\end{table}
\normalsize
\begin{table}
\begin{center}
\begin{tabular}{|c||c|c|c|}
\hline
$n$ & $\Delta^{(n)}_{exact}$ & $\Delta^{(n)}_{fitted}$ & $\delta_{rel} \Delta^{(n)}$ \\
\hline
0 & 3.464101615 & 3.464115090 & $3.9 \cdot 10^{-6}$  \\
\hline
1 &7.505553499 & 7.504893894 & $8.7 \cdot 10^{-5}$  \\
\hline
2 & -33.36441949 & -33.35019106 & $4.2 \cdot 10^{-4}$  \\
\hline
3 & 373.4996131 & 373.1565665 & $9.1 \cdot 10^{-4}$  \\
\hline
4 & - & -5914.704399 & $3.0 \cdot 10^{-3}$  \\
\hline
\end{tabular}
\bigskip
\caption{Comparison of the analytical predictions and the fitted values for $\Delta^{(n)}$ at $S=6$. 
\label{t1/6}}
\end{center}
\end{table}
\normalsize
\begin{table}
\begin{center}
\begin{tabular}{|c||c|c|c|}
\hline
$n$ & $\Delta^{(n)}_{exact}$ & $\Delta^{(n)}_{fitted}$ & $\delta_{rel} \Delta^{(n)}$ \\
\hline
0 & 4.0 & 4.000128998 & $3.2 \cdot 10^{-5}$  \\
\hline
1 & 11.5 & 11.49670954 & $2.8 \cdot 10^{-4}$  \\
\hline
2 & -62.63061568 & -62.54108289 & $1.4 \cdot 10^{-3}$  \\
\hline
3 & 876.3952895 & 873.6934855 & $3.0 \cdot 10^{-3}$  \\
\hline
4 & - & -17585.48981 & $5.0 \cdot 10^{-3}$  \\
\hline
\end{tabular}
\bigskip
\caption{Comparison of the analytical predictions and the fitted values for $\Delta^{(n)}$ at $S=8$. 
\label{t1/8}}
\end{center}
\end{table}
\normalsize
Though
the precision of the coefficients is not so high as it was in the Konishi case, the first four coefficients can be 
compared to the analytical predictions (\ref{D01}), (\ref{D2}) and (\ref{D3}). Our numerical data confirms the 
analytical predictions within the range of numerical errors.

In the higher spin cases, we also computed numerically the first few coefficients from the set of
 ${\mathfrak c}_{a,n}^{(0)}$ in (\ref{Cserpg}).
The fitting process went in exactly the same manner as in the case of the Konishi operator. The fitted values at 
different values of the spin are summarized in tables \ref{t2/4}., \ref{t2/6}., and \ref{t2/8}.
\begin{table}
\begin{center}
\begin{tabular}{|c||c|c|c|c|}
\hline
$n$ & ${\mathfrak c}_{1,n}^{(0)}$ & ${\mathfrak c}_{2,n}^{(0)}$ & $\mbox{Im}{\mathfrak c}_{3,n}^{(0)}$ & $\mbox{Im}{\mathfrak c}_{4,n}^{(0)}$ \\
\hline
0 & 1 & 1 & -210.5519430 & 0  \\
\hline
1 & 1.000000992 & 1.333333771 & -526.3809637 & -35.09247158  \\
\hline
2 & 1.000007109 & 1.333336921 & -526.3848403 & -140.3681036  \\
\hline
3 & 1.000009244 & 1.333341357 & -526.4039857 & -140.3688672  \\
\hline
4 & 0.999915635 & 1.333283501 & -526.3719528 & -140.3678449  \\
\hline
\end{tabular}
\bigskip
\caption{The numerical values of ${\mathfrak c}_{a,n}^{(0)}$ at $S=4$.
\label{t2/4}}
\end{center}
\end{table}
\normalsize

\begin{table}
\begin{center}
\begin{tabular}{|c||c|c|c|c|}
\hline
$n$ & ${\mathfrak c}_{1,n}^{(0)}$ & ${\mathfrak c}_{2,n}^{(0)}$ & $\mbox{Im}{\mathfrak c}_{3,n}^{(0)}$ & $\mbox{Im}{\mathfrak c}_{4,n}^{(0)}$ \\
\hline
0 & 1 & 1 & -473.7436596 & 0  \\
\hline
1 & 1.000001918 & 1.333334730 & -1184.362710 & -78.95735532  \\
\hline
2 & 1.000011630 & 1.333337367 & -1184.375396 & -315.8330460  \\
\hline
3 & 1.000024550 & 1.333342392 & -1184.301685 & -315.8459872  \\
\hline
4 & 0.999746232 & 1.333356285 & -1184.435398 & -315.8606964  \\
\hline
\end{tabular}
\bigskip
\caption{The numerical values of ${\mathfrak c}_{a,n}^{(0)}$ at $S=6$.
\label{t2/6}}
\end{center}
\end{table}
\normalsize

\begin{table}
\begin{center}
\begin{tabular}{|c||c|c|c|c|}
\hline
$n$ & ${\mathfrak c}_{1,n}^{(0)}$ & ${\mathfrak c}_{2,n}^{(0)}$ & $\mbox{Im}{\mathfrak c}_{3,n}^{(0)}$ & $\mbox{Im}{\mathfrak c}_{4,n}^{(0)}$ \\
\hline
0 & 1 & 1 & -842.2397513 & 0  \\
\hline
1 & 1.000038353 & 1.333329759 & -2105.636060 & -140.3852803  \\
\hline
2 & 0.999981037 & 1.333336736 & -2105.828532 & -561.5275144  \\
\hline
3 & 1.000424364 & 1.333489157 & -2106.476479 & -561.5864962  \\
\hline
\end{tabular}
\bigskip
\caption{The numerical values of ${\mathfrak c}_{a,n}^{(0)}$ at $S=8$.
\label{t2/8}}
\end{center}
\end{table}
\normalsize
Though the numerical values of the coefficients are not as accurate as they were in the case of the Konishi operator,
 one can see that the same structure shows up. Namely, for $n\geq 1+\delta_{a,4}$ the coefficients seem to be
 $n$-independent. Using the same train of thoughts, as
 it was done in the Konishi case, based on the numerical data of tables \ref{t2/4}., \ref{t2/6}., and
\ref{t2/8}., we made the following proposals
 for the exact values of the coefficients:
\begin{equation}\label{guess12s}
{\mathfrak c}_{1,n}^{(0)}=1, \qquad {\mathfrak c}_{2,n}^{(0)}=\frac 43, \qquad n=1,2,...
\end{equation}
\begin{eqnarray}\label{guess34s}
{\mathfrak c}_{3,0}^{(0)}=-\tfrac{4}{3} \pi^2 \,S^2 \,i, \qquad {\mathfrak c}_{3,n}^{(0)}=-\tfrac{10}{3} \pi^2 \,S^2 \, i,
 \qquad n=1,2,... \nonumber \\
{\mathfrak c}_{4,1}^{(0)}=-\tfrac{2}{9} \pi^2 \,S^2 i, \qquad {\mathfrak c}_{4,n}^{(0)}=-\tfrac{8}{9} \pi^2 \, S^2 \,i,
 \qquad n=2,3,...
\end{eqnarray}

We also constructed Pade-approximation like formulae to determine numerically $\Delta$ in the whole range the coupling constant.
The Pade-approximation like formulae for the cases $S=4,6,8$ can be found in appendix \ref{appE}. Unfortunately, these approximations are
not so accurate as that of the Konishi operator. The reason for that is two-fold. 
First, because we did not reach too large values of $g$ during our numerical work{\footnote{The largest values
of $g$ reached during the numerical work were 4.1, 3.5 and 2.74 in the cases $S=4,6,8$ respectively.}}. 
The second reason is the lower precision of the available numerical data. 
Nevertheless, according to our estimations, our Pade-approximation like formulae give the numerical values of $\Delta$ with 8-digits
of accuracy in the range, where numerical data are available, and with 4-5 digits of accuracy for
higher values of $g$.

The last problem, we studied in the higher spin cases, is the strong coupling behaviour of ${\bf p}_a$ functions
at the branch points. Without listing any fitted numerical data, we just note that our numerical results suggest
that close to the branch points the qualitative strong coupling behaviour of ${\bf p}_a$ functions is given
 by (\ref{branchbehaviour}). Thus, it is independent of the concrete value of 
the spin{\footnote{At least in case the spin is an even and positive integer number.}}.

\section{Summary}

In this paper, we solved numerically the QSC equations 
corresponding to some twist-2 single trace
operators from the $sl(2)$ sector of $AdS_5/CFT_4$ correspondence. Namely, we considered the twist-2 operators
with spins $S=2,4,6,8$. The primary purpose of the numerical study was to gain some information about the 
strong coupling behaviour of the solutions of the ${\bf P}\mu$-system.
 
We applied the numerical method  of \cite{QSCnum} to solve the QSC equations and we
wrote down all technical details, which were necessary to implement the numerical code in 
C++ language. Roughly speaking, the whole numerical algorithm consist of summations and of numerical solutions
of linear sets of equations. Both mathematical problems can be easily programmed in any fundamental programming
languages.

The most accurate numerical results were obtained in the case of the Konishi-operator. There,
$\lambda \sim 7737$ was the highest value of the 't Hooft coupling, which was reached by the numerical computations.
From our high precision numerical data, we could numerically confirm the analytical predictions of \cite{Gromov:2014bva}
for the first 4 coefficients of the strong coupling series of $\Delta.$ Moreover, due to the high precision
of the numerical data, we could give numerical predictions for 2 further coefficients in the strong coupling
expansion of $\Delta$. In the cases of $S=4,6,8$ the numerical data were less precise, nevertheless they proved to be
precise enough to confirm the analytical predictions of \cite{Gromov:2014bva}, though with much less precision.
We also constructed Pade--approximation like formulas which allow one to compute the anomalous dimensions of 
the states under consideration within short time and with satisfactory high precision. (See appendix \ref{appE}.)

Beyond the numerical determination of $\Delta$, we also focused our attention to determine the strong coupling
limit of the ${\bf p}_a$ functions. Since, in the numerical method the coefficients of their series representations
(\ref{p12}), (\ref{p34}) were the basic objects, we tried to determine the strong coupling behaviour of these 
coefficients.
 From the numerical data, we found that, at strong coupling, when $n\ll \sqrt{g}$ , the coefficients admit the series 
representations (\ref{Cserpg}) with $n_a$ given by (\ref{na}). The accurate numerical values obtained for 
the coefficients of (\ref{Cserpg}), inspired us to make analytical proposals
for the values of the leading order coefficients (\ref{guess12s},\ref{guess34s}).

For the Konishi operator, based on the high precision numerical data, we conjectured 
that the coefficients $c_{a,n}^{(k)}$ in (\ref{Cserpg})
are polynomials of order $2k\!+\!1$ in $n.$ This recognition led us to propose a strong coupling
series representation (\ref{p.a.}) for the ${\bf p}_a$-functions{\footnote{The fundamental 
functions ${\bf P}_a$ of the QSC method 
are connected to ${\bf p}_a$ by the simple formula (\ref{smallP}),
this why the results given for ${\bf p}_a$ in the previous sections, 
can be translated to the language of ${\bf P}_a$ in a straightforward manner. }}. 
We argued that (\ref{p.a.}) is
an appropriate strong coupling representation of ${\bf p}_a(u)$, if the rapidity $u$ lies
outside of an oval domain{\footnote{Here, the rapidity convention is the one, when the
branch points are scaled to be at $\pm 2.$}} containing the short real cut, such that its 
horizontal dimension is equal to
4 plus a number of order $\tfrac 1g$ and its vertical dimension is $\sim \tfrac{1}{\sqrt{g}}.$ 
(See figure \ref{Oval}.) 
Furthermore, outside of this domain (\ref{p.a.}) accounts for all power like contributions in $g$,
but neglects the exponentially small ones, which come from the index range $\sqrt{g} \lesssim n.$

Because of this restricted validity of (\ref{p.a.}), we also studied the behaviour of the
solutions close to the branch points. 
The result of this investigation can be summarized by the scaling behaviour given by (\ref{branchbehaviour}).

The strong coupling investigation of the numerical data suggested the strong coupling scaling
behaviour (\ref{canscalingK}) for the coefficients. This indicates that $\sqrt{g}$ is the relevant
scale of the problem at strong coupling and it tells us that there are 3 important
regimes of $n$ in the strong coupling limit. These are the $n\ll\sqrt{g}$, $n \sim \sqrt{g}$ and $n\gg\sqrt{g}$
regimes.  In the 3 different regimes the coefficients have different strong coupling behaviours.

We also discussed some general properties of the coefficients at fixed values of the coupling 
constant. 
If $c_{a,n}$ is considered as a continuous function of $n$, the numerical data implied that
\begin{itemize}
\item that $c_{a,n}$ has infinitely many zeros located periodically at large $n$, and
\item that $c_{a,n}$ decays as $ \sim n^{-{\epsilon}_a(g)} \,R^{-2 \, n}$ at large $n$, where
$R=|x_s(2+\tfrac{i}{g})|$ is the radius of convergence of the series (\ref{p12}), (\ref{p34}) 
and ${\epsilon}_a(g)$ is a numerical constant with approximate value ${\epsilon}_a(g) \sim 1.5\pm 0.2.$
\end{itemize}

Our numerical work contributes to the deeper understanding of the strong coupling
behaviour of the solutions of the QSC-equations and hopefully it will help in finding the
a method for the systematic analytical solution of the ${\bf P}\mu$-system in the strong coupling
limit.

\vspace{1cm}
{\tt Acknowledgements}

\noindent 
The authors thank Zolt\'an Bajnok and J\'anos Balog for useful discussions.
This work was supported by the J\'anos Bolyai Research Scholarship of the Hungarian Academy of
Sciences and OTKA K109312. The authors also would like to thank the support of an MTA-Lend\"ulet Grant.

\appendix

\section{Construction of initial values at strong coupling}\label{appA}

For small values of the coupling constant $g$, the numerical iterations can start from the
perturbative solution of the problem \cite{Marboe:2014gma}. This strategy works for $g \lesssim \frac14$.
For larger values of $g$, the good{\footnote{Good initial value means that the
numerical algorithm converges if the process starts from it.}} initial values should be composed of the previously obtained numerical data. 

In this appendix we describe, how to construct good initial values for the numerical iterations, provided
we have the numerical solution of the problem for several smaller values of $g$.
To construct good initial values, one should increase the value of $g$ in small steps.
We increased the value of $g$ uniformly at each step by $\Delta g=0.1,0.05, \, \mbox{or} \,0.02$. 
If we assume that every unknown coefficient is a smooth function of $g$, then
a good initial value of the numerical problem can be given by a numerical Taylor-series, constructed from the
numerical data belonging to previous values of $g$. Here, let $f$ a function of $g$. $f$ should be considered
here as the analog of any unknown coefficient of the numerical problem. E.g. $\Delta(g)$ is one of them.  

In case $\Delta g$ is small enough, a good initial value can be constructed as a second order
Taylor-series: 
\begin{equation}\label{Tay2}
f(g+\Delta g)=f(g)+f'(g) \, \Delta g+\frac12 \, f''(g) \, \Delta g^2+O(\Delta g ^3).
\end{equation}
For the numerical implementation of (\ref{Tay2}), one needs to compute the 
appropriately accurate numerical formulae for the derivatives:
\begin{equation}\label{d1}
f'(g)=\frac{f(g+\Delta g)-f(g-\Delta g)}{2 \, \Delta g}+O(\Delta g^2),
\end{equation}
\begin{equation}\label{d2}
f''(g)=\frac{f(g+\Delta g)+f(g-\Delta g)-2 \, f(g)}{\Delta g^2}+O(\Delta g^2).
\end{equation}
Inserting (\ref{d1}) and (\ref{d2}) into (\ref{Tay2}), and making the $g \to g-\Delta g$
substitution, one gets the formula:
\begin{equation}\label{three}
f(g)=3 \,f(g_1)-3 \, f(g_2)+f(g_3)+O(\Delta g^3),
\end{equation}
where for later convenience we introduced the notation:
$g_n=g-n \, \Delta g$.
By increasing the order of the Taylor-series method and using the same procedure, higher order
formulae can be derived. Here, we list them up to the sixth order.
The forms of the 4-, 5- and 6-order formulae take the form:
\begin{equation}\label{four}
f(g)=4 \,f(g_1)-6 \, f(g_2)+ 4\,f(g_3)-\,f(g_4)+O(\Delta g^4),
\end{equation}
\begin{equation}\label{five}
f(g)=5 \,f(g_1)-10 \, f(g_2)+ 10\,f(g_3)-5\,f(g_4)+\,f(g_5)+O(\Delta g^5),
\end{equation}
\begin{equation}\label{six}
f(g)=6 \,f(g_1)-15 \, f(g_2)+ 20\,f(g_3)-15\,f(g_4)+6\,f(g_5)-\, f(g_6)+O(\Delta g^6).
\end{equation}
Finally, we mention that, in case we had numerical data at least for six consecutive values of $g$,
then we used the 6-point rule (\ref{six}) to construct the initial values of the numerical 
algorithm for the next value of $g$.

\section{Chebyshev-polynomials}\label{appB}

In this appendix we summarize some useful properties and integral formulae of the 
Chebyshev-polynomials. The Chebyshev-polynomials of the first kind $T_n(u)$ 
form a sequence of orthogonal polynomials
 on $[-1,1]$ with respect to the weight function:
 $\frac{1}{\sqrt{1-u^2}}$. The orthogonality relation is given by the integral formula:
\begin{equation} \label{Tint}
\int\limits_{-1}^{1} \,du \, \frac{1}{\sqrt{1-u^2}} \, T_{n}(u) \, T_{m}(u)=
\delta_{nm} \, \frac{\pi}{2} \, (1+\delta_{n,0}), \qquad n,m=0,1,2,...
\end{equation}
The Chebyshev-polynomials can be given by the explicit formula:
\begin{equation}\label{Tn}
T_n(u)=\cos(n \, \arccos u), \qquad n=0,1,2,...
\end{equation}
For practical purposes, we define their slightly modified version:
\begin{equation}\label{That}
\hat{T}_n (u)=\left\{\begin{array}{ll}
\frac12, \qquad \qquad n=0, \\
T_{n}(u),  \qquad n=1,2,...
\end{array} \right.
\end{equation}
In the QSC method, close to the branch points, the relevant functions behave like $\sqrt{4 \, g^2-u^2}$.
This is why, in our numerical studies the Chebyshev-polynomials of the second kind $U_n(u)$ become
important, since they form an orthonormal basis on $[-1,1]$ with respect to the weight function
$\sqrt{1-u^2}$. They can be given by the explicit formula:
\begin{equation}\label{Un}
U_n(u)=\frac{\sin((n+1) \, \arccos u)}{\sin(\arccos u)}, \qquad n=0,1,2,...
\end{equation}
and the orthogonality relations they satisfy, read as:
 \begin{equation} \label{Uint}
\int\limits_{-1}^{1} \,du \, \sqrt{1-u^2} \, U_{n}(u) \, U_{m}(u)=
\delta_{nm} \, \frac{\pi}{2}  \qquad n,m=0,1,2,...
\end{equation}
The two kinds of Chebyshev-polynomials are related by a simple
recurrence relation:
\begin{equation}\label{TU}
\hat{T}_n(u)=\frac{U_n(u)-U_{n-2}(u)}{2}, \qquad n=0,1,2,...,
\end{equation}
where $U_n(u)$ for $n<0$ is zero by definition.
According to the theory of orthogonal polynomials, on $[-1,1]$ any smooth function $f$ can be represented
as a convergent series in either $T_n$ or $U_n$:
\begin{equation}\label{fTU}
f(u)=\sum\limits_{n=1}^{\infty} \, b_n\, \hat{T}_n(u)=\sum\limits_{n=1}^{\infty} \, a_n\, U_n(u),
\qquad u\in[-1,1].
\end{equation}
 As a consequence of (\ref{TU}), the coefficients are related by:
\begin{equation} \label{abn}
a_n=\frac{b_n-b_{n+2}}{2}, \qquad n=0,1,2....
\end{equation}
In our numerical approach, we expand our functions in terms of $U_n$. Nevertheless, 
in practice the coefficients of this expansion are determined via (\ref{abn}) from the coefficients of 
the expansion with respect to $T_n$. The reason is that during the numerical computations, we have
the values of the functions at discrete set of points and we should determine the coefficients
of the series from these discrete values. If the function under consideration is computed at
the positions of the zeros of the $l_c$th Chebyshev-polynomial $T_{l_c}$ with $l_c$ being a large
integer, then there are simple formulae in the literature to determine the first $l_c$ 
coefficients $b_n$ in (\ref{fTU}).  
Using the matrix (\ref{Cf}) 
they are given by:
\begin{equation}\label{bn}
b_{n}=\frac{2}{l_c} \, \sum\limits_{s=1}^{l_c} \, f(u_s) \, {\cal C}_{l_c-s+1,n+1}, \qquad n=0,1,...,l_c-1,
\end{equation}
where the discretization points are chosen to be zeros of $T_{l_c}$:
\begin{equation}\label{us}
u_s=-\, \cos\left( \frac{\pi \, (s-\frac12)}{l_c}\right), \, \qquad  T_{l_c}(u_s)=0, \qquad s=1,..,l_c. 
\end{equation}
Here, it is assumed that $l_c$ is so large that the coefficients with higher index are so small that they are
irrelevant up to the numerical precision required. Thus the series is truncated at the index $l_c$.

In our actual numerical computations, the following integral formulae for $U_n$ are important:
\begin{equation} \label{Upu}
\int\limits_{-1}^{1} \, du \, \frac{\sqrt{1-u^2} \, U_n(u)}{u-v}=-\frac{\pi}{x_s(2 \, v)^{n+1}}, \qquad 
v \in {\mathbb C} \setminus (-1,1),
\end{equation}
\begin{equation} \label{UpuPV}
- \! \! \! \! \! \!  \! \int\limits_{-1}^{1} \, du \, \frac{\sqrt{1-u^2} \, U_n(u)}{u-v}=-\pi \, \hat{T}_{n+1}(v), \qquad 
u \in (-1,1),
\end{equation}
where $x_s$ is given in (\ref{xs}) and (\ref{UpuPV}) contains a principal value integration.

\section{The derivation of formulae
(\ref{omegaija}) and (\ref{omegaregija})}\label{appC}

In this appendix we show, how to use the Chebyshev-expansions to the derivation of the formulae
(\ref{omegaija}) and (\ref{omegaregija}) for $\omega_{ij}$ and $\omega_{ij}^{reg}$. First,
we start with some remarks concerning the coefficients of the series representations (\ref{p12},\ref{p34}).

Let $f(u)$ be a function on ${\mathbb C}$ with the properties as follows:
\begin{itemize}
\item It has no poles,
\item It has a single branch cut at $[-2  g, 2 g]$ with square root type discontinuity.
\item The discontinuity on the branch cut is given by $i \, \rho(u)$.
\item The discontinuity becomes zero at the branch points, which means that 
it behaves like $\sim \sqrt{4 \, g^2 -u^2}$ at $\pm 2 g$.
\item $f$ decays at least as fast as $\tfrac 1 u$ at infinity.
\end{itemize}
Then $f(u)$ can be expressed by its discontinuity by the formula:
\begin{equation} \label{frho}
f(u)=\int\limits_{-2  g}^{2 g} \! \frac{dv}{2 \, \pi} \, \frac{\rho(v)}{v-u}.
\end{equation}
Moreover, since $\rho(\pm 2 g)=0$, it can be represented as:
\begin{equation} \label{rr0}
\rho(u)=\sqrt{4 \, g^2 - u^2} \, \rho_0(u), \qquad u \in [-2 g, 2 g],
\end{equation}
where $\rho_0(u)$ is a smooth regular function on $[-2 g , \, 2 g]$. This is why it can be 
expanded in a convergent series with respect to $U_n$s:
\begin{equation}\label{rho0}
\rho_0(u)=\sum\limits_{n=0}^{\infty} \, a_n \, U_n(\tfrac{u}{2 \, g}).
\end{equation}
As a consequence of (\ref{frho}), (\ref{rr0}), (\ref{rho0}) and (\ref{Upu}) $f(u)$ admits
the convergent series representation as follows:
 \begin{equation}\label{fxser}
f(u)=-g \, \sum\limits_{n=0}^{\infty} \,a_n \, \frac{1}{x_s(\tfrac{u}{g})}, \qquad u \in {\mathbb C}\setminus [-2 \, g , \, 2 \, g].
\end{equation}
 Consequently, we can conclude that the coefficients in the expansions (\ref{p12}) and (\ref{p34}) are nothing 
else, but the coefficients of the Chebyshev-series of the discontinuity functions{\footnote{In the sense of 
(\ref{rr0}) and (\ref{rho0})}} of ${\bf p}_a$s. In this sense the formulae (\ref{omegaija}) and (\ref{omegaregija}) 
are the periodic analogs of (\ref{fxser}).

Now we show, how to derive (\ref{omegaija}),(\ref{omegaregija}) 
and (\ref{Iij}) from (\ref{omega}), (\ref{omega0}), (\ref{omegac}). 
The derivation of (\ref{Iij}) goes as follows. One inserts (\ref{rho-U}) into (\ref{QQrho}) and the result into 
$I_{ij}$ of (\ref{omegac}). Then evaluating the integrals with the help of the appropriately 
scaled{\footnote{I.e. $u \to \tfrac{u}{2g}$ substitution in the integral.}} version (\ref{Uint}) taken at $m=0$, 
one ends up with (\ref{Iij}).

To derive (\ref{omegaija}), first one has to rephrase the kernel as an infinite sum:
\begin{equation}\label{cth}
\coth(\pi \, (u-v))=\frac{1}{\pi\, (u-v)}+\frac{1}{\pi} \,\sum\limits_{k=1}^{\infty} \, 
\left(  \frac{1}{u-v+ i \, k}  + \frac{1}{u-v- i \, k} \right).
\end{equation}
Then inserting (\ref{cth}), (\ref{QQrho}) and (\ref{rho-U}) into (\ref{omega0}) and evaluating the integrals
with the help of (\ref{Upu}) one ends up with (\ref{omegaija}).

To derive (\ref{omegaregija}), one should represent $\omega^{reg}_{ij}$ by the formula: \newline
$\omega^{reg}_{ij}(u+i\, 0)=\tfrac 12 (\omega_{ij}(u+i \, 0)+\omega_{ij}(u-i\,0))$.
The derivation of (\ref{omegaregija}) is very similar to that of (\ref{omegaija}).
The only difference comes from the $\sim \frac{1}{u-v}$ term of (\ref{cth}).
Now, the $\pm i \, 0$ prescriptions become important. 
 If they are treated by the Sokhotski-Plemelj formula, only the principal value part
remains. This principal value integral can be evaluated with the help of (\ref{UpuPV}),
which gives the term $T_{n+1}(u)$ in (\ref{omegaregija}).

\section{A method to compute (\ref{OMEGA}) numerically}\label{appD}

In the implementation of the numerical method for solving QSC equations, only such simple
mathematical operations appear, like summations and finding the solutions of some linear
equations. Both methods can be easily implemented in C++ language. There is only one
subtle quantity $\Omega_{A,n}(g)$ defined in (\ref{OMEGA}), which requires the accurate computation 
of an infinite sum. In this appendix, we describe, how to reduce the computation of 
this quantity to finite summations, provided one needs the result with a given numerical accuracy.
Here, we recall the definition of $\Omega_{A,n}(g)$,
\begin{equation}\label{OMEGA1}
\Omega_{A,n}(g)=\sum\limits_{k=1}^{\infty} \, \left(
\frac{1}{x_s(\tfrac{u_A-i \, k}{g})^{n}}+\frac{1}{x_s(\tfrac{u_A+i \, k}{g})^{n}}
\right), \qquad A,n=1,.. l_c
\end{equation}
where $u_A \in [-2\, g,2 \,g]$ are the discretization points. For the sake of simplicity, 
in the sequel we will omit the index $A$ from $u_A$.
First, we sketch the idea of the numerical computation and the deeper technical details will be given
in the subsequent paragraphs. For practical purposes, we introduce a short notation for the summand:
\begin{equation}\label{IX}
I^{(n)}_X(k,u)=
\frac{1}{x_s(\tfrac{u-i \, k}{g})^{n}}+\frac{1}{x_s(\tfrac{u+i \, k}{g})^{n}}.
\end{equation}
We introduce also an integer cutoff $\Lambda_X$ to write the infinite sum 
as a sum of two terms:
\begin{equation} \label{IXsum}
\sum\limits_{k=1}^{\infty} \, I^{(n)}_X(k,u)= \sum\limits_{k=1}^{\Lambda_X} \, I^{(n)}_X(k,u)+
\sum\limits_{k=\Lambda_X}^{\infty} \, I^{(n)}_X(k,u).
\end{equation}
The first term in the rhs. of (\ref{IXsum}) is a finite sum, so it can be evaluated numerically
by a computer. Since $\Lambda_X$ is chosen to be large, in the second term on the rhs. we can
use the large $k$ expansion of the summand. It defines a series in $1/k$, and the explicit sums of the $1/k$ 
powers can be expressed by the Riemann-zeta function. To reach a given accuracy, only a finite number of terms
 of the $1/k$ series needed to be taken into account.
If $\tfrac{1}{k^{N_x}}$ is the last term, which is summed in the
large $k$ series, then the magnitude of the numerical error is $\sim \frac{1}{\Lambda^{N_x}_X}$.

Unfortunately, this naive estimation needs to be corrected, when one takes a deeper look at the structure
of the summand (\ref{IX}). This is why, in the next paragraphs, we write down in more detail the numerical
computation of (\ref{OMEGA1}).

The first ingredient is the large $k$ expansion of the summand $I^{(n)}_X(k,u)$.
It can be obtained by inserting the following two series expansions into (\ref{IX}):
\begin{equation}\label{xsa}
x_s(\tfrac{u}{g})^{-\alpha}=\left( \frac{g}{u}\right)^{\alpha} \, \sum\limits_{s=0}^{\infty} \, \kappa_s^{(\alpha)} \,
\frac{g^{2 s}}{u^{2 s}},
\end{equation}
\begin{equation}\label{upik}
\frac{1}{(u+i\, k)^{n+2 s}}=\frac{1}{(i \, k)^{n+2 s}} \, \sum\limits_{m=0}^{\infty} \,\binom{-n-2s}{m} \,
\left( \frac{u}{i \, k}\right)^m,
\end{equation}
where $\kappa_s^{(\alpha)}$ is given by (\ref{kappans}).
The final form of the expansion takes the form:
\begin{equation}\label{IXser}
I^{(n)}_X(k,u)=\sum\limits_{p=0}^{\infty} \, \frac{1}{i^{p+n}} \, (1+(-1)^{p+n}) \, \frac{1}{k^{p+n}} \,
\sum\limits_{s=0}^{[p/2]} \, g^{n+2s} \, u^{p-2s} \,\kappa_s^{(n)} \, \binom{-n-2s}{p-2s},
\end{equation}
where $[...]$ stands for integer part. 

(\ref{IXser}) allows us to make the appropriate choice for
the cutoff parameters $\Lambda_X$ and $N_x$. For the sake of simplicity concentrate on the power like terms
in (\ref{IXser}). A typical such term looks like $\sim \frac{g^{n-q} \, u^q}{k^n}$. In the numerical
algorithm, we need to compute (\ref{IXsum}) at the discretization points, which lie in the interval $[-2g,2g]$.
This is why we can give an upper estimation for this typical power-like term:
 \begin{equation}\label{}
\bigg|\frac{g^{n-q} \, u^q}{k^n}\bigg| \lesssim \left( \frac{2 \,g}{k}\right)^n, \qquad u\in[-2 g,2g].
\end{equation}
This inequality tells us that, not the powers of $1/k$ determine the magnitudes of the
terms in the $1/k$ series, but the powers of $\tfrac{2\,g}{k}$. This means that, if $\tfrac{1}{k^{N_x}}$ is
the last term, we sum from $\Lambda_X$ to infinity in (\ref{IXsum}), than the numerical error can be estimated 
by $\left(\tfrac{2g}{\Lambda_X}\right)^{N_x}$ instead of the naively expected value 
$\left(\tfrac{1}{\Lambda_X}\right)^{N_x}$.

Now, we are in the position to make a choice for the values of $\Lambda_X$ and $N_x$. 
We require $N_c$ digits of accuracy for (\ref{IXsum}). This means that the estimated error term should be $\sim 10^{-N_c}$.
In accordance with the content of the previous paragraph, this requirement imposes an inequality
among the parameters $\Lambda_X$, $N_x$ and $N_c$.

\begin{equation}\label{NC1}
\left(\frac{2 g}{\Lambda_X}\right)^{N_x}\lesssim 10^{-N_c}.
\end{equation}
The value of $\Lambda_X$ is chosen to "maximize" the inequality:
 \begin{equation}\label{LX}
\Lambda_X \simeq 2 \, g \cdot10^{N_c/N_x}.
\end{equation}
Certainly, (\ref{LX}) does not allow to determine both $\Lambda_X$ and $N_x$.
One of them is free to choose and the other one is given by (\ref{LX}). 
In our actual numerical computations, we made the choices:
 \begin{equation}\label{LXchoice}
\Lambda_X =[200 \cdot g],
\end{equation}
and in accordance with (\ref{LX}):
\begin{equation}\label{NXchoice}
N_x =\left[1+N_c \cdot \tfrac{\ln 10}{\ln 100}\right]+\Delta N_x, \qquad \Delta N_x=0,1.
\end{equation}
Here, the value of $\Delta N_x$ is chosen in order for $N_x$ to be even. This makes the numerical implementation
a slightly simpler.
Since  the first term in the rhs. of (\ref{IXsum}) is straightforward to compute numerically,
we concentrate on the computation of the second term:
\begin{equation} \label{IXrem}
 \Omega^{(n)}_{\Lambda_X}(u)=\sum\limits_{k=\Lambda_X}^{\infty} \, I^{(n)}_X(k,u).
\end{equation}
From (\ref{IXser}) it can be seen that the summand is non-zero in case $n+p$ is even.
Thus, when $n$ is even, only the even values of $p$ enter the sum and in case $n$ is odd, only the
odd values of $p$ contribute. This is why, we write down separately the formulae for the $n$ even and odd
cases.

\underline{The even $n$ case}: 
 
Let $n=2 n_0, \quad n_0=1,2,...$, and $p=2 p_0, \quad p_0=0,1,2,...$.
Then (\ref{IXser}) takes the form:
 \begin{equation}\label{IXserPS}
I^{(2 n_0)}_X(k,u)=2\sum\limits_{p_0=0}^{\infty} (-1)^{p_0+n_0} \frac{1}{k^{2(p_0+n_0)}} 
\sum\limits_{s=0}^{p_0} g^{2(n_0+s)} \,  u^{2 (p_0-s)} \, \kappa_s^{(2 n_0)} \, \binom{-2(n_0+s)}{2(p_0-s)},
\end{equation}
If the $1/k$ series is truncated at $N_x$, then the sum in $p_0$ is also truncated 
as a consequence of the inequality: $2(p_0+n_0)\leq N_x$. Thus, the upper limit of the
summation becomes{\footnote{This was reason, why we choose $N_x$ to be even. Easier to program.}}: 
$p^{max}_0=\tfrac{N_x}{2}-n_0$.
Now the summation can be performed explicitly with the help of the Riemann-zeta function $\zeta(z)$.
Up to the required accuracy, the final result can be written as a finite sum:
\begin{equation}\label{OmegaPS}
\Omega^{(2 n_0)}_{\Lambda_X}(k,u)=2 \! \sum\limits_{p_0=0}^{p_0^{max}}  (-1)^{p_0+n_0} \,
 \zeta_{\Lambda_X}({2(p_0+n_0)}) \!
\sum\limits_{s=0}^{p_0}  g^{2(n_0+s)}  u^{2 (p_0-s)} \kappa_s^{(2 n_0)}  \binom{-2(n_0+s)}{2(p_0-s)},
\end{equation}
where $\zeta_{\Lambda_X}(z)=\zeta(z)-\sum\limits_{k=1}^{\Lambda_X} \frac{1}{k^z}$.

\underline{The odd $n$ case}: 
 
Again, we take the parametrizations: $n=2  n_0+1, \quad n_0=0,1,2,...$, and $p=2  p_0+1, \quad p_0=0,1,2,...$.
Then (\ref{IXser}) takes the form:
 \begin{equation}\label{IXserPTL}
I^{(2 n_0+1)}_X(k,u)=2 \! \sum\limits_{p_0=0}^{\infty} \frac{(-1)^{p_0+n_0+1}}{k^{2(p_0+n_0+1)}} \!
\sum\limits_{s=0}^{p_0}  g^{2(n_0+s)+1}  u^{2 (p_0-s)+1} \kappa_s^{(2 n_0+1)} \, \binom{-2(n_0+s)-1}{2(p_0-s)+1}.
\end{equation}
The $1/k$ series is truncated at $N_x$, thus the sum in $p_0$  becomes also truncated. 
From the inequality: $2(p_0+n_0+1)\leq N_x$, the upper limit of the
summation becomes: $\tilde{p}^{max}_0=\tfrac{N_x}{2}-n_0-1$,
and the final result becomes a finite sum again:
\begin{equation}\label{OmegaPTL}
\begin{split}
\Omega^{(2 n_0+1)}_{\Lambda_X}(k,u)=2 \sum\limits_{p_0=0}^{\tilde{p}_0^{max}} \, (-1)^{p_0+n_0+1} \,
 \zeta_{\Lambda_X}({2(p_0+n_0+1)}) \times \, \\
\sum\limits_{s=0}^{p_0} \, g^{2(n_0+s)+1} \, u^{2 (p_0-s)+1}
 \,\kappa_s^{(2 n_0+1)} \, \binom{-2(n_0+s)-1}{2(p_0-s)+1}.
\end{split}
\end{equation}

We close this appendix with a remark on the usage of the $\zeta$-function in C++.
During the development of our C++ code, we recognized that neither double nor long double
precisions are not enough to get accurate results at strong coupling. 
These built in precisions were not enough even to reach some kind of convergence.
This is why, we used an arbitrary precision package to C++, called CLN (Class Library of Numbers).
In the CLN library $\zeta(z)$ is a built in function and it could be used to our purposes.
If one uses pure C, or C++, it should be recognized that we need $\zeta(z)$ at a finite
number of integers. Thus one can compute the necessary values e.g. in Mathematica with high precision and than
they can be copied into the C-code and stored in a constant array.


\section{Pade-approximation like formulae for the anomalous dimensions}\label{appE}

In order for the readers to get some taste about the magnitude of the anomalous dimensions,
we begin this appendix with listing the numerical values of the anomalous dimensions at some values of the coupling
constant $g=\frac{\sqrt{\lambda}}{4 \,\pi}$. 
\begin{table}
\begin{center}
\begin{tabular}{|c|c||c|c|}
\hline
$g$ & $\Delta$ & $g$ & $\Delta$  \\
\hline
\hline
0.5 & 5.71272342478773903062 & 4.0 & 14.45378636296056157594  \\
\hline
1.0 & 7.60407071704738848334 & 4.5 & 15.29901169250471532720  \\
\hline
1.5 & 9.11375404891588560886 & 5.0 & 16.09983932145390471841  \\
\hline
2.0 & 10.40482174344050611272 & 5.5 & 16.7128504510418019769  \\
\hline
2.5 & 11.55154711104216029680 & 6.0 & 17.5923066098442921880  \\
\hline
3.0 & 12.59378147179885650906 & 6.5 & 18.2928791532391552907  \\
\hline
3.5 & 13.55582301629291387584 & 7.0 & 18.9675672851951075502  \\
\hline
\end{tabular}
\bigskip
\caption{Some numerical values of $\Delta$ for the Konishi operator.
\label{TD/2}}
\end{center}
\end{table}
\normalsize
\begin{table}
\begin{center}
\begin{tabular}{|c||c|c|c|}
\hline
$g$ & $\Delta_{S=4}$ & $\Delta_{S=6}$ & $\Delta_{S=8}$  \\
\hline
\hline
0.5 & 8.378286749267 & 10.805035317202 & 13.12115866686  \\
\hline
1.0 & 11.02483082714 & 13.965696581702 & 16.67666058421  \\
\hline
1.5 & 13.13499808832 & 16.498636307379 & 19.54186450481  \\
\hline
2.0 & 14.94093551777 & 18.673499820718 & 22.01043492694  \\
\hline
2.5 & 16.54666414765 & 20.611840708885 & 24.21585170200  \\
\hline
2.7 & 17.14616785384 & 21.336481745366 & 25.04143686845  \\
\hline
3.0 & 18.00750137760 & 22.378417558485 & -  \\
\hline
3.5 & 19.35706856273 & 24.012697674227 & -  \\
\hline
4.0 & 20.61764227985 & - & -  \\
\hline
4.1 & 20.86053885660 & - & -  \\
\hline
\end{tabular}
\bigskip
\caption{Some numerical values of $\Delta$ for the twist-2 operators with $S=4,6,8$.
\label{TD/468}}
\end{center}
\end{table}
\normalsize
\newpage
Apart from the numerical values we listed in the tables, the interested readers can find all the numerical data
we obtained, in the Mathematica notebook and text files attached to the text file of the paper.

Apart from fitting the strong coupling series coefficients of the the anomalous dimensions, 
we also used the numerical data to construct Pade-approximation like formulas in order 
to describe the anomalous dimensions of the  operators under consideration at all values of the 
coupling constant with satisfying numerical precision. 
Instead of the computation of an interpolating function composed of rational polynomials, we performed a nonlinear model fit to the data points.
This approach gave smooth approximants for real values of the coupling constant, and could inform us about the validity of the approximation as well. 

We found that fitting a naive rational polynomial approximation for $\Delta(g)$ does not give stable{\footnote{Here, 
by stability, we mean stability with respect to
increasing the order of polynomials in the fitted rational expressions.}}
values for the coefficients of the rational polynomial. 
This is not surprising, if one observes that in the perturbative expansion around $g=0$ only even powers are present, while in the strong coupling 
regime the leading term is $\sim \sqrt{g}$ and the corrections go as inverse powers of $g$.

To have an optimal form for the approximation, we basically followed the Ansatz used in \cite{Gromov:2009zb}: 

\begin{equation}\label{PadeS2}
\Delta(g)=(g^2+g_b^2)^{1/4} \frac{a_0 + a_1 h + \dots a_n h^n}{1+b_1 h+\dots b_n h^n}.
\end{equation}

Where $h=\frac{g^2}{\sqrt{g^2+\left( \frac{1}{4} \right ) ^2}}$ and $g_b$ is a suitable constant, 
whose value is chosen to be $2$ in the case of the $S=2,4,6,8$ twist-2 operators. 

In principle some analytical information  can be built into the Ansatz from the perturbative results \cite{Marboe:2014gma}, 
 by fixing some relations between coefficients.
For practical calculations however, we exploited only the known value{\footnote{I.e. $\Delta(0)=L+S$, where $L=2$ for twist-2 operators and $S=2$ for the Konishi-
state.}} of $\Delta(0)$ and the leading order strong coupling asymptotics of $\Delta(g)$
 given in (\ref{D01}). These data fixed $a_0$ and the ratio of $a_n$ and $b_n$.

Because of the high precision of the numerical data, an unusually high number of coefficients could be fitted.
For the Konishi operator, we stopped at $n=15$, where the coefficients seem to be still stable with respect to changing the value of $n$.

We performed the fits by Mathematica's build in {\textit{ NonlinearModelFit}} function, which provides "prediction 
bands"{\footnote{Interested readers can gain more information about this function in the help of Mathematica.}} allowing one to infer to the accuracy of the
Pade-approximation like formula, as well.

\begin{figure}[htb]
\begin{flushleft}
\hskip 15mm
\leavevmode
\epsfxsize=120mm
\epsfbox{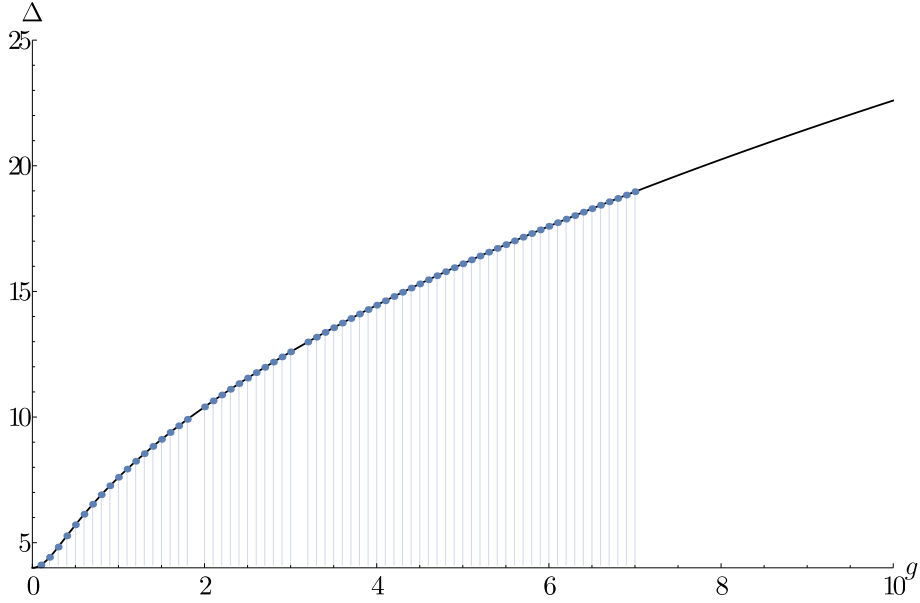}
\end{flushleft}
\caption{{\footnotesize
The plot of the Pade-approximation like formula and the data points for the anomalous dimension of the Konishi operator. 
}}
\label{PadeK}
\end{figure}

The measured points and the fitted curve are shown in figure \ref{PadeK}.

\begin{figure}[htb]
\begin{flushleft}
\hskip 15mm
\leavevmode
\epsfxsize=120mm
\epsfbox{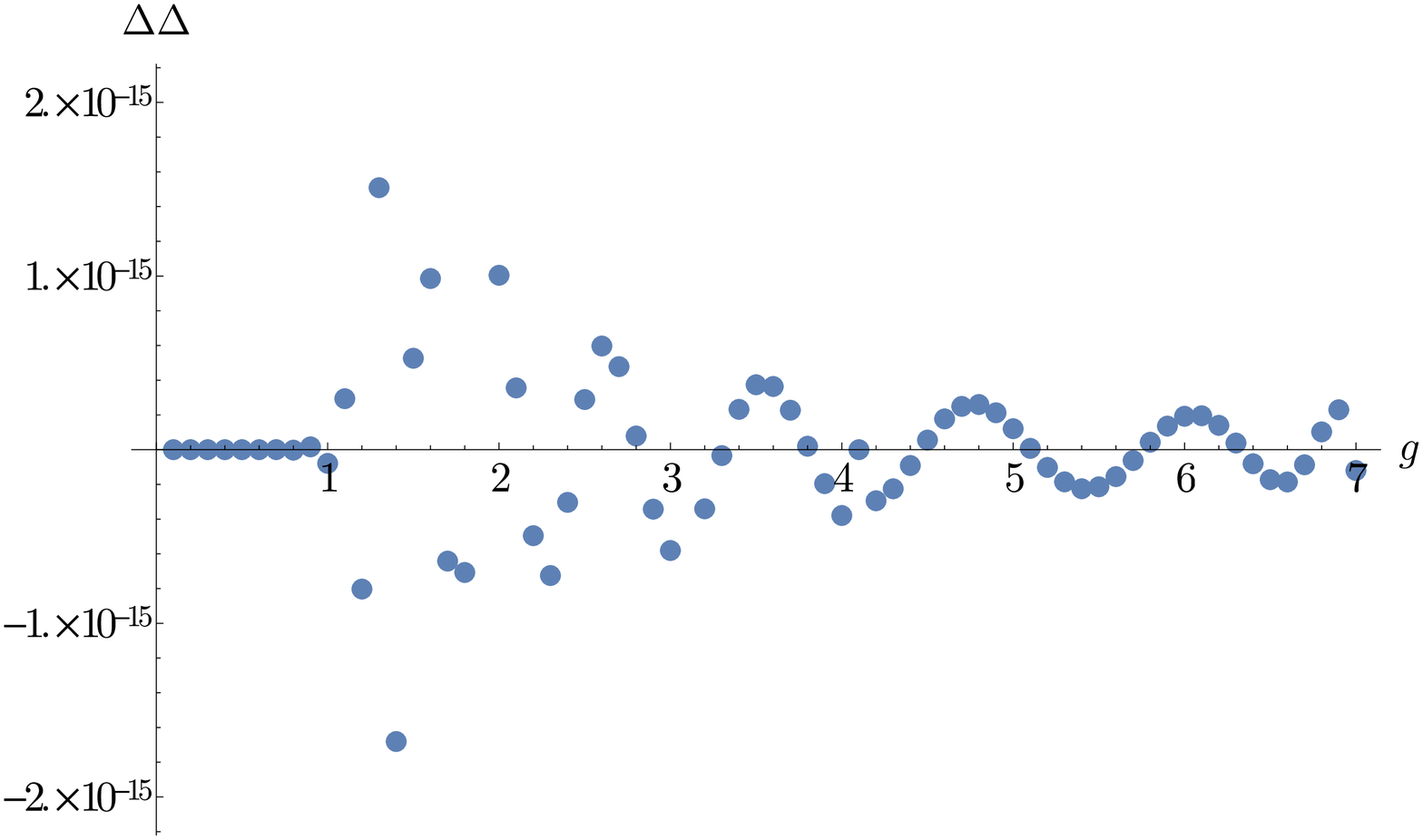}
\end{flushleft}
\caption{{\footnotesize
The plot of the difference of the Pade-approximation like formula and the data points for the anomalous dimension of the Konishi operator.
}}
\label{Res}
\end{figure}

Because of the small magnitude of the deviations, we show separately the residual plot of the data in figure \ref{Res}.
Figure \ref{Res} shows that the data points are so close to the fitted curve that the data points are approximated with the 
Pade-approximation like formula with $14$ digits of accuracy.

To predict the accuracy of the fitted curve beyond the measured interval, we used Mathematica's build in ``MeanPredictionBands''
function and we set the confidence level to $99 \%$.
 \begin{figure}[htb]
\begin{flushleft}
\hskip 15mm
\leavevmode
\epsfxsize=120mm
\epsfbox{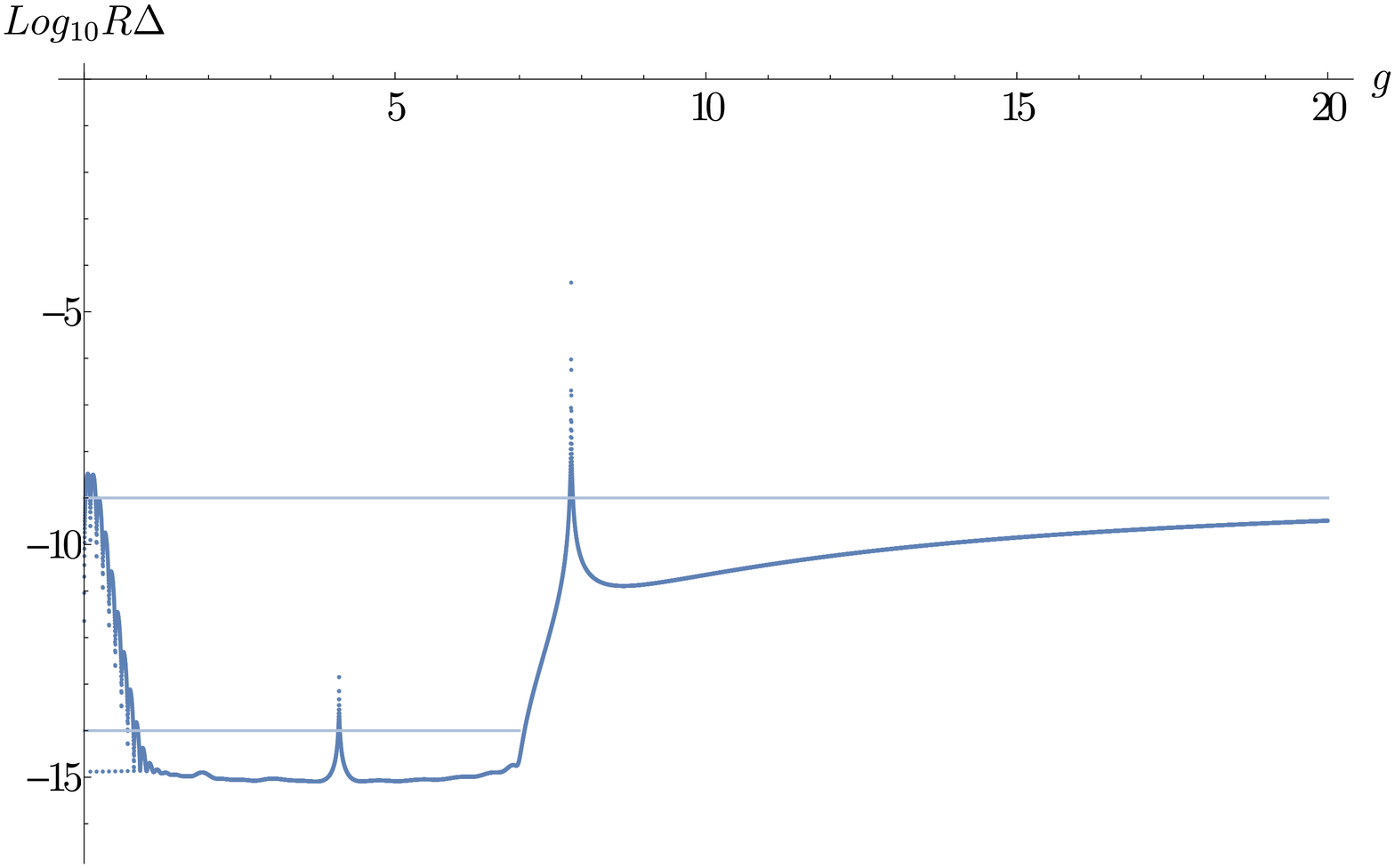}
\end{flushleft}
\caption{{\footnotesize
Magnitude of the confidence interval radius calculated from the mean
prediction bands at confidence level 99\% for the anomalous dimension of the Konishi operator.
}}
\label{bands}
\end{figure} 
Figure \ref{bands} shows that even outside of the range of available numerical data, the fitted Pade-approximation like formula can be taken 
seriously up to 9 digits of accuracy.

Analogously to (\ref{PadeS2}), Pade-approximation like formulas were constructed for the $S=4,6,8$ cases, as well. 
The structure of the approximation formulae are the same as that of the Konishi operator, the only difference is the actual form of the
rational $h$-dependent factor in (\ref{PadeS2}). 
To close this appendix, we list the $h$-dependent, rational expressions in Mathematica form{\footnote{The approximation formulae below can be copied
into a Mathematica notebook.}} 
for the Konishi and as well as for the $S=4,6,8$ cases.

\subsection{The rational part of (\ref{PadeS2}) for the Konishi operator}
{\tiny\begin{verbatim}
RationalS2[h_]:=(2 Sqrt[2] - 
    2.6520147223242266547552680472866143425726560990113810253190736097524135619263795142808621172969503855070056242414474 h + 
    17.5053989863052977462679119715560442550467128618303236093862249612574669472139903537690996820102090256443104021085147 h^2 - 
    1.2547732530898415105233577167280056559859191264714403490388988416580013925430642973848653446168030927925027598786689 h^3 + 
    34.6514155305087017638548001069378602676924996412031669624257920330407182512961117853308208719006894789605999774772507 h^4 + 
    35.7918745808862778915963373460843645929305215870352438937522529437745261787553200714511087390056497702966327025712285 h^5 + 
    32.3549022276784564470462831676296871662541698749575113619680099895915934549068390870901696487552638628104642034688071 h^6 + 
    73.1758283776132528902571673999357547953666208400320387039947370484301270692270495791157261398555343872469442164974097 h^7 + 
    2.6949333620817741405638722550041469443586016742208206647986645629580585493645510441754553304346761786969994170499848 h^8 + 
    103.2123610565218996868544472707216414762852999332521583443305087891429174132426636415540272025959415437532338533859126 h^9 - 
    12.2556140720479692096065594761952722407538447346745711988592818382946828110658094073886166859456943372981154820910009 h^10 + 
    45.9768281288524657158627218890027685317910891849771585617045462627990583545826686247002595913695792962175070597340142 h^11 - 
    12.4657903836422148450128994792829336995477268447109457157688355684807160528079337277577909880515796610214371961741174 h^12 + 
    6.6613022268228142721160548904106306670226279371208399446856632178705161052393217157347225139850110466324011345628654 h^13 - 
    2.5492305159012280125147073465166979710495646478359959190217032330445192495302576371972254239739412996943448772158454 h^14 + 
    0.2306376734329059954130555094648114347526366121683596491855270321697442352690672250026223973769961374976411239528838 h^15)/(1 - 
    1.672002301887886311555742881718840687068612011663886232471054280787757199501820790133258717017293562845012976808224 h + 
    6.7092566618601967671732950451739427229245702729064665235444884091811689897321211790613901482815473354062258857301414 h^2 - 
    3.9354910217683768105881824629978724835907321776401501456366957770874046406999435894828720463962473848069865801952085 h^3 + 
    11.3506065684148509270352527737365447876573525112949516867144319562332523148251081160005064723660335515102003738190245 h^4 + 
    6.8789617532632002135368896235486036402230565501161159253931472216647845739542589188770509066619227011776162009451864 h^5 + 
    3.6233034279889036095004718940751123943532203346060721386713899627299678403928976081750503190694187238122292054076278 h^6 + 
    18.1630120838794873202857314430872276104491963742299399660417809458505024914374697739012417188698324980778467209928236 h^7 - 
    1.5654562487134359202417255992131074530489575011412886142296752895588892268202058063865130800295603760089615914666873 h^8 + 
    20.055185995570302168497848623144207622335393661798675713461200527410642524304932918424767015628114081575641438876176 h^9 - 
    3.5500284557858645882886961812801734575341736039473950227308597039018417562080378769039019257782813870047631742016664 h^10 + 
    7.5591907474432678278130255239683440372143464765956516898665141307926471683316189587153378759409880349249896021689324 h^11 - 
    2.2011122049622993191169854651893374055546350249671512624823975704826480362234188364535576735683195924452165368825033 h^12 + 
    1.0012291659309748504496638743519224824349772121349865612365239825194137725975325374847748365328820392629121391282851 h^13 - 
    0.3621510477970165199650618921211905081611408443974185649480466473697471165969490684457209077245264368756546751119809 h^14 + 
    0.0325308432311336619150609994532287310782123239309891377394286096978699961309654143765755495046853985883065275665346 h^15)
\end{verbatim}
}
\newpage

\subsection{The rational part of (\ref{PadeS2}) in the $S=4$ case: }
{\tiny\begin{verbatim}
RationalS4[h_] := (3 Sqrt[2] - 
    6.0735292955536499384449440677672111327969796337303894779849891967611645550334407529076996579692175572863620178727903 h + 
    20.9467933494418163913230970551441351168498508829384039915177742691582139430484234339250223543567441724472131810284825 h^2 - 
    10.0681383828884601744966259410434382442845406754246760487756504323139335985695947452201037467597018198408690278389216 h^3 + 
    25.1054172992957617887549438850612770562231874819295535209150005503362794456404753432994166228748761560066540155453437 h^4 + 
    20.4864428981550592869209081099333046323060267990860434789353319113568525358723789289794506870927324776924993178444422 h^5 - 
    1.0627484008504069221300644575764933560407579668857517281881507348834223581810961764444105206114239080891266064493934 h^6 + 
    22.1093071369970405052804649412916310330104785583996114496819007614581916462944002467159316341446553227179201755043342 h^7 - 
    36.1341855722375991947776209926831534456455686318985316635449695456074561040546863892482620663219651316522104880791167 h^8 + 
    19.3892894018665541110986482318690524577836883391175828147334665185443459282059846465414416437578480855688563386628219 h^9 - 
    11.6927311318967390686182570772643607007848958553535653998982211726917518307157072550118800007147722794833287977371964 h^10 + 
    5.1069424835060129343883272898622087280882746157424978157792916219600790404770304752826795340509683611625057242115722 h^11 - 
    0.7242411150005901932718121683656569978581366651039941564128950074113597328749964576706357514166448627043099344221624 h^12)/(1 - 
    2.1103643808010977070381661007894200507322503082191915742243388397940396408822166558392367322656279782526890869256005 h + 
    5.7361283045310020095492617153169468779341623377263765625328177668489429703109375155632338872955121047397229192906774 h^2 - 
    4.6968768659535270592730136806686783133130248289368839240410251040561376125015510876984504345536633256709817000595383 h^3 + 
    6.1392532179188652330507574948996625613683401590786339876397165245923943722715861148526660971584531318865409535471163 h^4 + 
    2.6623713766134374282754364774311355701409103542004601076874596840235405698370944218418810553319537319706072614033461 h^5 - 
    2.5103984493907987133511139645576907136766573016782326983723434936765551250953008023242955870343604765237389579746429 h^6 + 
    3.9652544235491760630040615361379450686415972842381638264401819344611983190395953044956522974765279739855691245690758 h^7 - 
    5.0036600867562502054412394375827661482377773341678986212651066952139075634133837847646300971185112835456560783704238 h^8 + 
    2.6139915120942366968254471740149547970954588896924650699134271580328799314011900693540545174873359992316179296662784 h^9 - 
    1.302347112206723249577967634797300674354189665473241692345488576939465544860755842706189308556353212613462860443648 h^10 + 
    0.5179611070527746237211581811513231642098725368932521942351472356841170695712745787920770766906036255505750152721928 h^11 - 
    0.0722326004947029258295991916329695812282148490487616148020766546132983191630409413733553947306580381991771197132769 h^12)
\end{verbatim}
}
\subsection{The rational part of (\ref{PadeS2}) in the $S=6$ case: }
{\tiny\begin{verbatim}
RationalS6[
  h_] := (8 - 
    100.9712422243275190849246229251993192741911794991540062968499704395372636939332333694566381947977229840807629717096092 h - 
    23.3878217883362855267214878275967483217293729718207129129861122994432096296131809263251557663336058186243611984114012 h^2 - 
    145.1882254228125998833420011988583662308971630526795173363017623537633426957590977697708337250986145811533694031787488 h^3 - 
    153.4627424359626297086828199708468485867974259382068738229246577003440718487730278346011592406936160192883110733666984 h^4 - 
    543.153017740347722254370104141714205373276356032178890050706511442080820562090823179755886308860324 h^5 + 
    258.1283653657998540350766788433812726165598657859732687165731966820296486383158809370687571592734669470660661478046599 h^6 - 
    586.207047648599301656267012195426277162498703581865197056213132384379749956256813462714814369321535 h^7 + 
    2239.214399144420640484386749639400209647726681077666199767446657981882640952564949946657750629955358 h^8 - 
    964.716257221944752767634637514015449139899152143271710786363380461763812264210878920540873477610383 h^9)/(1 - 
    13.1714800628694473470209611088151129160064616098240746764954554474036654726318422549534024615017052807027622774187683 h + 
    3.8876132788690317672841963418631862842958954208509595192962077098679480237629472184102536588192340736613771207222501 h^2 - 
    14.2812918672348646940114755347140336168055696329339317680263499070642055045681888866156693134775053340405028616663882 h^3 - 
    15.3964380469182890061261482285497707447181112863206727428503986348533962250576732981927152624661482193631053975890067 h^4 - 
    71.8530075650565057485082662847108276995449063829903127753893630532445131826554832482023173297629508565948247430064268 h^5 + 
    96.8884705469867150068033488932930677889311264639959558772521854786290968816379606428665332013951420041317441857729168 h^6 - 
    105.3631049532953653476325182133519842798501011893992512769353377964897206704346307609566267365917759874520198936622719 h^7 + 
    195.8538810517030842567254982071399417796297840854365927964212013517949040407170361566320790597678443070595984308263815 h^8 - 
    78.560464425370936304914882948725295477962637928049948694139544327659312852613985159203338956495986784784426450256568 h^9)
\end{verbatim}
}
\newpage
\subsection{The rational part of (\ref{PadeS2}) in the $S=8$ case: }
{\tiny\begin{verbatim}
RationalS8[
  h_] := (10 - 
    10.1801899905768901647030752734080101665539359648024172140522924280038834686697942538821686991402335725852737455896418 h + 
    45.1811096309267774272326527131054048271817759178003820593040031535225896768063206516604934983015117374706994527294724 h^2 - 
    19.2963145970787472037871288958078431258566428499696360465089607466919293351462964176179508004630078872059877763720239 h^3 + 
    86.9205113190357210524841908751351413508750709650058921224108119459227721176043390245353964869525616798453519807620974 h^4 + 
    12.694014605880073393730815508797593196951576758400933815785482566664912377574521750978103169833308295803466277748134 h^5 + 
    113.9655950504605500917831180255326456789896327232116693864711792687354527642324834829334744726708575441761609003354315 h^6 + 
    8.7815309528862062717572045952100187669365000288172446602895180470064784754241370251495362729261458354717327641306 h^7 + 
    74.3477309836338830179940185323079428377630885262116576450038405795782074111115826903545070578198527964276032429704048 h^8 - 
    15.6698903183360858249706280801499018759978537041104861147576437549614538982563379506034678464982979708737542695527083 h^9)/(1 - 
    1.4990903629755832242222350875666005650887953286853845829254320835282768414984727111470726430134537556153564819472741 h + 
    4.8641445388233955974565249875600735909207036310887497268385771417241562104364801726190184954607850539977826247396362 h^2 - 
    3.394607543945598115938687204246402586708655179783419529059163743920921735855580425691652303729126172155447914891898 h^3 + 
    8.8793954298443395537295411806036684674739877282614723594728367150937924862103561756038684037961134925431901971616927 h^4 - 
    1.5835929872589538438684924714863465339273432896295585876222422145455521696377482214435820143615761001592809992583181 h^5 + 
    10.1525581391148702817084153234957930826118213513921552107546141094482513473433706232696270011428769924076531496976586 h^6 - 
    1.013064033315107195686630779119844911805680349375347799058147670386997887583847863140171427009697374924315655563819 h^7 + 
    5.495492309509413653323751743394030752220231273021286279125853217399404510510285128434852010096027866522319183662199 h^8 - 
    1.1050986116176318066372278417595696761641728822488020671744615174384487011251562536164887931827333989754540872918397 h^9)
\end{verbatim}
}

\section{Various tables of numerical data}\label{appF}

This appendix contains some tables of numerical data which demonstrates that
the coefficients ${\mathfrak c}_{a,n}^{(k)}$ of (\ref{Cserpg}) are polynomials of order $2k$ in $n$.


\begin{table}
\begin{center}
\begin{tabular}{|c||c|c|c|}
\hline
$n$ & ${\mathfrak c}_{1,n}^{(2)}$ & $\alpha_{1,2}^{(n)}$ &  $\Delta P_{rel}$ \\
\hline
1 & 0.1804664578815959 & 0 & 0   \\
\hline
2 & 1.481738156681282 & 0.1804664578815959 & 0   \\
\hline
3 & 5.247995942046115 & 1.120805240918091 & 0   \\
\hline
4 & 13.30837177138386 & 1.344180845647056 & 0   \\
\hline
5 & 27.97694871345078 & 0.4849511117607040 & $1.5 \cdot 10^{-11}$   \\
\hline
6 & 52.05276094865590 & - &  $3.4 \cdot 10^{-11}$  \\
\hline
7 & 88.81979377072911 & - &  $3.4 \cdot 10^{-11}$  \\
\hline
8 & 142.0469835827955 & - &  $3.8 \cdot 10^{-11}$  \\
\hline
9 & 215.9882178816906 & - &  $1.3 \cdot 10^{-10}$  \\
\hline
10 & 315.3823355026930 & - &  $4.3 \cdot 10^{-10}$  \\
\hline
11 & 445.4531260569294 & - &  $8.0 \cdot 10^{-10}$  \\
\hline
12 & 611.9093290618418 & - &  $1.1 \cdot 10^{-9}$  \\
\hline
13 & 820.9446344509916 & - &  $2.3 \cdot 10^{-8}$  \\
\hline
14 & 1079.237672828109 & - & $7.4 \cdot 10^{-8}$   \\
\hline
\end{tabular}
\bigskip
\caption{Numerical values of ${\mathfrak c}_{1,n}^{(2)}$ and the estimated values of the
 coefficients $\alpha_{1,2}^{(n)}$ of the
 polynomial Ansatz (\ref{PAR}). $\Delta P_{rel}$ is the relative error measuring, how
 precise the polynomial description of the various coefficients.
\label{tcanka1k2}}
\end{center}
\end{table}
\normalsize


\begin{table}
\begin{center}
\begin{tabular}{|c||c|c|c|}
\hline
$n$ & ${\mathfrak c}_{1,n}^{(3)}$ & $\alpha_{1,3}^{(n)}$ &  $\Delta P_{rel}$ \\
\hline
1 & -0.006431714483032767 & 0 & 0   \\
\hline
2 & -0.6062295975751446 & -0.006431714483032767 & 0   \\
\hline
3 & -4.686060101968099 & -0.5933661686090790 & 0   \\
\hline
4 & -19.97352168384532 & -2.886666452691764 & 0   \\
\hline
5 & -62.52223158770824 & -4.840932003491656 & 0   \\
\hline
6 & -161.1215781979589 & -3.485086784826361 & 0   \\
\hline
7 & -362.6311389926113 & -0.9246655667562551 & $9.8 \cdot 10^{-11}$   \\
\hline
8 & -738.2397638237687 & - & $3.8 \cdot 10^{-11}$   \\
\hline
9 & -1390.649323143346 & - & $9.5 \cdot 10^{-10}$   \\
\hline
10 & -2462.183134098600 & - & $2.2 \cdot 10^{-9}$   \\
\hline
11 & -4143.819016121646 & - & $3.8 \cdot 10^{-9}$   \\
\hline
12 & -6685.146976936698 & - & $7.4 \cdot 10^{-9}$   \\
\hline
13 & -10405.25164211582 & - & $3.1 \cdot 10^{-8}$   \\
\hline
14 & -15704.51870008449 & - & $1.0 \cdot 10^{-7}$   \\
\hline
\end{tabular}
\bigskip
\caption{Numerical values of ${\mathfrak c}_{1,n}^{(3)}$ and the estimated values of the
 coefficients $\alpha_{1,3}^{(n)}$ of the
 polynomial Ansatz (\ref{PAR}). $\Delta P_{rel}$ is the relative error measuring, how
 precise the polynomial description of the various coefficients.
\label{tcanka1k3}}
\end{center}
\end{table}
\normalsize


\begin{table}
\begin{center}
\begin{tabular}{|c||c|c|c|}
\hline
$n$ & ${\mathfrak c}_{2,n}^{(1)}$ & $\alpha_{2,1}^{(n)}$ &  $\Delta P_{rel}$ \\
\hline
1 & -0.6227843696181658 & -0.6227843696181658 & 0   \\
\hline
2 & -1.793005777675029 & -1.170221408056863 & 0   \\
\hline
3 & -3.510664224170827 & -0.5474370384389350 & 0   \\
\hline
4 & -5.775759709104784 & - & $1.3 \cdot 10^{-13}$   \\
\hline
5 & -8.588292232476890 & - & $2.7 \cdot 10^{-13}$   \\
\hline
6 & -11.94826179428543 & - & $5.4 \cdot 10^{-13}$   \\
\hline
7 & -15.85566839449540 & - & $3.0 \cdot 10^{-12}$   \\
\hline
8 & -20.31051203322899 & - & $2.4 \cdot 10^{-13}$   \\
\hline
9 & -25.31279271044849 & - & $3.3 \cdot 10^{-12}$   \\
\hline
10 & -30.86251042620811 & - & $8.9 \cdot 10^{-12}$   \\
\hline
11 & -36.95966518654400 & - & $1.8 \cdot 10^{-10}$   \\
\hline
12 & -43.60425697643396 & - & $9.2 \cdot 10^{-11}$   \\
\hline
\end{tabular}
\bigskip
\caption{Numerical values of ${\mathfrak c}_{2,n}^{(1)}$ and the estimated values of the
 coefficients $\alpha_{2,1}^{(n)}$ of the polynomial Ansatz (\ref{PAR}). $\Delta P_{rel}$ is the relative error measuring, how
 precise the polynomial description of the various coefficients.
\label{tcanka2k1}}
\end{center}
\end{table}
\normalsize


\begin{table}
\begin{center}
\begin{tabular}{|c||c|c|c|}
\hline
$n$ & ${\mathfrak c}_{2,n}^{(2)}$ & $\alpha_{2,2}^{(n)}$ &  $\Delta P_{rel}$ \\
\hline
1 & 0.09585846497288947 & 0.09585846497288947 & 0   \\
\hline
2 & 1.021217323436306 & 0.9253588584634161 & 0   \\
\hline
3 & 4.167297736496755 & 2.220721554597033 & 0   \\
\hline
4 & 11.56177286163683 & 2.027673157482597 & 0   \\
\hline
5 & 25.87891733851396 & 0.6466014821748191 & 0   \\
\hline
6 & 50.43960728878038 & - & $3.5 \cdot 10^{-12}$   \\
\hline
7 & 89.21132031318082 & - & $4.3 \cdot 10^{-11}$   \\
\hline
8 & 146.8081355087541 & - & $6.8 \cdot 10^{-14}$   \\
\hline
9 & 228.4907334402296 & - & $3.3 \cdot 10^{-11}$   \\
\hline
10 & 340.1663961625466 & - & $6.9 \cdot 10^{-11}$   \\
\hline
11 & 488.3890078651079 & - & $1.4 \cdot 10^{-9}$   \\
\hline
12 & 680.3590519535390 & - & $6.5 \cdot 10^{-10}$   \\
\hline
\end{tabular}
\bigskip
\caption{Numerical values of ${\mathfrak c}_{2,n}^{(2)}$ and the estimated values of the
 coefficients $\alpha_{2,2}^{(n)}$ of the polynomial Ansatz (\ref{PAR}). $\Delta P_{rel}$ is the relative error measuring, how
 precise the polynomial description of the various coefficients.
\label{tcanka2k2}}
\end{center}
\end{table}
\normalsize


\begin{table}
\begin{center}
\begin{tabular}{|c||c|c|c|}
\hline
$n$ & $\mbox{Im}{\mathfrak c}_{3,n}^{(1)}$ & $\mbox{Im}\alpha_{3,1}^{(n)}$ &  $\Delta P_{rel}$ \\
\hline
0 & 5.524784188107441 & - & - \\
\hline
1 & 64.74543991933578 & -13.32477173346874 & 0   \\
\hline
2 & 196.8455216110168 & 78.07021165280452 & 0   \\
\hline
3 & 382.9754733415742 & 54.02987003887646 &  0  \\
\hline
4 & 623.1352951111397 & - & $2.1 \cdot 10^{-13}$   \\
\hline
5 & 917.3249869194190 & - & $1.1 \cdot 10^{-13}$   \\
\hline
6 & 1265.544548767033 & - & $4.2 \cdot 10^{-13}$   \\
\hline
7 & 1667.793980652089 & - & $2.9 \cdot 10^{-13}$   \\
\hline
8 & 2124.073282581139 & - & $1.7 \cdot 10^{-12}$   \\
\hline
9 & 2634.382454575155 & - &  $1.3 \cdot 10^{-11}$  \\
\hline
10 & 3198.721496788098 & - &  $7.6 \cdot 10^{-11}$  \\
\hline
11 & 3817.090409134617 & - &  $1.4 \cdot 10^{-10}$  \\
\hline
12 & 4489.489191747794 & - & $2.4 \cdot 10^{-10}$   \\
\hline
\end{tabular}
\bigskip
\caption{Numerical values of ${\mathfrak c}_{3,n}^{(1)}$ and the estimated values of the
 coefficients $\alpha_{3,1}^{(n)}$ of the polynomial Ansatz (\ref{PAR}). $\Delta P_{rel}$ is the relative error measuring, how
 precise the polynomial description of the various coefficients.
\label{tcanka3k1}}
\end{center}
\end{table}
\normalsize


\begin{table}
\begin{center}
\begin{tabular}{|c||c|c|c|}
\hline
$n$ & $\mbox{Im}{\mathfrak c}_{3,n}^{(2)}$ & $\mbox{Im}\alpha_{3,2}^{(n)}$ &  $\Delta P_{rel}$ \\
\hline 
0 & 5.325801411122541 & - & - \\
\hline
1 & -2.288843244285942 & 6.431972167002477 & 0   \\
\hline
2 & -118.7316609148375 & -8.720815411288419 & 0   \\
\hline
3 & -498.9700840592895 & -107.7220022592632 & 0   \\
\hline
4 & -1362.894724197043 & -156.0736032146372 & 0   \\
\hline
5 & -2994.213201152263 & -63.81700830476387 & 0   \\
\hline
6 & -5740.450143161579 & - & $1.9 \cdot 10^{-11}$   \\
\hline
7 & -10012.94718658865 & - & $2.5 \cdot 10^{-11}$   \\
\hline
8 & -16286.86297686484 & - & $6.9 \cdot 10^{-11}$   \\
\hline
9 & -25101.17316925544 & - & $2.0 \cdot 10^{-10}$   \\
\hline
10 & -37058.67044144858 & - & $7.6 \cdot 10^{-10}$   \\
\hline
11 & -52825.96445690946 & - & $1.2 \cdot 10^{-9}$   \\
\hline
12 & -73133.48191514628 & - & $1.8 \cdot 10^{-9}$   \\
\hline
\end{tabular}
\bigskip
\caption{Numerical values of ${\mathfrak c}_{3,n}^{(2)}$ and the estimated values of the
 coefficients $\alpha_{3,2}^{(n)}$ of the polynomial Ansatz (\ref{PAR}). $\Delta P_{rel}$ is the relative error measuring, how
 precise the polynomial description of the various coefficients.
\label{tcanka3k2}}
\end{center}
\end{table}
\normalsize


\begin{table}
\begin{center}
\begin{tabular}{|c||c|c|c|}
\hline
$n$ & $\mbox{Im}{\mathfrak c}_{4,n}^{(1)}$ & $\mbox{Im}\alpha_{4,1}^{(n)}$ &  $\Delta P_{rel}$ \\
\hline
1 & 7.173847962732688 & 9.268243065061945 & -   \\
\hline
2 & 28.16886240030072 & 18.90061933523877 & 0   \\
\hline
3 & 61.47744707921176 & 14.40796534367226 &  0  \\
\hline
4 & 109.1939971017951 & - & 0   \\
\hline
5 & 171.3185124681082 & - & $3.4 \cdot 10^{-13}$   \\
\hline
6 & 247.8509931781603 & - & $7.3 \cdot 10^{-13}$   \\
\hline
7 & 338.7914392317561 & - & $5.2 \cdot 10^{-13}$   \\
\hline
8 & 444.1398506280317 & - & $1.8 \cdot 10^{-12}$   \\
\hline
9 & 563.8962273730128 & - & $5.7 \cdot 10^{-12}$   \\
\hline
10 & 698.0605694634556 & - & $1.3 \cdot 10^{-11}$   \\
\hline
11 & 846.6328769665629 & - & $9.9 \cdot 10^{-11}$   \\
\hline
12 & 1009.613149902535 & - & $1.3 \cdot 10^{-10}$   \\
\hline
\end{tabular}
\bigskip
\caption{Numerical values of ${\mathfrak c}_{4,n}^{(1)}$ and the estimated values of the
 coefficients $\alpha_{4,1}^{(n)}$ of the polynomial Ansatz (\ref{PAR}). $\Delta P_{rel}$ is the relative error measuring, how
 precise the polynomial description of the various coefficients.
\label{tcanka4k1}}
\end{center}
\end{table}
\normalsize


\begin{table}
\begin{center}
\begin{tabular}{|c||c|c|c|}
\hline
$n$ & $\mbox{Im}{\mathfrak c}_{4,n}^{(2)}$ & $\mbox{Im}\alpha_{4,2}^{(n)}$ &  $\Delta P_{rel}$ \\
\hline
1 & -2.022552224999431 & -0.9871619756651118 & -   \\
\hline
2 & -9.130909582033406 & -8.143747606368294 & 0   \\
\hline
3 & -45.43588371906081 & -28.16122653065911 & 0   \\
\hline
4 & -149.3926321550942 & -39.49054776834688 & 0   \\
\hline
5 & -377.5095715450139 & -17.01786888653341 & 0   \\
\hline
6 & -803.3129874302336 & - & 0   \\
\hline
7 & -1517.347034225805 & - & $8.5 \cdot 10^{-12}$    \\
\hline
8 & -2627.173735175549 & - & $4.2 \cdot 10^{-11}$   \\
\hline
9 & -4257.372983040239 & - & $6.4 \cdot 10^{-11}$   \\
\hline
10 & -6549.542538591222 & - & $1.2 \cdot 10^{-10}$   \\
\hline
11 & -9662.298038790211 & - & $8.8 \cdot 10^{-10}$   \\
\hline
12 & -13771.27298617740 & - & $2.0 \cdot 10^{-9}$   \\
\hline
\end{tabular}
\bigskip
\caption{Numerical values of ${\mathfrak c}_{4,n}^{(2)}$ and the estimated values of the
 coefficients $\alpha_{4,2}^{(n)}$ of the polynomial Ansatz (\ref{PAR}). $\Delta P_{rel}$ is the relative error measuring, how
 precise the polynomial description of the various coefficients.
\label{tcanka4k2}}
\end{center}
\end{table}

\end{document}